\documentclass[preprint,12pt]{elsarticle}

\usepackage{caption}
\usepackage{amsmath}
\usepackage{amssymb}
\usepackage{amsthm}
\usepackage{mathrsfs}

\usepackage{subcaption}

\usepackage{booktabs} 
\usepackage{multirow} 

\usepackage{url}
\usepackage{hyperref}
\usepackage{xcolor}
\usepackage{cleveref} 

\biboptions{numbers,sort&compress}

\journal{Physics of Fluids}

\begin{document}

\begin{frontmatter}



\title{A three-dimensional unified gas-kinetic wave-particle solver for flow computation in all regimes}


\author[a]{Yipei Chen}
\author[c]{Yajun Zhu}
\author[a,b]{Kun Xu\corref{cor1}}

\cortext[cor1] {Corresponding author.}
\ead{makxu@ust.hk}

\address[a]{Department of Mathematics, Hong Kong University of Science and Technology, Hong Kong, China}
\address[b]{Department of Mechanical and Aerospace Engineering, Hong Kong University of Science and Technology, Hong Kong, China}
\address[c]{National Key Laboratory of Science and Technology on Aerodynamic Design and Research, Northwestern Polytechnical University, Xi’an, Shaanxi 710072, China}

\begin{abstract}
In this paper, the unified gas-kinetic wave-particle (UGKWP) method has been constructed on three-dimensional unstructured mesh with parallel computing for multiscale flow simulation.
Following the direct modeling methodology of the unified gas-kinetic scheme (UGKS), the UGKWP method models the flow dynamics uniformly in different regime and gets the local cell's Knudsen number dependent numerical solution directly without the requirement of kinetic scale cell resolution. The UGKWP method is composed of evolution of deterministic wave and stochastic particles. With the dynamic wave-particle decomposition, the UGKWP method is able to capture the continuum wave interaction and rarefied particle transport under a unified framework and achieves
 the high efficiency in different flow regime. The UGKWP flow solver is validated by many three-dimensional test cases of different Mach and Knudsen numbers, which include 3D shock tube problem, lid-driven cavity flow, high-speed flow passing through a cubic object, and hypersonic flow around a space vehicle. The parallel performance has been tested on the Tianhe-2 supercomputer, and reasonable parallel performance has been observed up to one thousand core processing. Due to wave-particle formulation, the UGKWP method has great potential in solving three-dimensional multiscale transport with the co-existence of continuum and rarefied flow regimes, especially for the high-speed rarefied and continuum flow around space vehicle in near space flight.
\end{abstract}


\end{frontmatter}


\section{Introduction}\label{sec:Introduction}
Non-equilibrium flow appears in a wide range of applications, such as the re-entry of spacecraft in upper planetary atmospheres, vacuum devices,
and fluid-structure interaction in Microelectromechanical systems. For example, for a vehicle in a near-space flight at Mach number $6$ and Reynolds number $5000$, the local Knudsen number defined by $Kn_{local} = l|\nabla \rho|/\rho$ with the mean free path $l$ can cover a wide range of values with five orders of magnitude difference\cite{jiang2019implicit}. To simulate such a multiscale problem, it requires numerical algorithm to capture both equilibrium and non-equilibrium flow in different regime, such as the hydrodynamic regime in the highly compressible leading edge, the whole transition regime across the vehicle surface, and the rarefied regime in the highly expanding trailing edge.

The well-known Euler and Navier-Stokes-Fourier (NSF) equations are hydrodynamic equations, which are valid for the continuum flow. They become inaccurate in the near continuum and transition regimes. On the other hand, the Boltzmann equation models the gas dynamics in the kinetic scale
of particle mean free path and collision time. The solution in all flow regimes can be obtained by solving the Boltzmann equation
under the kinetic scale resolution. However, the high-dimensionality of the equation, nonlinearity of collision term, and its integro-differential nature make the deterministic Boltzmann solver extremely expensive in memory requirement and computational cost. Instead, for practical high-speed non-equilibrium flow computation, the direct simulation Monte Carlo (DSMC) method \cite{bird1994molecular} becomes the main choice due to its high efficiency for solving the Boltzmann equation from the stochastic particle approach.
Similar to the modeling in the derivation of the Boltzmann equation, the separation of particle transport and collision in DSMC enforces  the numerical mesh size and time step to be less than the particle mean free path and collision time, i.e., the so-called kinetic scale of DSMC modeling.

Developed at Sandia National Laboratories, stochastic parallel rarefied-gas time-accurate analyzer (SPARTA)\cite{gallis2015stochastic} is an open source 2 \& 3D DSMC simulator optimized for exascale parallel computing and embed with both static and dynamic load balancing across processors. Particles in SPARTA advect through a hierarchical oct-tree based Cartesian grid that overlays the simulation box. Additionally, dsmcFoam\cite{scanlon2010open} and its upgrade release dsmcFoam+\cite{white2018dsmcfoam+} have been developed within the framework of OpenFOAM\cite{weller1998tensorial,greenshields2015openfoam,OpenFOAM}, which notably features with dynamic load balancing on arbitrary 2D/3D polyhedron mesh, molecular vibrational and electronic energy modes, chemical reactions and gravitational force. Other DSMC codes such as MONACO\cite{dietrich1996scalar}, SMILE\cite{ivanov1998statistical}, DAC\cite{lebeau1999parallel} with different mesh topologies and collision treatments can be found in the literature. Apart from stochastic solvers, the deterministic numerical scheme for Boltzmann and kinetic model equations with discrete velocity points have been extensively studied in the last several decades. Nesvetay-3D\cite{titarev2014construction} is an implicit solver on unstructured mesh developed by Titarev et al. with both physical and velocity space decomposed parallelization.
Recently, Zhu et al. has implemented discrete unified gas kinetic scheme (DUGKS)\cite{guo2013discrete,guo2015discrete} with the Shakhov collision model\cite{shakhov1968generalization} named dugksFoam\cite{zhu2017dugksfoam}. Unlike the traditional DVM method, DUGKS is a multiscale solver, and the time step is not restricted by the particle collision time due to the coupled treatment of particle transport and collision. Besides the traditional physical space decomposition parallel strategy, dugksFoam features a parallel computing ability based on the velocity space decomposition.

The recently developed unified gas-kinetic wave-particle (UGKWP) method is a multiscale method for all flow regimes\cite{liu2020unified,zhu2019unified}, and is used in other multiscale transport simulation as well, such as  photon transport\cite{li2020unified}. The UGKWP is constructed under the unified gas-kinetic scheme (UGKS) framework \cite{xu2010unified}. Instead of using discrete velocity method in UGKS, the UGKWP uses both hydrodynamic wave and stochastic particles to model the flow evolution, where a time and scale-dependent flux function is constructed through the coupled wave and particle transport across a cell interface in order to update both macroscopic flow variables and microscopic gas distribution function inside each control volume. Due to the adaptive wave-particle decomposition, the hydrodynamic equilibrium flow and the kinetic non-equilibrium particle free transport can be simulated efficiently by their separate
representations and their dynamical coupling  according to the local cell Knudsen number, i.e., $Kn_c=\tau/\Delta t$ with the particle relaxation time $\tau$ and numerical time step $\Delta t$. The UGKWP method has unified preserving property\cite{guo2019unified} to present the physical solution in all flow regime from the kinetic scale transport to the Navier-Stokes wave propagation without the constraint on the numerical cell size and time step being less than the particle mean free path and collision time.

The particle number in UGKWP is proportional to $\exp{(-1/Kn_c)}$, which is a function of the cell Knudsen number. In the continuum flow regime, the particle number will be significantly reduced due to the small cell Knudsen number. In the highly rarefied regime, similar to DSMC, the particle will play a dominant role in the flow evolution with the association of statistical noise. The steady-state solution can be obtained from the averaging of time-accurate evolution solution. Due to the decoupled treatment of particle transport and collision, DSMC requires the cell size to be a fraction of the particle mean free path and becomes very expensive in the transition and near continuum flow regime. In contrast, the UGKS and UGKWP have no such requirement with coupled particle transport and collision process in the gas evolution and flux construction\cite{xu2015direct}.
Moreover, the DSMC method handles the collision process by selecting particle collision pairs. In the low Knudsen number case, intensive collisions have to be dealt with, which make DSMC impractical in the near continuum flow simulation, such as the flying vehicle at an altitude below $80km$.
In contrast, the UGKWP method removes collisional particles and re-samples them from the updated macroscopic flow variables.
The modeling of collective effect of particles' collision within a time step in UGKWP doesn't require the time step and cell size to be less than the particle collision time and mean free path. As a result, the UGKWP is suitable for multiscale flow computation. Furthermore, with the implementation of particles in UGKWP, the ray effect\cite{zhu2019ray}, which is observed in DVM-type schemes in the highly non-equilibrium regime due to the inadequate numerical resolution in the particle velocity space, can be totally avoided. A series of 1D/2D test cases covering a wide range of Mach and Knudsen numbers have been conducted extensively to validate the scheme \cite{liu2020unified,zhu2019unified}. In this work, the UGKWP method will be extended to 3D unstructured mesh with parallel computing via spatial decomposition, and the code is potentially applicable to 3D flow simulation with arbitrary geometries in all flow regimes. The development from a 2D to a 3D code makes great effort to solve all problems related to the complex geometry, reconstruction, multidimensional flux from wave-particle decomposition, particle tracking, and parallelization.

The rest of the paper is organized as follows. In \cref{sec:Numerical_method}, the numerical procedure of the UGKWP method on unstructured mesh is presented. \Cref{sec:Code_framework} covers the construction of 3D UGKWP on parallel framework. In \cref{sec:Numerical_tests}, several numerical examples, including the 3D Sod shock tube inside a square-column, Lid-driven cubic cavity flow, and the high-speed flow around a cube and space vehicle, will be computed to demonstrate the performance of the current algorithm in multiscale flow simulation. Conclusion and further developments are given in the last section.

\section{3D Unified gas-kinetic wave-particle method}\label{sec:Numerical_method}
In this section, the unified gas-kinetic wave-particle (UGKWP) method will be introduced. In the classical kinetic theory, the Boltzmann equation reads
\begin{equation}
	\frac{\partial f}{\partial t} + \mathbf{u} \cdot \nabla_\mathbf{x} f = Q(f),
\end{equation}
where $f(\mathbf{x},\mathbf{u},t)$ is the gas distribution function, which depends on the particle velocity $\mathbf{u}\in\mathbb{R}^3$, physical space position $\mathbf{x}\in\mathbb{R}^3$, and time $t\in\mathbb{R}^+$. $Q(f)$ is the nonlinear Boltzmann collision operator. In many applications, the collision term is usually simplified by other relaxation-type collision models $S(f)$, such as Bhatnagar-Gross-Krook (BGK)\cite{bhatnagar1954model}, the ellipsoidal statistical BGK (ES-BGK)\cite{holway1966kinetic}, and the Shakhov model\cite{shakhov1968generalization}. In general, the kinetic model can be written as
\begin{equation}\label{eq:kinetic_model_eq}
	\frac{\partial f}{\partial t}+\mathbf{u}\cdot\nabla_\mathbf{x}f=S(f).
\end{equation}
Throughout this paper, the BGK relaxation model
\begin{equation}
	f_t+\mathbf{u}\cdot\nabla_\mathbf{x}f=\frac{g-f}{\tau}
\end{equation}
will be used to construct the UGKWP method. Here $\tau$ denotes the relaxation time, which is related the dynamic viscosity coefficient $\mu$ and the pressure $p$, i.e., $\tau=\mu/p$. The local equilibrium state $g$ is the Maxwellian distribution
\begin{equation}
	g=\rho\left(\frac{\lambda}{\pi}\right)^{\frac{3+K}{2}}\exp[-\lambda((\mathbf{u}-\mathbf{U})^2+\boldsymbol{\xi}^2)],
\end{equation}
with density $\rho$, macroscopic velocity $\mathbf{U}$, internal degree of freedom $K$, and the internal variable $\boldsymbol{\xi} = (\xi_1,\dotsc,\xi_K)$. $\lambda$ is related to the temperature $T$ by $\lambda=m/(2k_BT)=1/(2RT)$. Here, $m$ and $k_B$ represent the molecular mass and the Boltzmann constant, respectively. $R = k_B/m$ is the specific gas constant. Typically, the relaxation parameter in the kinetic model can be calculated through
\begin{equation}\label{eq:tau_via_T_ref}
	\tau = \frac{\mu}{p} = \frac{\mu_{ref}}{p}\left(\frac{T}{T_{ref}}\right)^\omega,
\end{equation}
where $\mu_{ref}, T_{ref}$ are the reference viscosity coefficient and temperature, and $\omega$ is power index, which related to Variable Hard Sphere (VHS) or Variable Soft Sphere (VSS) Models.

\subsection{Unified gas kinetic framework}\label{subsec:Unified_gas_kinetic_framework}
The unified scheme is direct modeling in the discretized space $\sum_i \Omega_i \subset \mathbb{R}^3$ and time $t^n\in\mathbb{R}^+$ \cite{xu2015direct}. The cell averaged conservative flow variables $\mathbf{W}_i=(\rho_i,(\rho\mathbf{U})_i,(\rho E)_i)$ on a physical cell $\Omega_i$ is defined as
\begin{equation}
	\mathbf{W}_i = \frac{1}{|\Omega_i|}\int\limits_{\Omega_i} \mathbf{W(\mathbf{x})} \mathrm{d}\mathbf{x},
\end{equation}
and the cell averaged distribution function $f_i$ on physical cell $\Omega_i$ is defined as
\begin{equation}
	f_i = \frac{1}{|\Omega_i|}\int\limits_{\Omega_i} f(\mathbf{x}) \mathrm{d}\mathbf{x}.
\end{equation}
In terms of conservative flow variables, from $t^n$ to $t^{n+1}$ on cell $\Omega_i$, the discretized conservation laws for $\mathbf{W}_i$ and $f_i$ are
\begin{equation}\label{eq:FVmacroEq}
	\mathbf{W}_i^{n+1} = \mathbf{W}_i^n - \frac{1}{|\Omega_i|}\sum\limits_{j\in N(i)} \mathbf{F}_{ij}|S_{ij}|,
\end{equation}
and
\begin{equation}\label{eq:FVmicroEq}
	f_i^{n+1} = f_i^n - \frac{1}{|\Omega_i|} \sum\limits_{j\in N(i)} \mathcal{F}_{ij} |S_{ij}| + \int_{t^n}^{t^{n+1}} S(f_i) \mathrm{d} t,
\end{equation}
where $N(i)$ denotes the set of the interface-adjacent neighboring cells of cell $i$, and cell $j$ is one of the neighbors. The interface between cells $i$ and $j$ is represented by the subscript $ij$. Hence, $|S_{ij}|$ and $\mathbf{n}_{ij}$ are referred to the area of the interface $ij$ and the unit normal vector of the interface $ij$ pointing from cell $i$ to cell $j$. $\mathbf{F}_{ij}$ and $\mathcal{F}_{ij}$ denotes the macroscopic and microscopic fluxes across the interface, respectively. $|\Omega_i|$ is the volume of cell $i$, $\Delta t = t^{n+1}-t^n$ denotes the discretized time step.

It should be noted that Eqs.\eqref{eq:FVmacroEq} and \eqref{eq:FVmicroEq} are the fundamental physical laws on the scale of mesh size and time step, which describe the conservations of macroscopic flow variables and microscopic gas distribution function.
The macroscopic conservative flow variables, their fluxes, and the flux for the particle transport are related to the  moments of the gas distribution function through
\begin{equation}
	\mathbf{W}_i = \int f_i \boldsymbol{\psi} \mathrm{d}\Xi,
\end{equation}
\begin{equation}
	\mathbf{F}_{ij} = \int_{t^n}^{t^{n+1}} \int \mathbf{u}\cdot\mathbf{n}_{ij} f_{ij}(t) \boldsymbol{\psi} \mathrm{d}\Xi \mathrm{d} t,
\end{equation}
and
\begin{equation}
\mathcal{F}_{ij} = \int_{t^n}^{t^{n+1}} \mathbf{u}\cdot\mathbf{n}_{ij} f_{ij}(t) \mathrm{d} t,
\end{equation}
where $f_{ij}(t)$ is the time-dependent distribution function on the cell interface, $\boldsymbol{\psi}=(1,\mathbf{u},\frac{1}{2}(\mathbf{u}^2+\boldsymbol{\xi}^2))$ is collision invariants, $\mathrm{d}\Xi=\mathrm{d}\mathbf{u}\mathrm{d}\boldsymbol{\xi}$, $\mathrm{d}\mathbf{u}=\mathrm{d}u\mathrm{d}v\mathrm{d}w$, and $\mathrm{d}\boldsymbol{\xi}=\mathrm{d}\xi_1 \mathrm{d}\xi_2 \dotsm \mathrm{d}\xi_K$. The BGK relaxation term satisfies the compatibility condition
\begin{equation}
	\int S(f) \boldsymbol{\psi} \mathrm{d}\Xi = \int \frac{g-f}{\tau} \boldsymbol{\psi} \mathrm{d}\Xi = \mathbf{0}
\end{equation}
for the mass, momentum, and energy conservations during the particle collision process.

The multiscale flow evolution in the unified algorithm relies on the construction of the flux function at the cell interfaces. The time-dependent gas distribution function $f_{ij}(t)$ couples particle free streaming and collision determines the flow physics in different regime, which is based on the integral solution of the BGK model
\begin{equation}\label{eq:integral_solution}
	f(\mathbf{x}_0,t) = \frac{1}{\tau} \int_{t_0}^t g(\mathbf{x}',t')e^{-(t-t')/\tau} \mathrm{d}t' + e^{-(t-t_0)/\tau} f_0(\mathbf{x}_0-\mathbf{u}(t-t_0)),
\end{equation}
where $\mathbf{x}_0$ is the point for the evaluation of the local gas distribution function, $\mathbf{x}'=\mathbf{x}_0 - \mathbf{u} (t-t')$ is the particle trajectory. Typically, $\mathbf{x}_0$ is denoted as $\mathbf{x}_{ij}$, the center of a cell interface for flux evaluation. $f_0(\mathbf{x})$ is the initial distribution function around $\mathbf{x}_0$ at the beginning of each step $t_0 = t^n$, and $g(\mathbf{x}, t)$ is the equilibrium state distributed around $\mathbf{x}_0$ and $t_0$. Specifically, for second-order accuracy, with transformation $t = t - t_0$, $\mathbf{x} = \mathbf{x}-\mathbf{x}_0$, the local expansions are
\begin{equation}
	g(\mathbf{x},t) = g_0 + g_\mathbf{x} \cdot \mathbf{x} + g_t t,
\end{equation}
and
\begin{equation}\label{eq:initial_distribution_function}
	f_0(\mathbf{x}) = f_0 + f_\mathbf{x} \cdot \mathbf{x}.
\end{equation}
The time-dependent distribution function at the center of cell interface $\mathbf{x}_{ij}$ can be constructed as
\begin{equation}\label{eq:time_dependent_distribution_function_at_the_cell_interface}
	f_{ij}(t) = \underbrace{c_1 g_0 + c_2 g_\mathbf{x}\cdot\mathbf{u} + c_3 g_t}_{f_{ij}^{eq}(t)} + 				\underbrace{c_4 f_0 + c_5 f_\mathbf{x}\cdot\mathbf{u}}_{f_{ij}^{fr}(t)},
\end{equation}
with the coefficients
\begin{equation}
	\begin{aligned}
		c_1 &= 1 - e^{-t/\tau}, \\
		c_2 &= te^{-t/\tau}-\tau(1-e^{-t/\tau}), \\
		c_3 &= t - \tau(1-e^{-t/\tau}), \\
		c_4 &= e^{-t/\tau}, \\
		c_5 &= -te^{-t/\tau}.
	\end{aligned}
\end{equation}
Note that $f_{ij}^{eq}(t)$ and $f_{ij}^{fr}(t)$ are the terms related to the evolution of the local equilibrium state $g(\mathbf{x},t)$ and the initial distribution function $f_0(\mathbf{x})$, respectively. The initial gas distribution function $f_0$ in Eq. \eqref{eq:time_dependent_distribution_function_at_the_cell_interface} is reconstructed from the updated gas distribution function at $t^n$, which has the form
\begin{equation}
	f_0(\mathbf{x}_{ij},t) =
		\begin{cases}
			f_i^n+(\nabla_{\mathbf{x}}f)_i^n\cdot(\mathbf{x}_{ij}-\mathbf{x}_{i})-\mathbf{u}\cdot(\nabla_{\mathbf{x}}f)_i^n (t-t_n), &\mathbf{n}_{ij}\cdot\mathbf{u}\geq0, \\
			f_j^n+(\nabla_{\mathbf{x}}f)_j^n\cdot(\mathbf{x}_{ij}-\mathbf{x}_{j})-\mathbf{u}\cdot(\nabla_{\mathbf{x}}f)_j^n (t-t_n), &\mathbf{n}_{ij}\cdot\mathbf{u} < 0,
		\end{cases}
\end{equation}
where $f_i^n$ and $f_j^n$ are the initial distribution functions at neighboring cells around the cell interface $ij$. Here $(\nabla_{\mathbf{x}}f)_i^n$ is the spatial gradient of the initial distribution function inside the cell $i$ and can be reconstructed via least square with Venkatakrishnan's limiter\cite{venkatakrishnan1995convergence} or Barth and Jespersen limiter\cite{barth1989design}.

The local equilibrium state $g_0$ in Eq. \eqref{eq:time_dependent_distribution_function_at_the_cell_interface} is computed from the compatibility condition
\begin{equation}
	\mathbf{W}_0 = \int g_0 \boldsymbol{\psi} \mathrm{d}\Xi = \int f_0 \boldsymbol{\psi} \mathrm{d}\Xi,
\end{equation}
and the spatial and temporal derivatives of the equilibriums state can be obtained through the micro-macro relationship
\begin{equation}
	\begin{aligned}
		\mathbf{W}_\mathbf{x} &= \int g_\mathbf{x} \boldsymbol{\psi} \mathrm{d}\Xi = \int f_\mathbf{x} \boldsymbol{\psi} \mathrm{d}\Xi, \\
		\mathbf{W}_t		  &= -\int \mathbf{u}\cdot g_\mathbf{x} \boldsymbol{\psi} \mathrm{d}\Xi.
	\end{aligned}
\end{equation}

\Cref{eq:integral_solution,eq:time_dependent_distribution_function_at_the_cell_interface} present a transition process from the initial non-equilibrium distribution function to the equilibrium one with the increment of particle collision. It shows an evolution process from the kinetic to the hydrodynamic scale, and the real solution depends on the local
parameter $\tau/\Delta t$, i.e., the local cell Knudsen number. Specifically, the integrated microscopic flux over a time step gives
\begin{equation}\label{eq:microscopic_fluxes_across_the_interface}
	\begin{aligned}
		\mathcal{F}_{ij} &= \int_0^{\Delta t} \mathbf{u}\cdot\mathbf{n}_{ij} f_{ij}(t) \mathrm{d}t \\
					 	 &= \underbrace{\mathbf{u}\cdot\mathbf{n}_{ij} (q_1 g_0 + q_2 g_\mathbf{x}\cdot\mathbf{u} + q_3 g_t)}_{\mathcal{F}_{ij}^{eq}} + \underbrace{\mathbf{u}\cdot\mathbf{n}_{ij} (q_4 f_0 + q_5 f_\mathbf{x}\cdot\mathbf{u})}_{\mathcal{F}_{ij}^{fr}},
	\end{aligned}
\end{equation}
where $\mathcal{F}_{ij}^{eq}$ and $\mathcal{F}_{ij}^{fr}$ are the equilibrium microscopic flux and the free transport microscopic flux, respectively. Similarly, the macroscopic fluxes for conservative variables are splitting into the equilibrium flux $\mathbf{F}_{ij}^{eq}$ and the free streaming flux $\mathbf{F}_{ij}^{fr}$
\begin{equation}\label{eq:macroscopic_fluxes_across_the_interface}
	\mathbf{F}_{ij} = \int_0^{\Delta t}\int \mathbf{u}\cdot\mathbf{n}_{ij} f_{ij}(t) \boldsymbol{\psi}\mathrm{d}\Xi\mathrm{d}t = \underbrace{\int \mathcal{F}_{ij}^{eq} \boldsymbol{\psi}\mathrm{d}\Xi}_{\mathbf{F}_{ij}^{eq}} + \underbrace{\int \mathcal{F}_{ij}^{fr} \boldsymbol{\psi}\mathrm{d}\Xi}_{\mathbf{F}_{ij}^{fr}},
\end{equation}
with the coefficients
\begin{equation}
	\begin{aligned}
		q_1 &= \Delta t - \tau(1-e^{-\Delta t/\tau}), \\
		q_2 &= 2\tau^2(1-e^{-\Delta t/\tau}) - \tau \Delta t - \tau \Delta t e^{-\Delta t/\tau}, \\
		q_3 &= \frac{\Delta t^2}{2} - \tau \Delta t + \tau^2(1-e^{-\Delta t/\tau}), \\
		q_4 &= \tau(1-e^{-\Delta t/\tau}), \\
		q_5 &= \tau\Delta t e^{-\Delta t/\tau} - \tau^2(1-e^{-\Delta t/\tau}).
	\end{aligned}
\end{equation}
With the variation of $\tau/\Delta t$, Eq. \eqref{eq:microscopic_fluxes_across_the_interface} and \eqref{eq:macroscopic_fluxes_across_the_interface} can provide multiscale flow evolution solution. When $\Delta t \gg \tau$, only the terms $\mathcal{F}_{ij}^{eq}$ with $q_1 \approx \Delta t$ and $q_3 \approx \Delta t^2/2$ are remained for equilibrium wave interaction; when $\Delta t \ll \tau$, $\mathcal{F}_{ij}^{fr}$  with $q_4 \approx \Delta t$ and $q_5 \approx −\Delta t^2/2$ are left for non-equilibrium particle free transport.

In deterministic UGKS \cite{xu2010unified}, the cell averaged distribution function $f_i$ is further discretized in the particle velocity space with discrete velocity points $\mathbf{u}_k$ to capture the non-equilibrium distribution function. Compared with many other DVM with separate particle free-streaming and collision, the mesh size and time step in UGKS are not limited by the particle mean free path and collision time due to their coupled evolution solution for the flux evaluation. Moreover, the NS solutions can be obtained automatically by UGKS in the continuum regime even with $\Delta t \gg \tau$, such as for the laminar boundary layer solution at high Reynolds number. For UGKWP, instead of discretizing the particle velocity space, the particle will be used directly to represent the non-equilibrium gas distribution function.

\subsection{Particle evolution in UGKWP}\label{subsec:Particle_evolution}
The integral solution of the kinetic model equation \eqref{eq:integral_solution} can be rewritten as
\begin{equation}\label{eq:multiscale_evolution_equation_for_simulation_particle}
	f(\mathbf{x},t) = (1-e^{-t/\tau})g_p(\mathbf{x},t) + e^{-t/\tau}f_0(\mathbf{x}-\mathbf{u}t),
\end{equation}
where
\begin{equation}
	g_p = g_0 + \left(\frac{te^{-t/\tau}}{1-e^{-t/\tau}}-\tau\right)\mathbf{u}\cdot\nabla_\mathbf{x}g + \left(\frac{t}{1-e^{-t/\tau}}-\tau\right) \partial_t g.
\end{equation}
\Cref{eq:multiscale_evolution_equation_for_simulation_particle} states that the distribution function at time $t$ is a combination of the initial distribution function $f_0$ and the modified equilibrium state $g$. The probability for the particle without suffering collision at time $t$ is $e^{-t/\tau}$. Otherwise, it will collide with other particle and the post-collision distribution is determined by the  distribution $g_p$. The cumulative distribution for particle free streaming at time $t_f$ is given by
\begin{equation}\label{eq:cumulative_distribution_function}
	\mathscr{F}(t) = (t_f \leq t) = e^{-t/\tau}.
\end{equation}
 A particle $P_k(m_k,\mathbf{x}_k,\mathbf{u}_k,e_k)$ can be represented by its mass $m_k$, position $\mathbf{x}_k$, velocity $\mathbf{u}_k$, and internal energy $e_k$. Its free transport time is
\begin{equation}\label{eq:sampling_of_free_transport_time}
	t_f = \min(-\tau\ln(\eta),\Delta t),
\end{equation}
where $\eta$ is a random number generated from a uniform distribution on the interval $(0,1)$, i.e., $\eta\sim U(0,1)$. Moreover, the location $\mathbf{x}^*$ of the particle free transport up to time $t_f$ can be accurately tracked,
\begin{equation}
	\mathbf{x}_k^* = \mathbf{x}_k^n + \mathbf{u}_k t_f,
\end{equation}
where the particle velocity $\mathbf{u}_k$ keeps the same value.

According to the time $t_f$ assigned to each particle, these particles with $t_f = \Delta t$ are called \textbf{collisionless particles} $P_f$, and the particles with $t_f < \Delta t$ are called \textbf{collisional particles} $P_c$. The collisional particles $P_{c,k}$ should be deleted at the collision time $t_f$ and re-sampled from distribution function $g_p$ with the updated macroscopic quantities $\mathbf{W}_i^{n+1}$ via sampling, i.e.,
\begin{equation}
	\begin{aligned}
		\mathbf{x}_k^{n+1} &\sim U(\Omega_i), \\
		\mathbf{u}_k^{n+1} &\sim g_p(\mathbf{W}_i^{n+1}).
	\end{aligned}
\end{equation}
The position of the re-sampled particle $\mathbf{x}_k^{n+1}$ is uniformly distributed inside the cell $\Omega_i$ where the collision happens. Similarly to the DSMC method, the internal energy $e_k^{n+1}$ is sampled according to the temperature and internal degree of freedom $K$. The particles mass $m_k$ can be prescribed and will be discussed later. The above scheme is the unified gas-kinetic particle (UGKP) method.

Theoretically, in the next time step, the re-sampled equilibrium particles will be reclassified into collisionless and collisional particles again according to the free transport time $t_f$, and only the collisionless particles will be retained at the end of the next time step. Since the collisional particle will disappear in the next time step, it is not necessary to re-sample it, and its dynamic impact, such as the contribution to the flux, can be calculated analytically. Therefore, in order to reduce the noise variance in near continuum regime and avoid re-sampling collisional particles repeatedly, only the collisionless particles in the \textbf{hydrodynamic wave} $\mathbf{W}_i^h = \mathbf{W}_i - \mathbf{W}_i^p$ need to be re-sampled at the beginning of each time step. This is the basic idea of the unified gas-kinetic wave-particle (UGKWP) method. Here, $\mathbf{W}_i^p$ is the total conservative quantities of collisionless particles remained in cell $\Omega_i$ at the end of each time step,
\begin{equation}\label{eq:total_conservative_quantities_of_collisionless_particles}
	\mathbf{W}_i^p =  \frac{1}{|\Omega_i|} \sum_{\mathbf{x}_k \in \Omega_i}\boldsymbol{\phi}_k,
\end{equation}
where the vector $\boldsymbol{\phi}_k = m_k(1,\mathbf{u}_k,\frac{1}{2}(\mathbf{u}_k^2+e_k))$ denotes the mass, momentum, and energy carried by the particle $P_k$.

Based on the cumulative distribution Eq.\eqref{eq:cumulative_distribution_function}, the proportion of the collisionless particles can be evaluated in each cell at the beginning of each time step. The total mass density of the re-sampled collisionless particle takes a portion of the updated hydrodynamic wave density $\rho^h$ from the previous time step
\begin{equation}
	\rho^{hp} = e^{-\Delta t/\tau} \rho^h.
\end{equation}
Based on this observation, the particle evolution procedure of the UGKWP method can be summarized as
\begin{enumerate}
	\item Obtain free streaming time $t_{f,k}$ for the remaining particles $P_{f,k}^{n}$.
	\item Sample the collisionless particles $P_{f,k}^{n}$ from hydrodynamic wave with distribution $g_p(\mathbf{W}^{n})$. Note that the collisionless particles with total mass density $\rho^{hp,n} = e^{-\Delta t/\tau} \rho^{h,n}$ have the free streaming time $t_f = \Delta t$.
	\item Stream all the particles and classified into two categories, i,e, collisionless particles $P_{f,k}^{n+1}$ and collisional particles $P_{c,k}^{*}$.
	\item Keep collisionless particles $P_{f,k}^{n+1}$, and remove collisional particles $P_{c,k}^{*}$. Calculate total conservative quantities of the remained collisionless particles $\mathbf{W}^{p,n+1}$ according to Eq. \eqref{eq:total_conservative_quantities_of_collisionless_particles}. The conservative quantities of collisional particles $\mathbf{W}^{h,n+1}$ are obtained from the updated total conservative quantities $\mathbf{W}^{n+1}$ in Eq.\eqref{eq:FVmacroEq} as $\mathbf{W}^{h,n+1} = \mathbf{W}^{n+1}-\mathbf{W}^{p,n+1}$. The detailed formulation for the update of $\mathbf{W}^{n+1}$ will be presented in the next subsection.
\end{enumerate}

\begin{figure}[htbp!]
    \centering
    \begin{subfigure}[b]{0.24\textwidth}
        \includegraphics[width=\textwidth]{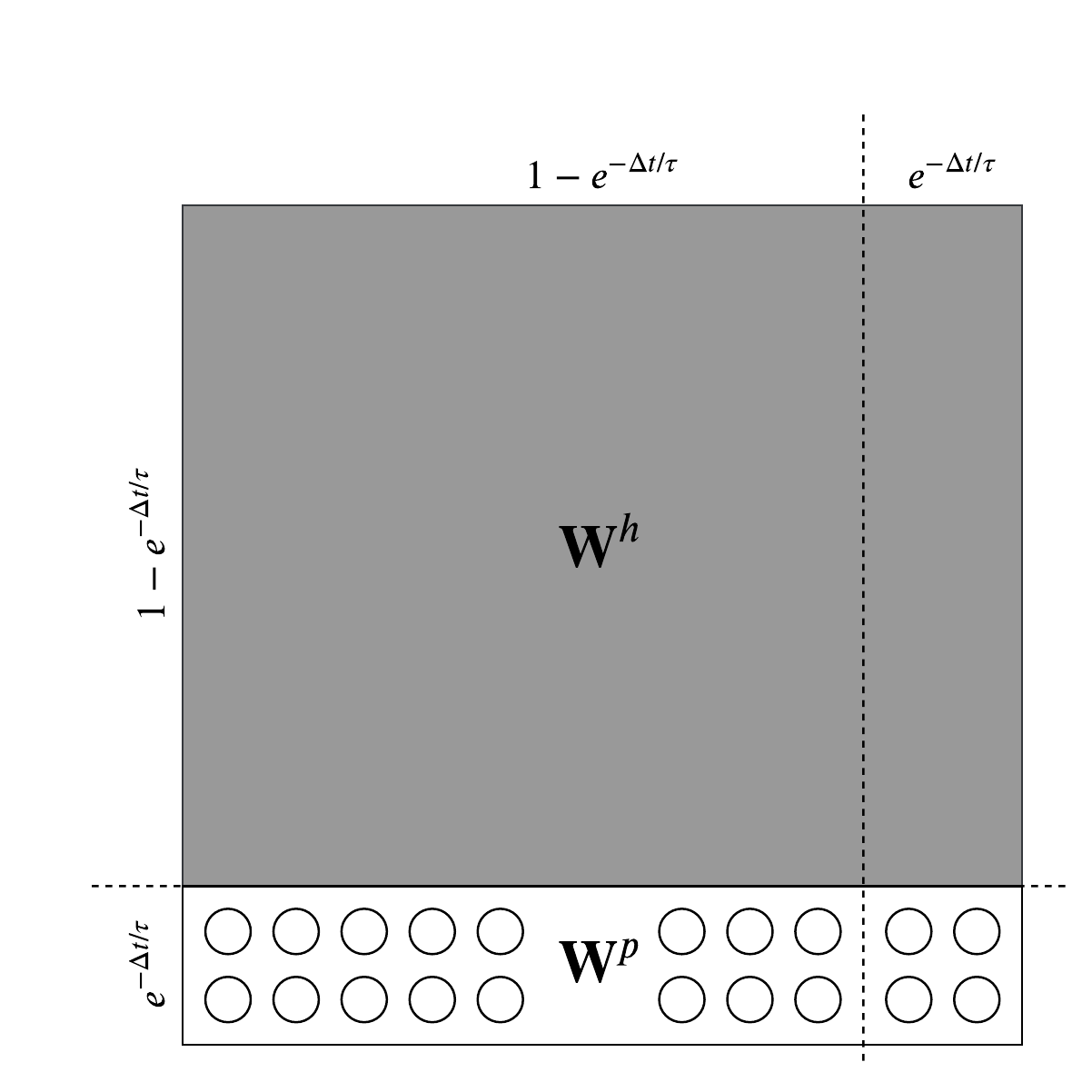}
        \caption{}
        \label{fig:evolution_of_waves_and_particles_a}
    \end{subfigure}
    \begin{subfigure}[b]{0.24\textwidth}
        \includegraphics[width=\textwidth]{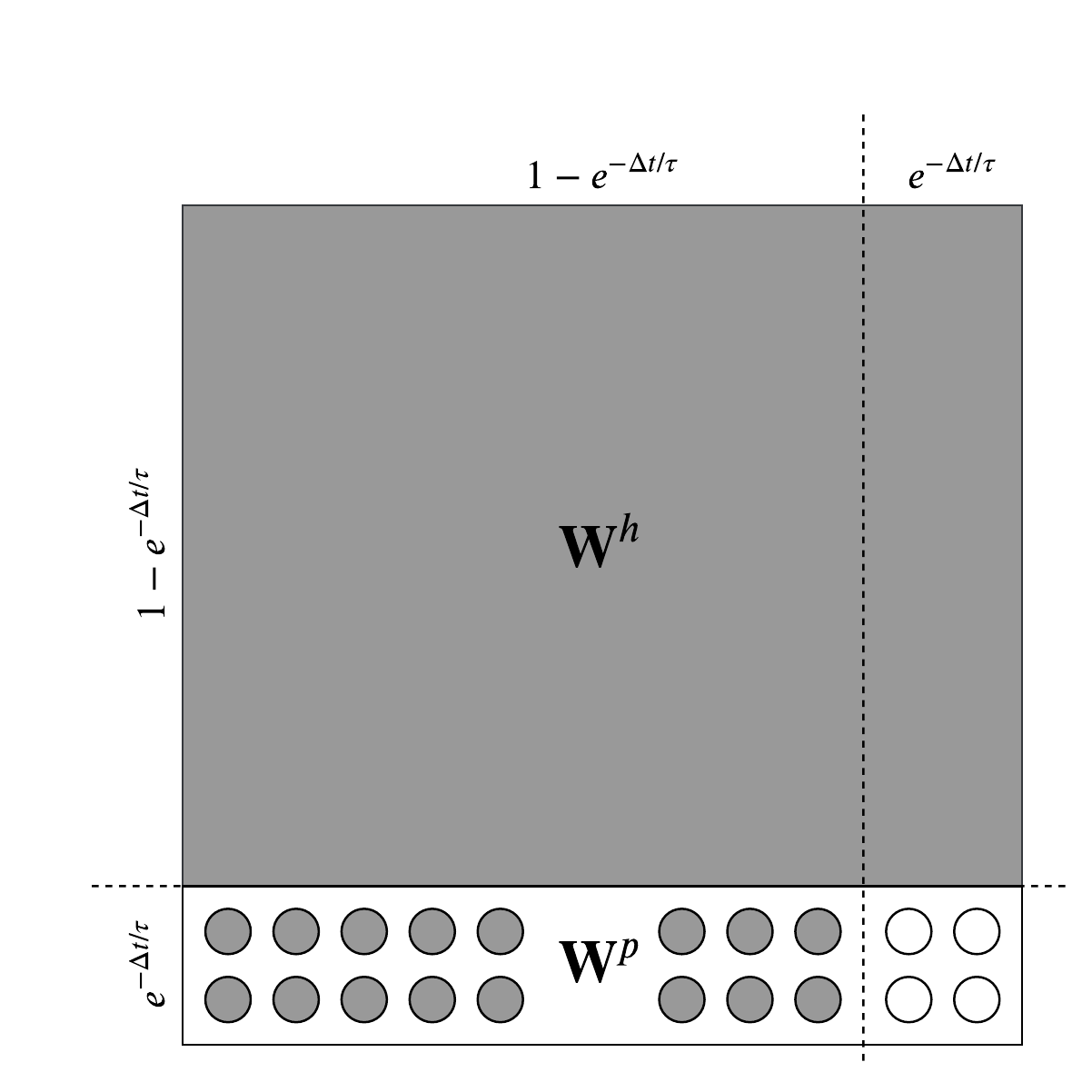}
		\caption{}
        \label{fig:evolution_of_waves_and_particles_b}
    \end{subfigure}
    \begin{subfigure}[b]{0.24\textwidth}
        \includegraphics[width=\textwidth]{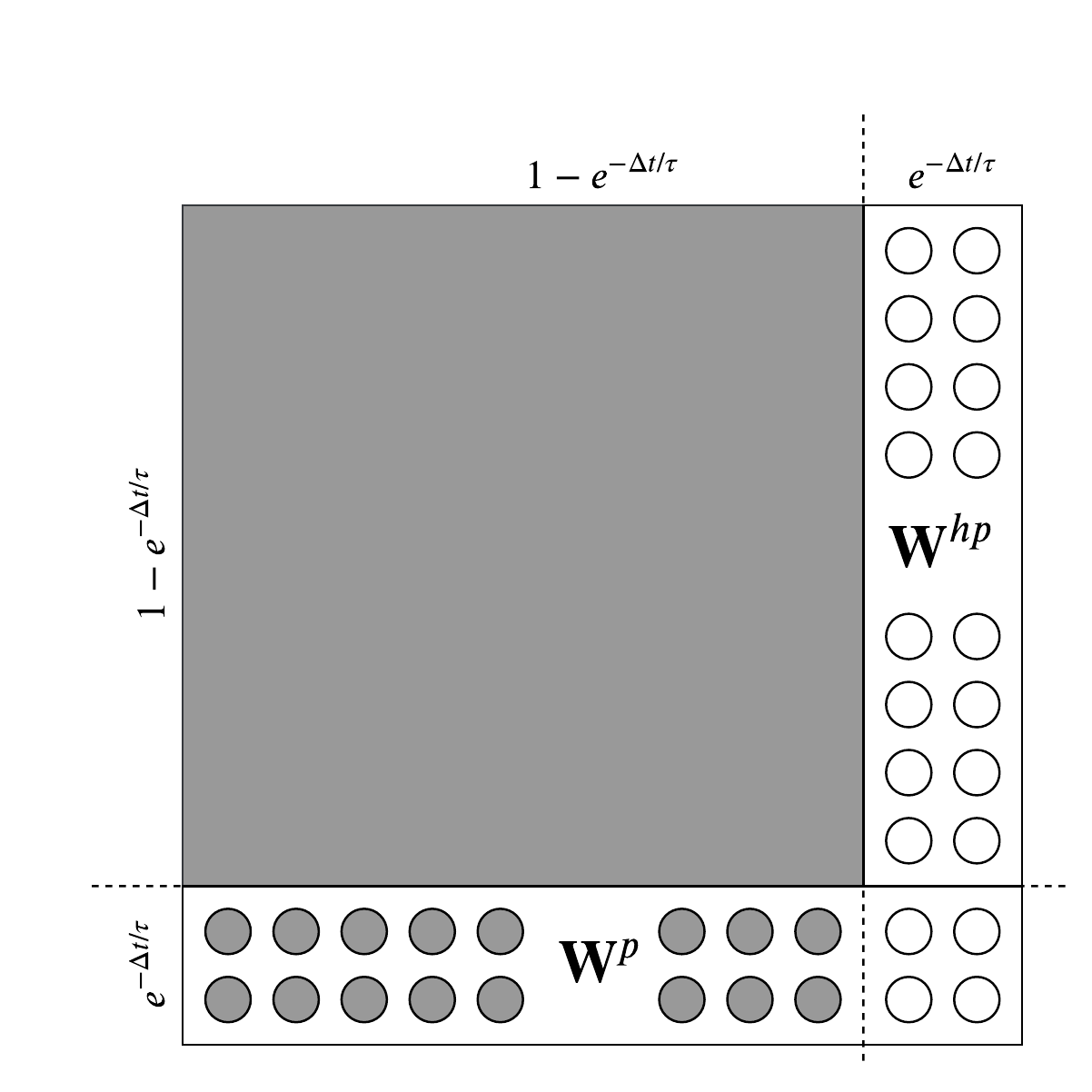}
        \caption{}
        \label{fig:evolution_of_waves_and_particles_c}
    \end{subfigure}
    \begin{subfigure}[b]{0.24\textwidth}
        \includegraphics[width=\textwidth]{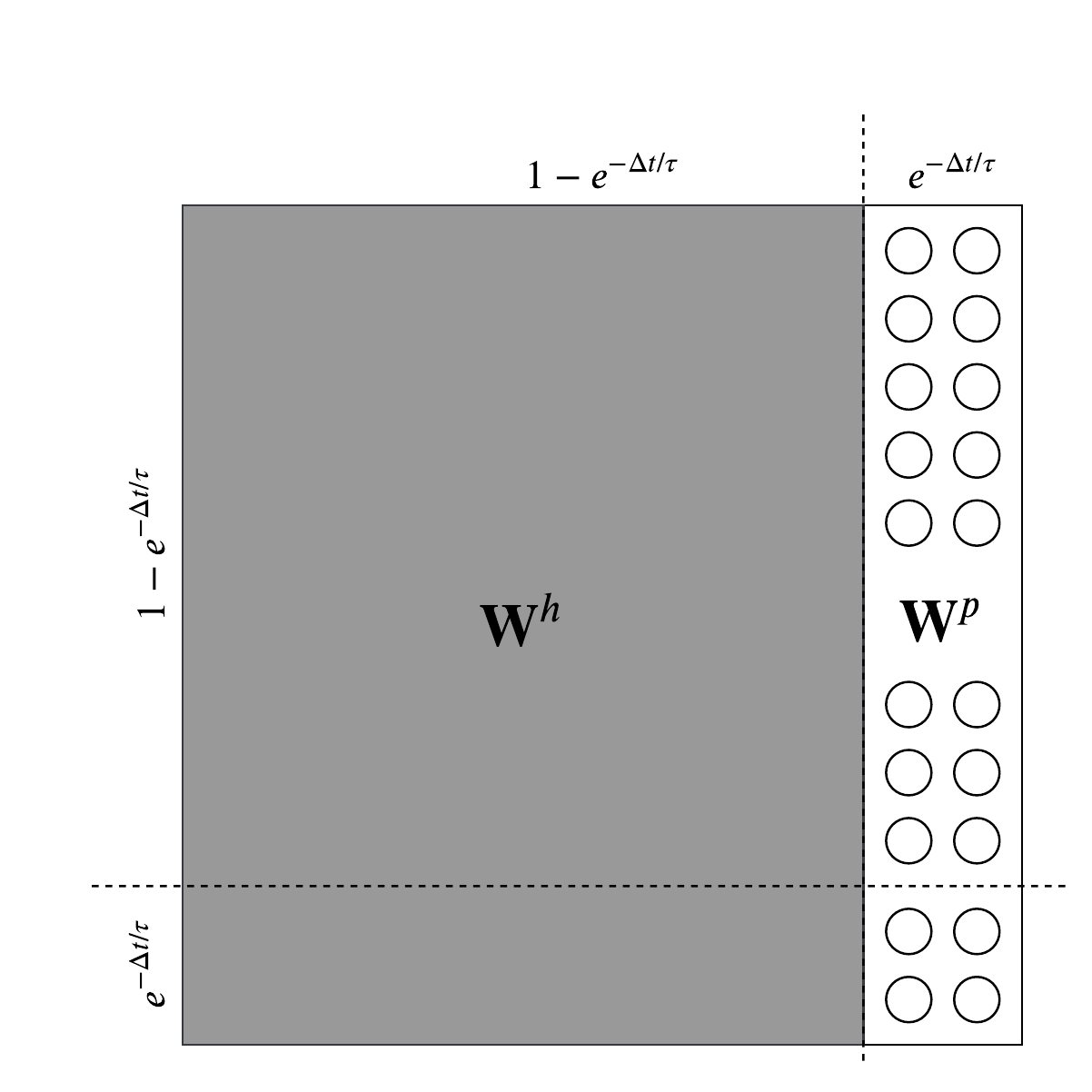}
        \caption{}
        \label{fig:evolution_of_waves_and_particles_d}
    \end{subfigure}
    \caption{The diagram illustrates the interplay of waves and particles in the UGKWP method. Grey block: waves, hollow circle: collisionless particles, solid circle: collisional particles. (a) Initial field; (b) Classification of the collisionless particles and collisional particles for the part of $\mathbf{W}^p$ according to the free transport time $t_f$; (c) Sample collisionless particles $\mathbf{W}^{hp}$ from hydrodynamic waves $\mathbf{W}^h$; (d) Update on both macroscopic and microscopic level.}\label{fig:evolution_of_waves_and_particles}
\end{figure}

The interplay of waves (collisional) and particles (collisionless) in the UGKWP method is illustrated through a series of figures in \cref{fig:evolution_of_waves_and_particles}. From the diagram, the multi-efficiency property of UGKWP\cite{liu2020unified} is clearly indicated, i.e., the computational efficiency of UGKWP goes to the high efficient approach in the corresponding regime. For example, in near continuum regime, i.e., $\tau\to 0$, the proportion of collisionless particle decreases exponentially. The UGKWP becomes a scheme without particles, and its computational cost is comparable to a traditional NS solver. On the other hand, for highly non-equilibrium hypersonic flow, such as $\tau \gg \Delta t$, the particles will play a dominant role to capture the non-equilibrium transport, and the efficiency of the scheme will go to the particle method, such as DSMC.

It has been shown in \cite{liu2020unified} that the UGKWP method is a kinetic equation solver in the rarefied regime and preserves the Navier-Stokes solution in the continuum regime with the particles re-sampled from the first-order approximation of $g_p$. Even though the particles are sampled uniformly inside the control volume, the spatial accuracy can be still kept in the near continuum regime, because the portion of particles $e^{-\Delta t/\tau}$ is minimal and the hydrodynamic wave evolution is dominant by the updated $\mathbf{W}$, which is computed analytically with second-order accuracy. The DSMC method requires $\Delta t$ to be less than the particle mean collision time, which is equivalent to $t_f = \Delta t$. The free transport time in UGKWP method is obtained from the sampling process in Eq. \eqref{eq:sampling_of_free_transport_time}, where the particle collisional effect, such as evolving to the equilibrium distribution $g_p$, has been modeled in the scheme through the evolution solution in \eqref{eq:multiscale_evolution_equation_for_simulation_particle}. Without using this evolution solution with time accumulating particle collision effect, or any other equivalent form, it is impossible to design a multiscale method, which can recover the NS solution in the continuum flow regime.

\subsection{Macroscopic variable update}\label{eq:Macroscopic_evolution}
The UGKWP updates the macroscopic variables for each control volume in Eq. \eqref{eq:FVmacroEq}. The equilibrium part flux $\mathbf{F}_{ij}^{eq}$ is directly calculated from the macroscopic flow field as given by Eq. \eqref{eq:macroscopic_fluxes_across_the_interface}. In UGKWP, the calculation of the free streaming flux $\mathbf{F}_{ij}^{fr}$ will be divided into two parts. The free streaming flux from collisional hydrodynamic waves of $(1-e^{-\Delta t/\tau})\mathbf{W}^h$ can be calculated analytically. The other free streaming flux from collisionless particles of hydrodynamic waves $e^{-\Delta t/\tau}\mathbf{W}^h$ and the remained particles $\mathbf{W}^p$ can be evaluated by counting the particles passing through the cell interface during a time step. The free streaming flux contributed from the collisional hydrodynamic waves of $(1-e^{-\Delta t/\tau})\mathbf{W}^h$ on the cell interface $ij$ is
\begin{equation}
	\begin{aligned}
		\mathbf{F}_{ij}^{fr,wave} &= \mathbf{F}_{ij}^{fr,UGKS}(\mathbf{W}^h) - \mathbf{F}_{ij}^{fr,DVM}(\mathbf{W}^{hp}) \\
								&= \int \mathbf{u}\cdot\mathbf{n}_{ij} \left[ (q_4 g_0^h + q_5 \mathbf{u}\cdot g_\mathbf{x}^h) - e^{-\Delta t/\tau}\int_0^{\Delta t} (g_0^h-t\mathbf{u}\cdot g_\mathbf{x}^h) \mathrm{d}t \right] \boldsymbol{\psi}\mathrm{d}\Xi \\
								&=\int \mathbf{u}\cdot\mathbf{n}_{ij} \left[(q_4-\Delta t e^{-\Delta t/\tau})g_0^h + (q_5+\frac{\Delta t^2}{2}e^{-\Delta t/\tau}) \mathbf{u}\cdot g_\mathbf{x}^h \right] \boldsymbol{\psi} \mathrm{d}\Xi,
	\end{aligned}
\end{equation}
where $g_0^h$ is the Maxwellian distribution with temperature and average velocity determined by total macroscopic variables $\mathbf{W}$, but the density by $\mathbf{W}^h$. $g^h_{\mathbf{x}}$ is the spatial derivative of the Maxwellian distribution, which can be obtained from the reconstruction of $\mathbf{W}$ and $\mathbf{W}^h$. The free streaming flux $\mathbf{F}_{ij}^{fr}$ in Eq. \eqref{eq:macroscopic_fluxes_across_the_interface}
is computed partially by particles and partially by the contribution of $g_0^h(\mathbf{W}^h)$ analytically. In addition, the subtraction of $\mathbf{F}_{ij}^{fr,DVM}(\mathbf{W}^{hp})$ from $\mathbf{F}_{ij}^{fr,UGKS}(\mathbf{W}^h)$ aims to remove the free transport fluxes which are still calculated  by the collisionless particle $P_f$ sampled from $\mathbf{W}^{hp}$. The total non-equilibrium free streaming flux $\mathbf{F}_{ij}^{fr}$ also includes the contribution from the remaining particles $P_k$ from the previous time step. During the free transport process, the contribution to the numerical fluxes of cell $i$ can be obtained by counting the particles across the cell interfaces,
\begin{equation}
	\mathbf{F}_i^{fr,p} = \sum_{\mathbf{x}_k^{n+1},\mathbf{x}_k^* \in \Omega_i}\boldsymbol{\phi}_k - \sum_{\mathbf{x}_k^n \in \Omega_i}\boldsymbol{\phi}_k.
\end{equation}
Finally, the updates of the conservative flow variables in the UGKWP method are
\begin{equation}\label{eq:update_macro_variables}
	\begin{aligned}		
	\mathbf{W}_i^{n+1} &= \mathbf{W}_i^n - \frac{1}{|\Omega_i|}\sum\limits_{j\in N(i)} \mathbf{F}_{ij}^{eq}|S_{ij}| - \frac{1}{|\Omega_i|}\sum\limits_{j\in N(i)} \mathbf{F}_{ij}^{fr}|S_{ij}|, \\
					   &= \mathbf{W}_i^n - \frac{1}{|\Omega_i|}\sum\limits_{j\in N(i)} \mathbf{F}_{ij}^{eq}|S_{ij}| - \frac{1}{|\Omega_i|}\sum\limits_{j\in N(i)} \mathbf{F}_{ij}^{fr,wave}|S_{ij}| + \frac{\mathbf{F}_{i}^{fr,p}}{|\Omega_i|}.
	\end{aligned}
\end{equation}

\subsection{Miscellaneous details}\label{subsec:Miscellaneous_details}

\noindent{(a) Time step on unstructured mesh}

Follow the implementation in \cite{vijayan19943d}, the time step for unsteady flow simulation is obtained from
\begin{equation}
	\Delta t = C \min_i \frac{\Omega_i}{\Lambda_i^x+\Lambda_i^y+\Lambda_i^z},
\end{equation}
with Courant number $C$ typically satisfied $0<C<1$ and convective spectral radii of cell $i$
\begin{equation}
	\begin{aligned}
		\Lambda_i^x &= (|U_i|+c)\Delta S_i^x,\\
		\Lambda_i^y &= (|V_i|+c)\Delta S_i^y,\\
		\Lambda_i^z &= (|W_i|+c)\Delta S_i^z,
	\end{aligned}
\end{equation}
where $c = 3\sigma_i = 3\sqrt{RT_i}$ is approximately the sound speed, $\mathbf{U}_i = (U_i, V_i, W_i)$ is the macroscopic velocity. The variables $\Delta S_i^x$, $\Delta S_i^y$, and $\Delta S_i^z$, respectively, represent projections of the control volume on the y-z-, x-z-, and x-y-plane, which are given by
\begin{equation}
	\begin{aligned}
		\Delta S_i^x &= \frac{1}{2} \sum_{j\in N(i)} |S_{ij}^x|,\\
		\Delta S_i^y &= \frac{1}{2} \sum_{j\in N(i)} |S_{ij}^y|,\\
		\Delta S_i^z &= \frac{1}{2} \sum_{j\in N(i)} |S_{ij}^z|,
	\end{aligned}
\end{equation}
where $S_{ij}^x$, $S_{ij}^y$, and $S_{ij}^z$ denote the x-, y-, and the z-component of the face vector $\mathbf{S}_{ij} = |S_{ij}|\mathbf{n}_{ij}$.

\vspace{1em}
\noindent{(b) Particle sampling}

At the beginning of each time step, the collisionless particles of hydrodynamic waves will be sampled in pairs from Maxwellian distribution function $g(\mathbf{W}^n)$. Specifically, given with the macroscopic velocity $\mathbf{U}=(U,V,W)$, temperature $T$, and vector $\mathbf{X}$ that sampled from the normal distribution, a pair of particles with microscopic velocities $\mathbf{u}=\mathbf{U}+\sqrt{RT}\mathbf{X}$ and $\mathbf{•}{u}'=\mathbf{U}-\sqrt{RT}\mathbf{X}$ will be sampled. To determine the sampling particle number $N_{sam}$ in the cell, a prescribed preference number $N_{ref}$ is required. Further, the reference mass $m_{ref}$ can be determined from the total particle mass and the reference number $N_{ref}$
\begin{equation}
	m_{ref} = \frac{(\rho^p+\rho^{hp})|\Omega|}{N_{ref}} = \frac{(\rho^p+e^{-\Delta t/\tau}\rho^h)|\Omega|}{N_{ref}}.
\end{equation}
Once the reference mass $m_{ref}$ is available, the number of particles to be sampled symmetrically is determined by
\begin{equation}
	N_{sam} = 2 \left\lceil{\frac{\rho^{hp}|\Omega|}{2 m_{ref}}}\right\rceil =2 \left\lceil{\frac{e^{-\Delta t/\tau}\rho^h|\Omega|}{2 m_{ref}}}\right\rceil.
\end{equation}
If the reference mass $m_{ref}$ is the same for all cells, then the total number of particles per cell would be exactly equal to $N_{ref}$. In this way, the total number of particles in each cell can be controlled around the given reference number $N_{ref}$ in near continuum regime regardless of mesh distribution. Moreover, the minimum number of particles $N_{min}$ per cell can be prescribed to adjust the sampled particles' number such that
\begin{equation}
	N_{sam} = \max\{N_{sam},N_{min}-N_{left}\},
\end{equation}
where $N_{left}$ is the collisionless particles left at the initial of each time step. Finally, the sampled mass weight $m_{sam}$ for each sampled particle is
\begin{equation}
	m_{sam} = \frac{\rho^{hp}|\Omega|}{N_{sam}} = \frac{e^{-\Delta t/\tau}\rho^h|\Omega|}{N_{sam}},
\end{equation}
which guarantees that the total sampled mass is exactly equal to $\rho^{hp}|\Omega|$.

\vspace{1em}
\noindent{(c) Time averaging}

For steady-state solution, the flow field $\mathbf{\bar{W}}$ starts to be averaged after a given time step $N_{avg}$,
\begin{equation}
	\mathbf{\bar{W}} = \frac{\sum\limits_{n>N_{avg}} \Delta t^n\mathbf{W}^n}{\sum\limits_{n>N_{avg}} \Delta t^n}
\end{equation}
where $\Delta t^n = t^n-t^{n-1}$.
The averaged flow field $\mathbf{\bar{W}}$ is assumed to be convergent if the relative change in two-successive steps is less than a given tolerance, such as $\varepsilon = 10^{-8}$. Then, the flow variables, such as the temperature $\bar{T}$ and macroscopic velocity $\mathbf{\bar{U}}$, can be obtained from the averaged conservative flow variables $\mathbf{\bar{W}}$.

\vspace{1em}
\noindent{(d) Numerical dissipation}

The UGKWP targets the continuum and rarefied flow. In the continuum flow regime, the strong shock structure is usually unresolved by the mesh size. Therefore, numerical dissipation is added through relaxation time to enlarge the shock thickness to the mesh size scale,
\begin{equation}
	\tau_{num} = \frac{\mu}{P}+C_2 \frac{|P_l-P_r|}{|P_l+P_r|}\Delta t,
\end{equation}
where $P_l$ and $P_r$ are the reconstructed pressures at the left and right side of the cell interface and
$C_2$ is a constant, such as $C_2 = 10$ for strong shock in the continuum regime.

\vspace{1em}
\noindent{(e) Boundary condition}

The proper treatment of boundary condition is crucial for a numerical scheme. For a diffusive wall condition with normal direction $\mathbf{n}$ pointing toward the computational domain, the incoming distribution function $f_{in}(t)$ at boundary is given by Eq.\eqref{eq:time_dependent_distribution_function_at_the_cell_interface}. The distribution function of emitted particles from the wall has a Maxwellian distribution
\begin{equation}
	g_w = \rho_w\left(\frac{1}{2\pi RT_w}\right)^{\frac{3+K}{2}}\exp\left[-\frac{(\mathbf{u}-\mathbf{U}_w)^2+\boldsymbol{\xi}^2}{2RT_w}\right],
\end{equation}
where $T_w$ and $\mathbf{U}_w$ are prescribed wall temperature and velocity. Based on the non-penetration condition, $\rho_w$ in the above Maxwellian is given by
\begin{equation}
\int_0^{\Delta t} \int_{\mathbf{n}\cdot(\mathbf{u}-\mathbf{U}_w)<0} \mathbf{n}\cdot(\mathbf{u}-\mathbf{U}_w) f_{in}(t) \mathrm{d}\mathbf{u} \mathrm{d} t = \Delta t \int_{\mathbf{n}\cdot(\mathbf{u}-\mathbf{U}_w)\geq 0} \mathbf{n}\cdot(\mathbf{u}-\mathbf{U}_w) g_w \mathrm{d}\mathbf{u}.
\end{equation}

\section{3D UGKWP code and parallelization}\label{sec:Code_framework}
UGKWP solver is constructed under a finite volume framework on 3D unstructured mesh. It includes not only the reconstruction and flux evaluation module as in the traditional finite volume solver, but also the particle sampling and tracking module as in pure particle method.

\subsection{Structure of the solver}\label{subsec:Structure of the solver}
\begin{figure}
	\centering
	\includegraphics[width=0.75\textwidth]{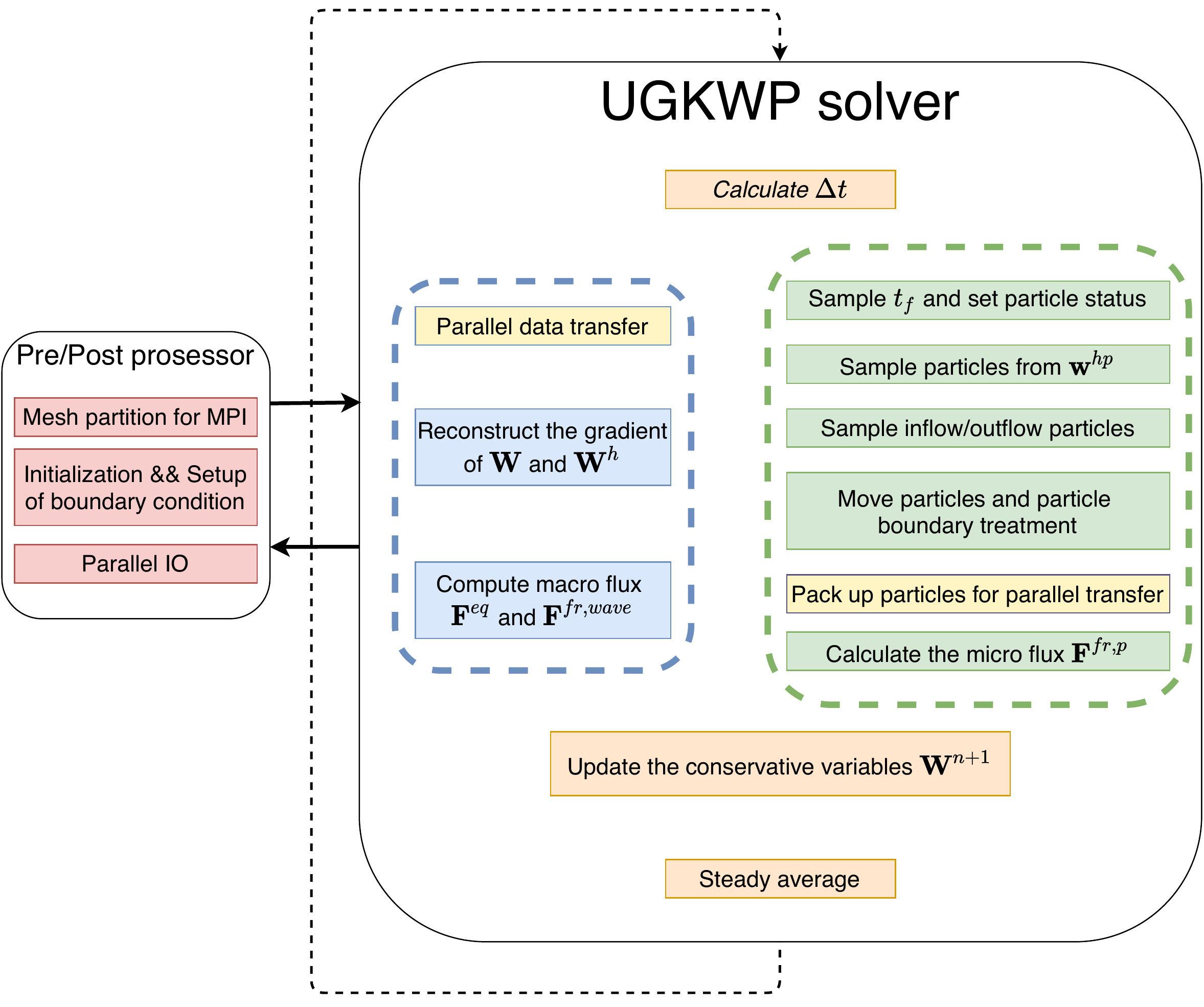}
    \caption{The structure and main components of the UGKWP solver}
    \label{fig:structure_ugkwp3D}
\end{figure}

The main components of UGKWP solver are sketched in \cref{fig:structure_ugkwp3D}. As shown in the diagram, the program starts with the pre- and post-processor module, where the mesh partition, initialization, setup of boundary condition, and parallel IO are handled inside. In UGKWP solver, the numerical procedures are organized into a macroscopic field level and a microscopic particle level. Accordingly, the macroscopic components surrounded by the blue dash line consist of the parallel data transfer, reconstruction of the macroscopic gradient, and macroscopic flux calculation with boundary treatment. The green dash block contains the components for microscopic particles which are stored in the doubly-linked list.
There are frequent operations, such as tracking particles and calculating the macroscopic fluxes. The insert/delete operation is efficient in the scenario of parallel transfer and sampling/elimination of the particles.

\subsection{Parallelization}\label{subsec:Parallelization}
The parallelization of the current code adopts Message Passing Interface (MPI) based on the physical mesh decomposition. Every MPI process deals with a non-overlapping sub-domain, and the information like conservative variables and particles are communicated with the neighboring domain through corresponding boundaries. Since the macroscopic solver has second-order accuracy, no padding area of sud-domain is required.

The code has been tested on Tianhe-2, located in the National Supercomputer Center in Guangzhou, China. Tianhe-2 contains 16,000 nodes that each of which possesses one Intel Xeon E5-2692 12 Cores @2.2 GHz CPU and 88 gigabytes of memory (64 used by the Ivy Bridge processors, and 8 gigabytes for each of the Xeon Phi processors). The computing nodes of Tianhe-2 are interconnected by TH Express-2 network.

To test the parallel efficiency and scalability of the UGKWP, the code is compiled using Intel C/C++ compiler of version 18.0.0 with -O3 optimization flag, and are linked to the MPICH2 with a customized GLEX channel. Since the scaling is problem-specific (depending on Knudsen number and preference number of particles per cells $N_{ref}$), multiple test cases and several factors affecting the performance will be analyzed in a series of three-dimensional lid-driven cavity flow tests.

\subsubsection{Macroscopic field computation}
Firstly, only the parallel speedup of pure macroscopic field computation using different MPI processes is measured to eliminate the computation of particles and communication of particle parallel transfer. Without the involvement of particle generation and particle transportation, the UGKWP degenerates to the gas-kinetic scheme (GKS) for the continuum flow computation \cite{prendergast1993numerical}. As it becomes a deterministic solver, for simplicity, the Knudsen number is fixed at $10^{-4}$ in the following parallel computation. The averaged running time (wall clock time) of a single iteration step is measured, and the measurement is ensured to be over $100$ seconds, and no IO time is counted.

To investigate the Amdahl’s law (strong scaling) at different fixed problem size, the physical domain is discretized as $D^3$, where $D$ is the number of cells along with each direction with the values $64, 128, 256$ separately. To test the Gustafson’s law (weak scaling), we concern the speedup for a scaled problem size to the number of processors. Hence, $D=64$ on one node with 24 cores is chosen as the baseline, i.e., keeping the number of cells per processors as $64^3/24$, and the grid size increased simultaneously as the increment of the number of processors $P$. The corresponding speedup is measured based on the averaged single-node simulation time, i.e., $S_P = 24T_p/T_{24}$.

\begin{figure}
	\centering
	\includegraphics[width=0.5\textwidth]{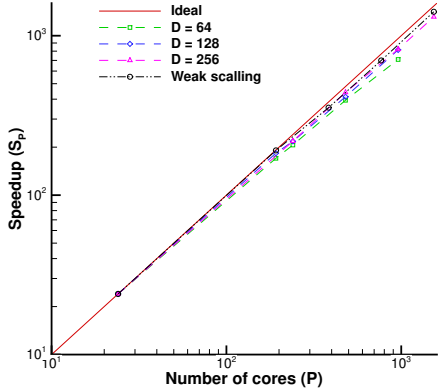}
    \caption{Strong and weak scaling analysis without the involvement of particles}
    \label{fig:parallel_speedup_noP}
\end{figure}

Both strong and weak scaling analysis is plotted in \cref{fig:parallel_speedup_noP}. The solid red line represents the ideal linear speedup. From the diagram, the overall computational time of the various physical grid sizes scales well with the number of processing cores (or MPI processes). Although strong scaling is very sensitive towards the serial fraction of the program and the communication overhead (e.g., synchronization) could further degrade performance, the worst efficiency still has $73.8\%$ for the case $D = 64$ with $P = 960$ number of processors. Moreover, it is observed that strong scaling performance increases considerably as the increase of grid size. The strong scaling parallel efficiency for the largest problem size $D = 256$ can reach $85.4\%$ despite the usage of $P = 1536$ number of processing cores. Finally, the weak scaling is also verified by increasing both the job size and the number of processing cores, and a satisfactory weak scaling efficiency up to $92\%$ has been achieved even with $P = 1536$ number of processing cores.

\subsubsection{Involvement of microscopic particles}
Next, the particles are included in the experiment to explore the scalability and parallel efficiency of the implementation. Problems with different numbers of cells $D^3$, Knudsen number $Kn$, and preference number of particles per cell $N_{ref}$ are run across different nodes. The maximum problem size is limited by the total memory available on each node, and no IO time is recorded. The average computing time (wall clock time) for a single iteration step is averaged from $1001$ to $1100$ steps to ensure sufficient running time and reach a steady-state solution.

\begin{figure}[htbp!]
    \centering
    \begin{subfigure}[b]{0.49\textwidth}
        \includegraphics[width=\textwidth]{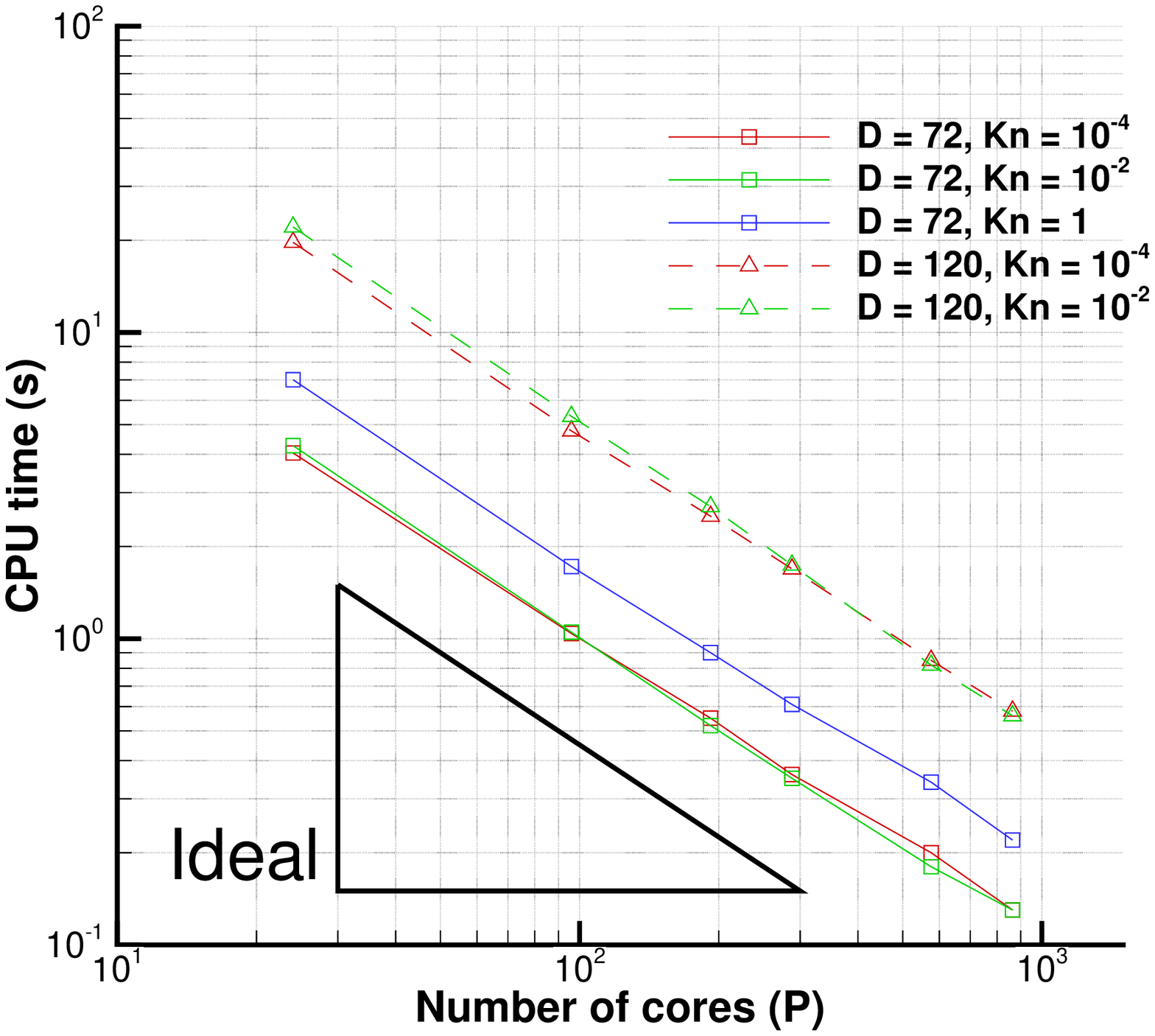}
        \caption{}
        \label{fig:parallel_CPU_time_50P}
    \end{subfigure}
    \begin{subfigure}[b]{0.49\textwidth}
        \includegraphics[width=\textwidth]{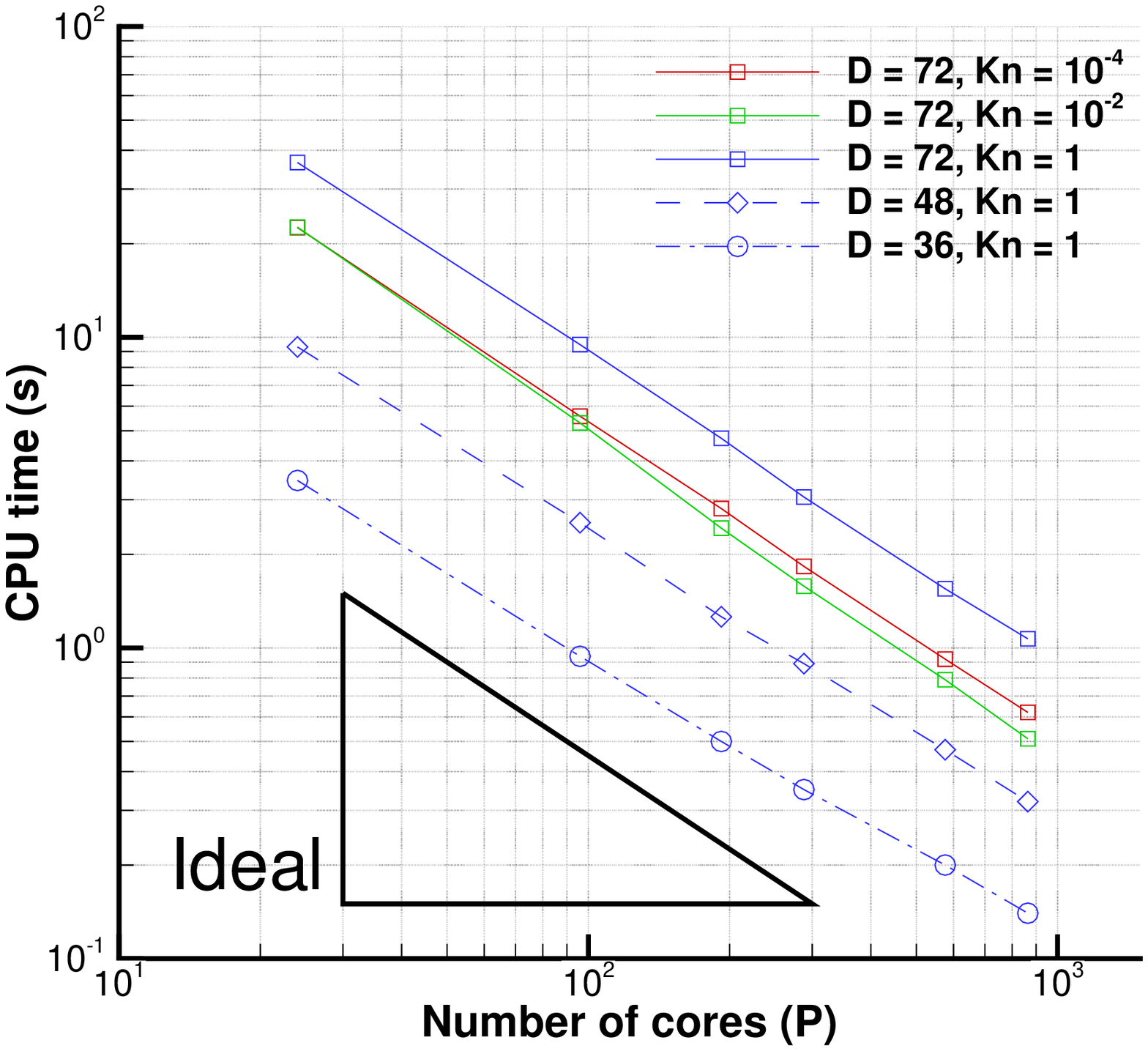}
        \caption{}
        \label{fig:parallel_CPU_time_500P}
    \end{subfigure}
    \caption{Performance scaling on Tianhe-2 where each node has 24 cores (a) $N_{ref} = 50$ and (b) $N_{ref} = 500$}\label{fig:parallel_CPU_time_P}
\end{figure}

The averaged CPU time (second) per step against the number of cores is shown in \cref{fig:parallel_CPU_time_P}. The results indicate that the scaling performance is good for all simulations because the CPU time decreases almost linearly as the increase in the number of processing cores. Actually, from \cref{tab:parallel_efficiency_50P} and \cref{tab:parallel_efficiency_500P}, the parallel efficiency $E_P = S_P/P$ for cases with $D \geq 72 $ is over $86\%$ even with $P = 864$ number of processing cores.

Furthermore, several interesting patterns have been observed. First, for the cases of both $N_{ref} = 50$ and $N_{ref} = 500$, we can observe that the absolute CPU time raises not only as of the increment of grid size but also the Knudsen number. The reason is that the mean free path of the particles becomes large at a high Knudsen number, which produces the unbalanced distribution of particles among different sub-domains as well as in the processing cores. Secondly, the scaling performance deteriorates as the shrinkage of grid size, especially in high-Knudsen number cases, because the proportion of communication time would increase as the number of cells per core declined, and the uneven effect of particles distributed among processing cores would also be amplified. Accordingly, the worst parallel efficiency $E_{864} = 68.7\%$ is observed for the case $D = 36$ at $Kn = 1$ with respect to $P = 864$ number of cores. Lastly, the parallel efficiency would increase as the reference number of particles becomes larger.

Another interesting phenomenon is that maximum parallel efficiency can be greater than one. For instance, the maximum parallel efficiency observed is $E_{864} = 123.2\%$ and achieves at the cases of $D = 72, Kn = 10^{-2}$ and $ N_{ref} = 500$ using $P = 864$ cores. Actually, the parallel efficiency in transition regime $Kn = 10^{-2}$ is even higher than that in the near continuum regime $Kn = 10^{-4}$ with the same grid size $D$ and reference number of particles $N_{ref}$. Besides, the worst parallel efficiency is even larger than $86.3\%$ for all cases at $Kn = 10^{-4}, 10^{-2}$. This counterintuitive parallel efficiency in the transition regime is probably due to the doubly-linked list data structure for storing particles. In the transition or near continuum regime, the collision between particles is intensive, and the frequent elimination/resampling of particles involves frequent delete/insert operation in memory. Nonetheless, the bottleneck caused by the implemented data structure can be alleviated through the replacement of a sequence container, like the STL vector.

In summary, intensive parallel tests of cavity flow show satisfactory strong and weak scaling performance of the UGKWP code.
The current implementation becomes a valuable tool for simulating complex flow problems across thousands of processing cores in parallel computation. The actual parallel efficiency might vary for particular simulation setup.

\begin{table}[htbp!]
\caption{Parallel efficiency for cases with $N_{ref} = 50$ on Tianhe-2}
\label{tab:parallel_efficiency_50P}
\resizebox{\columnwidth}{!}{
\begin{tabular}{@{}ccccccc@{}}
\toprule
\multirow{2}{*}{Nodes} & \multirow{2}{*}{Cores} & \multicolumn{3}{c}{$D = 72$}                & \multicolumn{2}{c}{D = 120}     \\ \cmidrule(l){3-7}
                       &                        & $Kn = 10^{-4}$ & $Kn = 10^{-2}$ & $Kn = 1$  & $Kn = 10^{-4}$ & $Kn = 10^{-2}$ \\ \midrule
1                      & 24                     & $100\%$        & $100\%$        & $100\%$   & $100\%$        & $100\%$        \\
4                      & 96                     & $97.1\%$       & $101.7\%$      & $101.9\%$ & $103.2\%$      & $103.9\%$      \\
8                      & 192                    & $91.8\%$       & $102.6\%$      & $97.4\%$  & $98.1\%$       & $102.3\%$      \\
12                     & 288                    & $93.5\%$       & $101.7\%$      & $95.8\%$  & $97.1\%$       & $105.8\%$      \\
24                     & 576                    & $84.2\%$       & $98.8\%$       & $85.9\%$  & $96.6\%$       & $112.3\%$      \\
36                     & 864                    & $86.3\%$       & $91.2\%$       & $88.5\%$  & $94.3\%$       & $109.6\%$      \\ \bottomrule
\end{tabular}
}
\end{table}

\begin{table}[htbp!]
\caption{Parallel efficiency for cases with $N_{ref} = 500$ on Tianhe-2}
\label{tab:parallel_efficiency_500P}
\begin{tabular}{@{}ccccccc@{}}
\toprule
\multirow{2}{*}{Nodes} & \multirow{2}{*}{Cores} & \multicolumn{3}{c}{$D = 72$}               & $D = 48$ & $D = 36$ \\ \cmidrule(l){3-7}
                       &                        & $Kn = 10^{-4}$ & $Kn = 10^{-2}$ & $Kn = 1$ & $Kn = 1$ & $Kn = 1$ \\ \midrule
1                      & 24                     & $100\%$        & $100\%$        & $100\%$  & $100\%$  & $100\%$  \\
4                      & 96                     & $101.2\%$      & $106.7\%$      & $96.4\%$ & $92\%$   & $92\%$   \\
8                      & 192                    & $100.3\%$      & $116.3\%$      & $96.5\%$ & $92.4\%$ & $86.5\%$ \\
12                     & 288                    & $102.7\%$      & $119.3\%$      & $99.5\%$ & $87.2\%$ & $82.4\%$ \\
24                     & 576                    & $102.1\%$      & $119.3\%$      & $98.2\%$ & $82.5\%$ & $72.1\%$ \\
36                     & 864                    & $101\%$        & $123.2\%$      & $94.8\%$ & $80.8\%$ & $68.7\%$ \\ \bottomrule
\end{tabular}
\end{table}

\section{Numerical examples}\label{sec:Numerical_tests}
In this section, the accuracy and computational efficiency of the UGKWP solver will be evaluated through many test cases with a wide range of Knudsen and Mach numbers. The numerical Sod shock tube problem in 3D, lid-driven cubic cavity flow, high-speed flow passing through a cube, and the flow around a space vehicle, are tested. The results are compared with those from UGKS/DUGKS and DSMC. Without a special statement, the diffusive boundary condition with full accommodation is applied for the isothermal walls. The code is compiled with GCC version 7.5.0, and all computations are carried out on a workstation with [Dual CPU] Intel\textregistered Xeon(R) Platinum 8168 @ 2.70GHz with 48 cores and 270 GB memory unless indicated otherwise.

\subsection{Sod shock tube inside a square-column}\label{subsec:Sod_shock_tube_inside_a_square-column}
The Sod shock tube problem is simulated inside a square-column for diatomic gas at different Knudsen numbers to validate the current UGKWP method, and the result is compared with the 1D UGKS solution.

In this test case, the following non-dimensionalization is used
\begin{gather*}
	\hat \rho=\frac{\rho}{\rho_\infty},\ \hat U =\frac{U}{C_\infty},\ \hat V =\frac{V}{C_\infty},\ \hat W =\frac{W}{C_\infty},\ \hat T=\frac{T}{T_\infty},\ \hat P=\frac{P}{\rho_\infty C_\infty^2},\\
    \hat t=\frac{t}{t_\infty},\ \hat x=\frac{x}{L},\ C_\infty=\sqrt{\frac{2 k_B T_\infty}{m}},\ t_\infty=\frac{L}{C_\infty},
\end{gather*}
and the initial condition for the non-dimensional variables is
\begin{equation}
	(\hat \rho,\hat U,\hat V,\hat W,\hat P) = \left\{
		\begin{array}{lr}
			(1,0,0,0,1),			& 0<\hat x<0.5, \\
			(0.125,0,0,0,0.1),		& 0.5<\hat x<1.
		\end{array}
	\right.
\end{equation}
For UGKWP simulation, the physical domain is a $[0,1]\times[-0.1,0.1]\times[-0.1,0.1]$ square-column tube, which is discretized by $100 \times 5 \times 5$ uniform mesh points. The preset reference numbers of particles are $N_{ref} = 200, 400, 1000, 2000$, $3200, 3200$ for the cases at $Kn = 10^{-4}, 10^{-3}, 10^{-2}, 0.1, 1, 10$ respectively. Least square reconstruction with Venkatakrishnan limiter is utilized. For the UGKS simulation, the 1D physical domain $[0,1]$ is discretized uniformly with $100$ cells. Composed Newton-Cotes quadrature with $101$ velocity points in range $[-6,6]$ is fixed to discretize the one-dimensional velocity space. van Leer limiter is used for the reconstruction of both conservative variables and discrete distribution function. The left and right boundaries are treated as far-field, and the others are treated as symmetric planes. The CFL number for both UGKWP and UGKS simulation is $0.9$, and the reference viscosity is given in Eq.\eqref{eq:tau_via_T_ref} with $\omega = 0.74$. The results at the time $t=0.12$ in all flow regimes are presented.

\begin{figure}[htbp!]
    \centering
    \begin{subfigure}[b]{0.3\textwidth}
        \includegraphics[width=\textwidth]{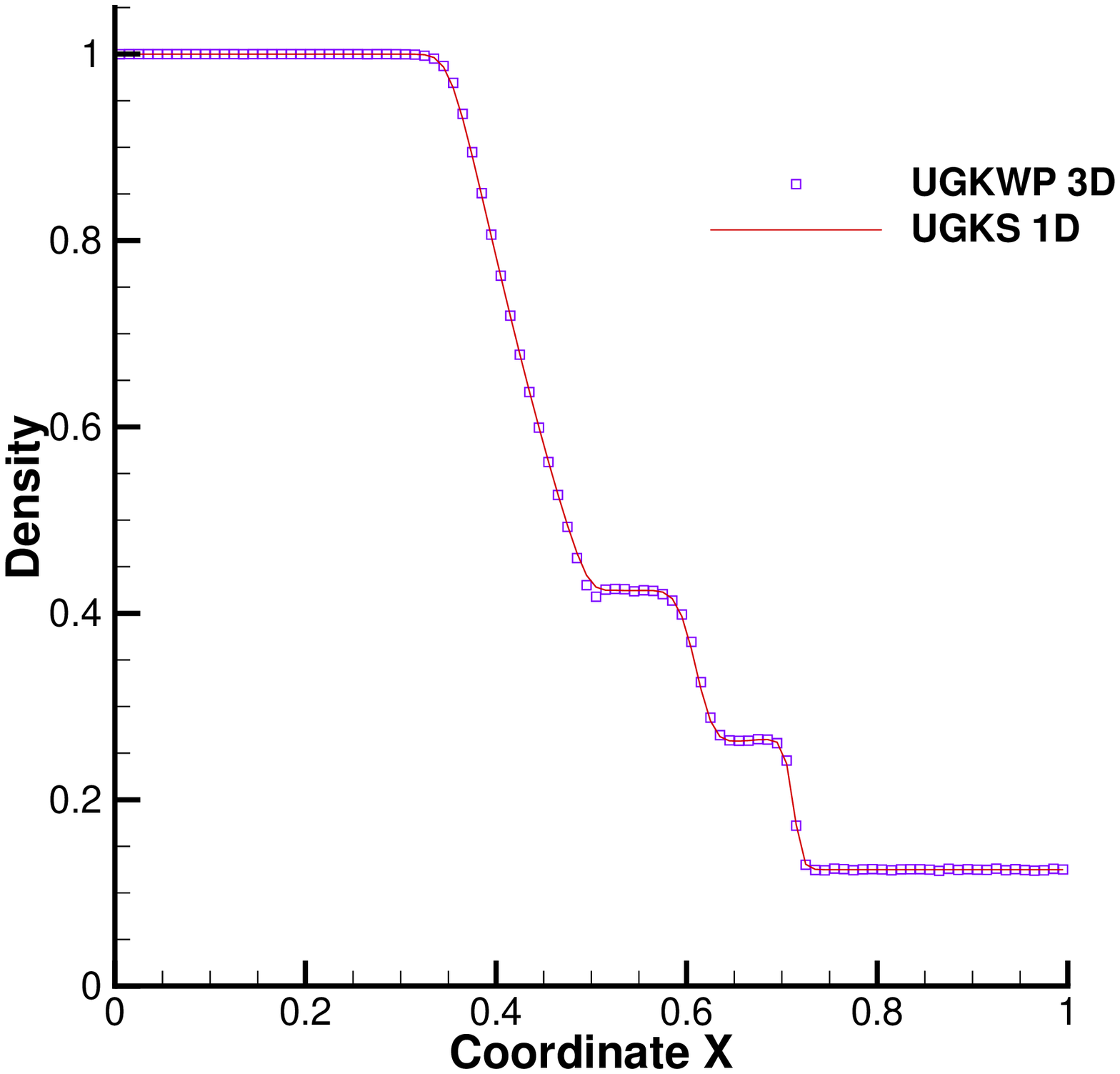}
        \caption{}
        \label{fig:SodShockTube_kn1e-4_Density}
    \end{subfigure}
    \begin{subfigure}[b]{0.3\textwidth}
        \includegraphics[width=\textwidth]{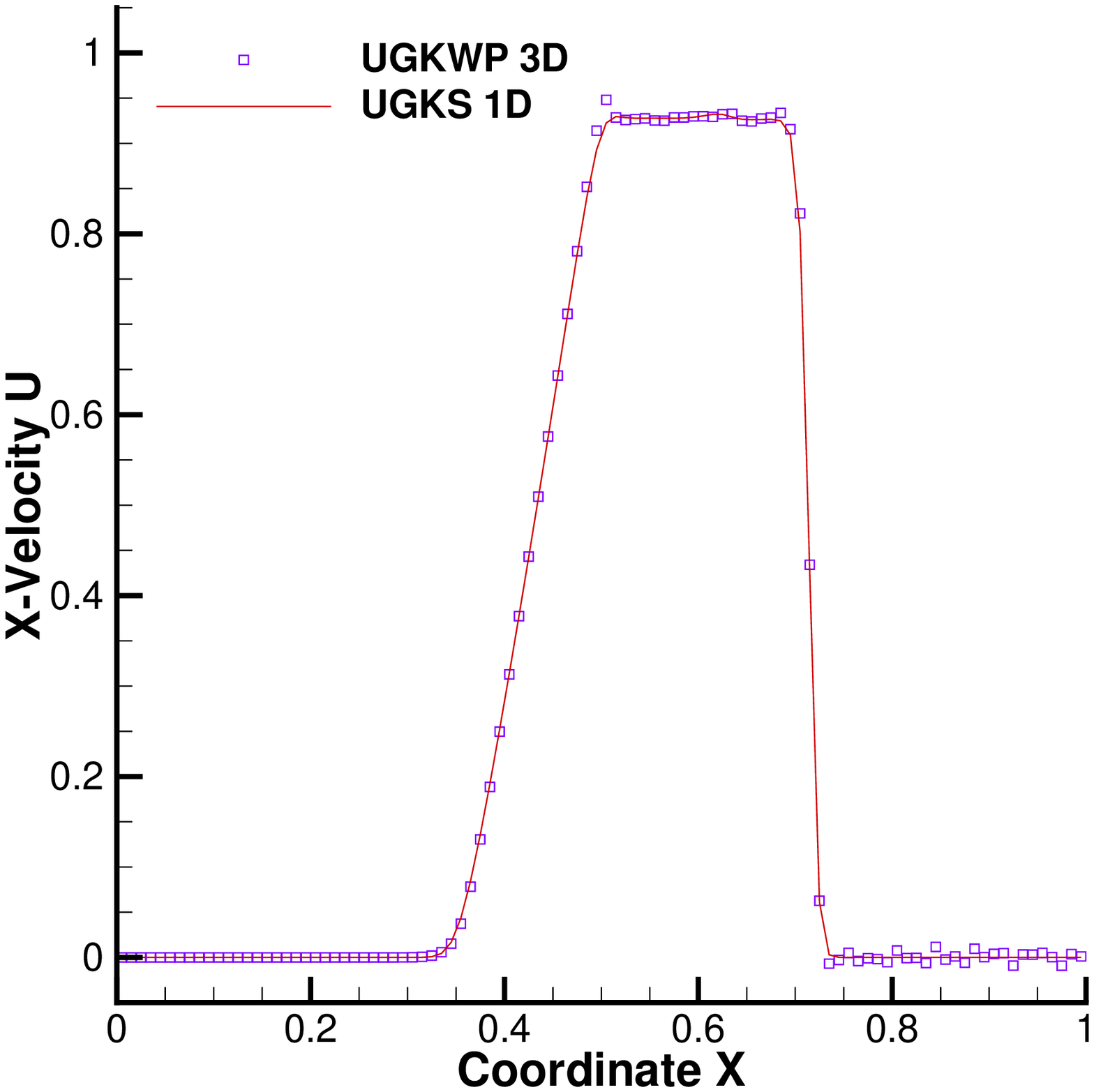}
		\caption{}
        \label{fig:SodShockTube_kn1e-4_U}
    \end{subfigure}
    \begin{subfigure}[b]{0.3\textwidth}
        \includegraphics[width=\textwidth]{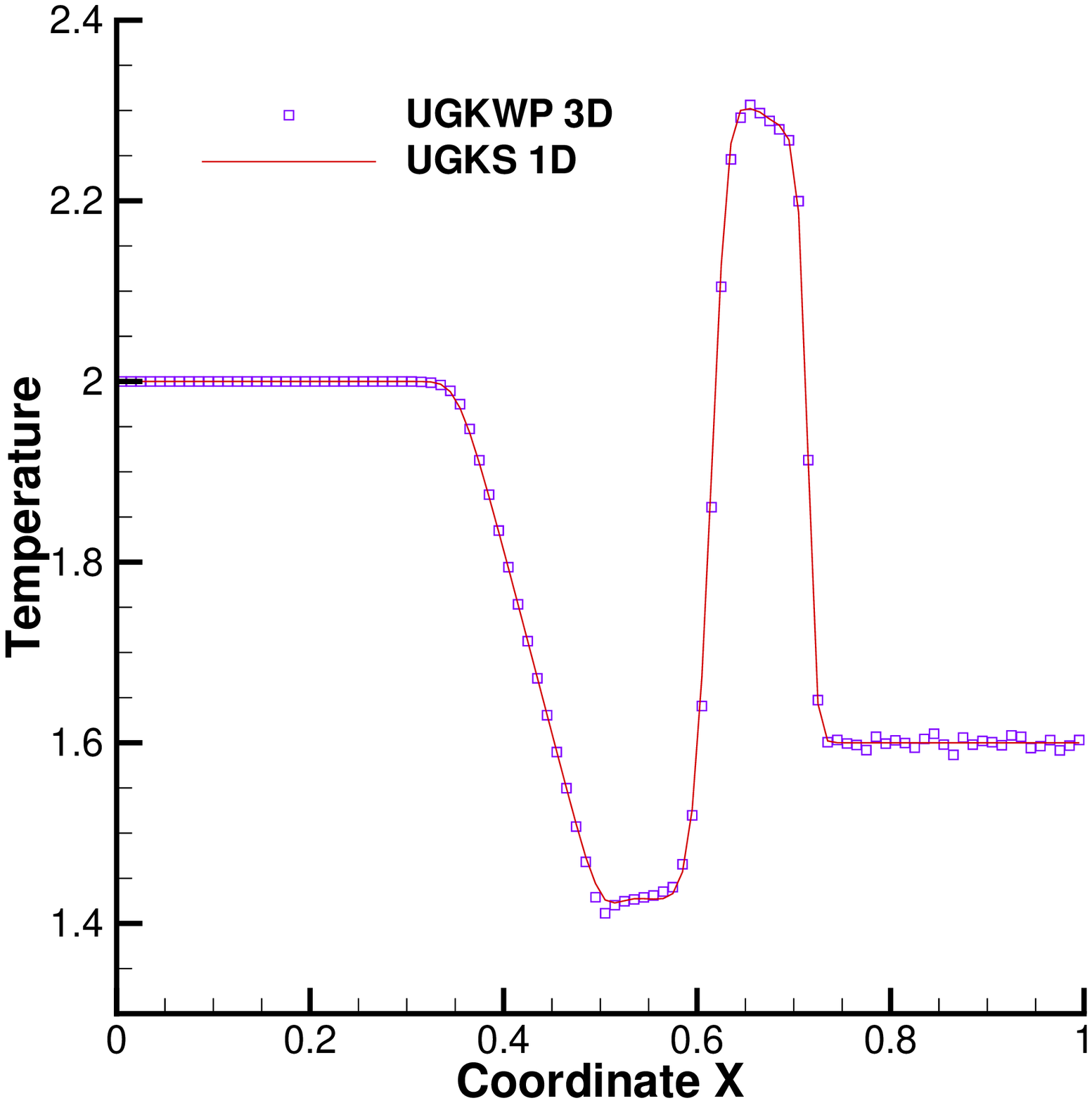}
        \caption{}
        \label{fig:SodShockTube_kn1e-4_Temperature}
    \end{subfigure}
    \caption{Sod shock tube at $Kn = 10^{-4}$. (a) Density, (b) X-Velocity U, and (c) Temperature.}\label{fig:SodShockTube_kn1e-4}
\end{figure}

\begin{figure}[htbp!]
    \centering
    \begin{subfigure}[b]{0.3\textwidth}
        \includegraphics[width=\textwidth]{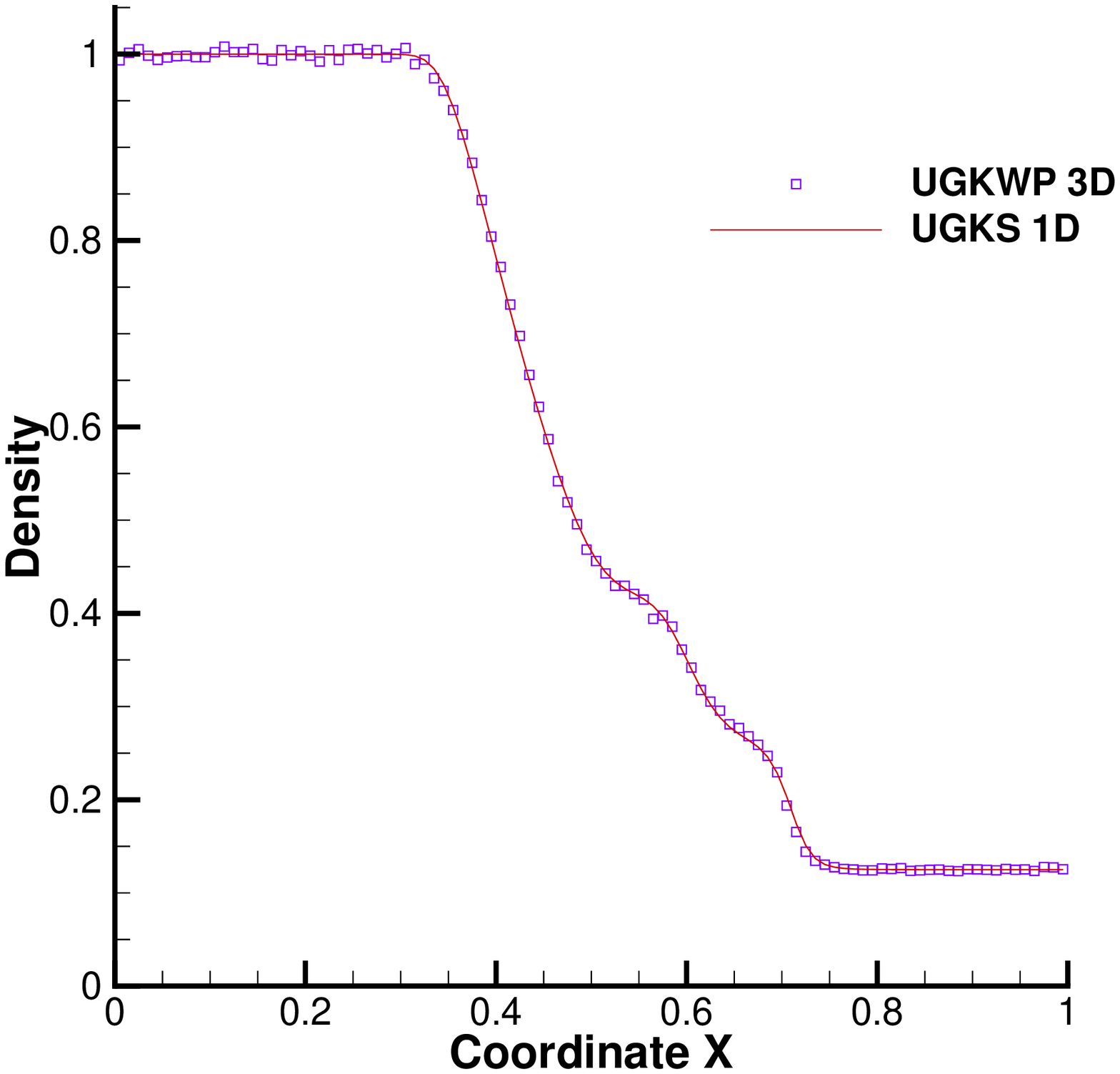}
        \caption{}
        \label{fig:SodShockTube_kn1e-3_Density}
    \end{subfigure}
    \begin{subfigure}[b]{0.3\textwidth}
        \includegraphics[width=\textwidth]{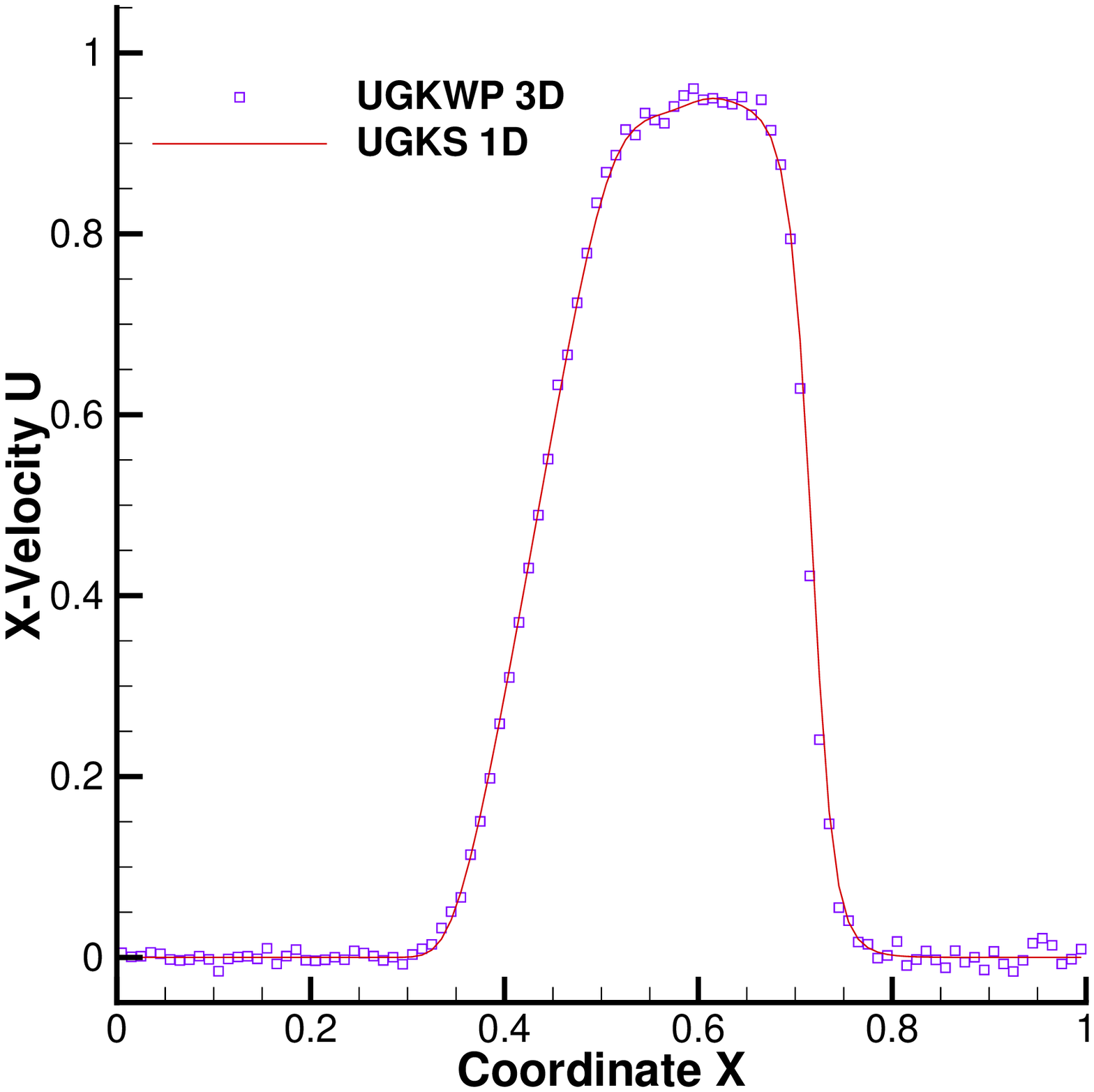}
		\caption{}
        \label{fig:SodShockTube_kn1e-3_U}
    \end{subfigure}
    \begin{subfigure}[b]{0.3\textwidth}
        \includegraphics[width=\textwidth]{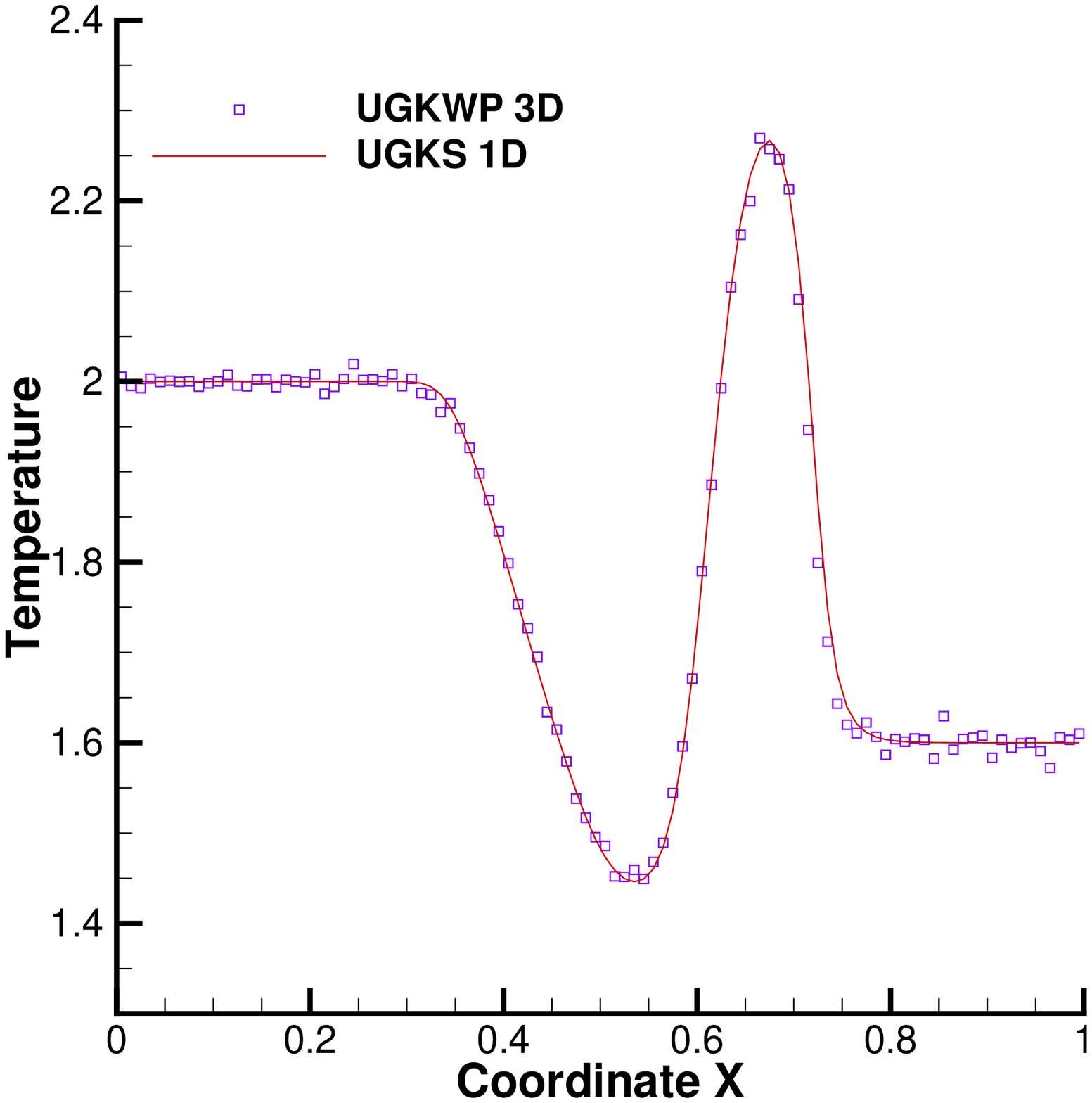}
        \caption{}
        \label{fig:SodShockTube_kn1e-3_Temperature}
    \end{subfigure}
    \caption{Sod shock tube at $Kn = 10^{-3}$. (a) Density, (b) X-Velocity U, and (c) Temperature.}\label{fig:SodShockTube_kn1e-3}
\end{figure}

\begin{figure}[htbp!]
    \centering
    \begin{subfigure}[b]{0.3\textwidth}
        \includegraphics[width=\textwidth]{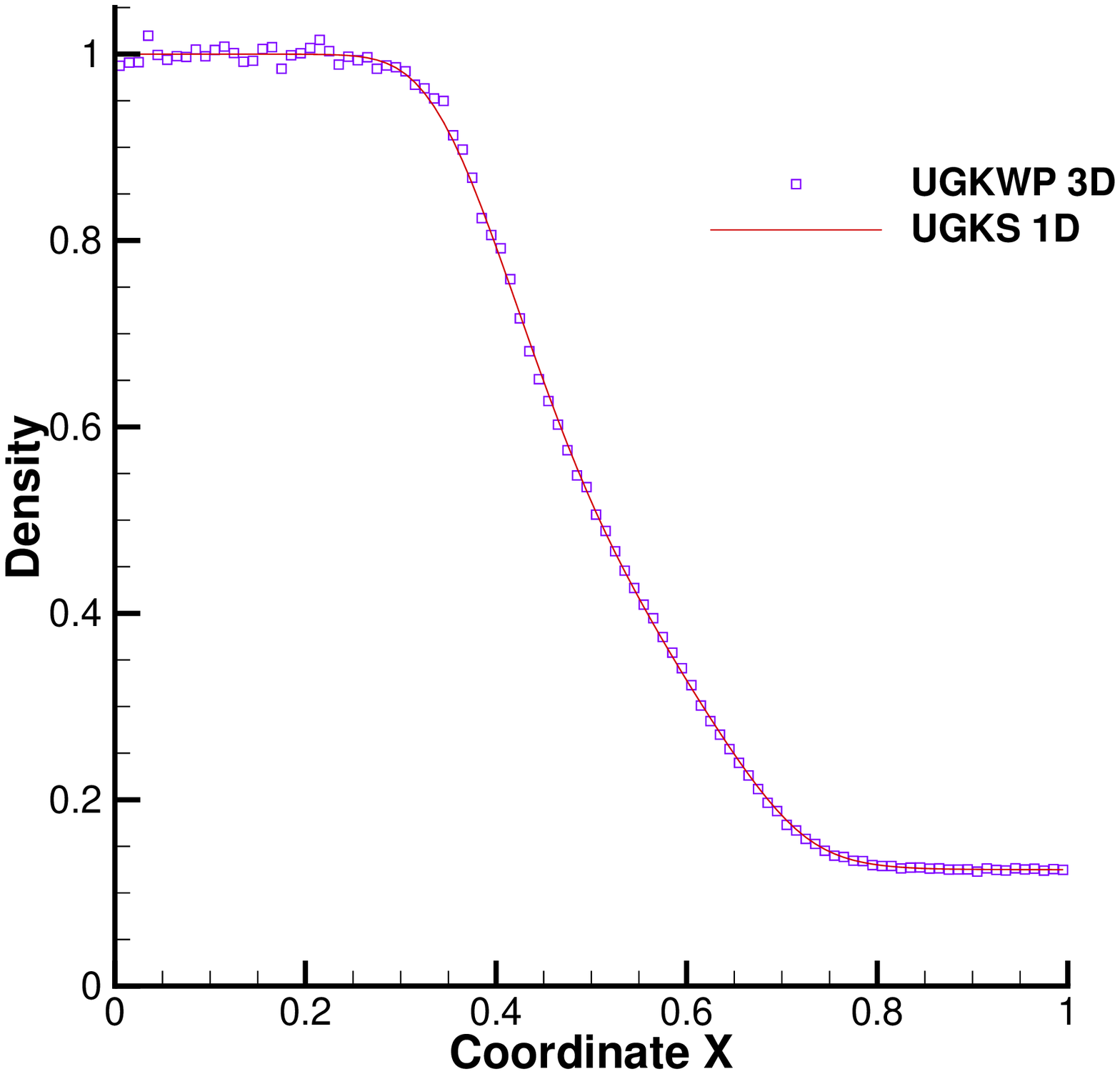}
        \caption{}
        \label{fig:SodShockTube_kn1e-2_Density}
    \end{subfigure}
    \begin{subfigure}[b]{0.3\textwidth}
        \includegraphics[width=\textwidth]{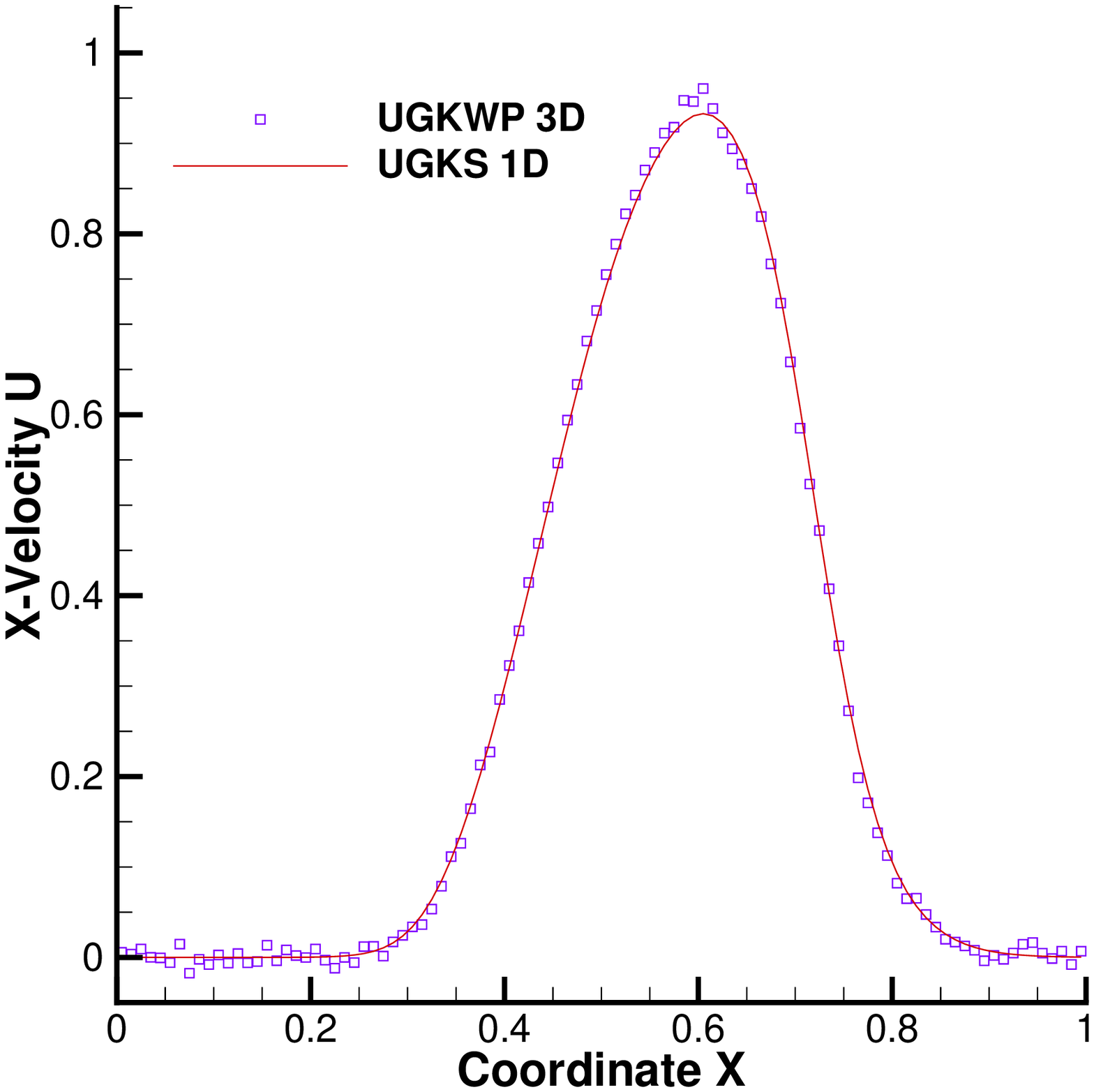}
		\caption{}
        \label{fig:SodShockTube_kn1e-2_U}
    \end{subfigure}
    \begin{subfigure}[b]{0.3\textwidth}
        \includegraphics[width=\textwidth]{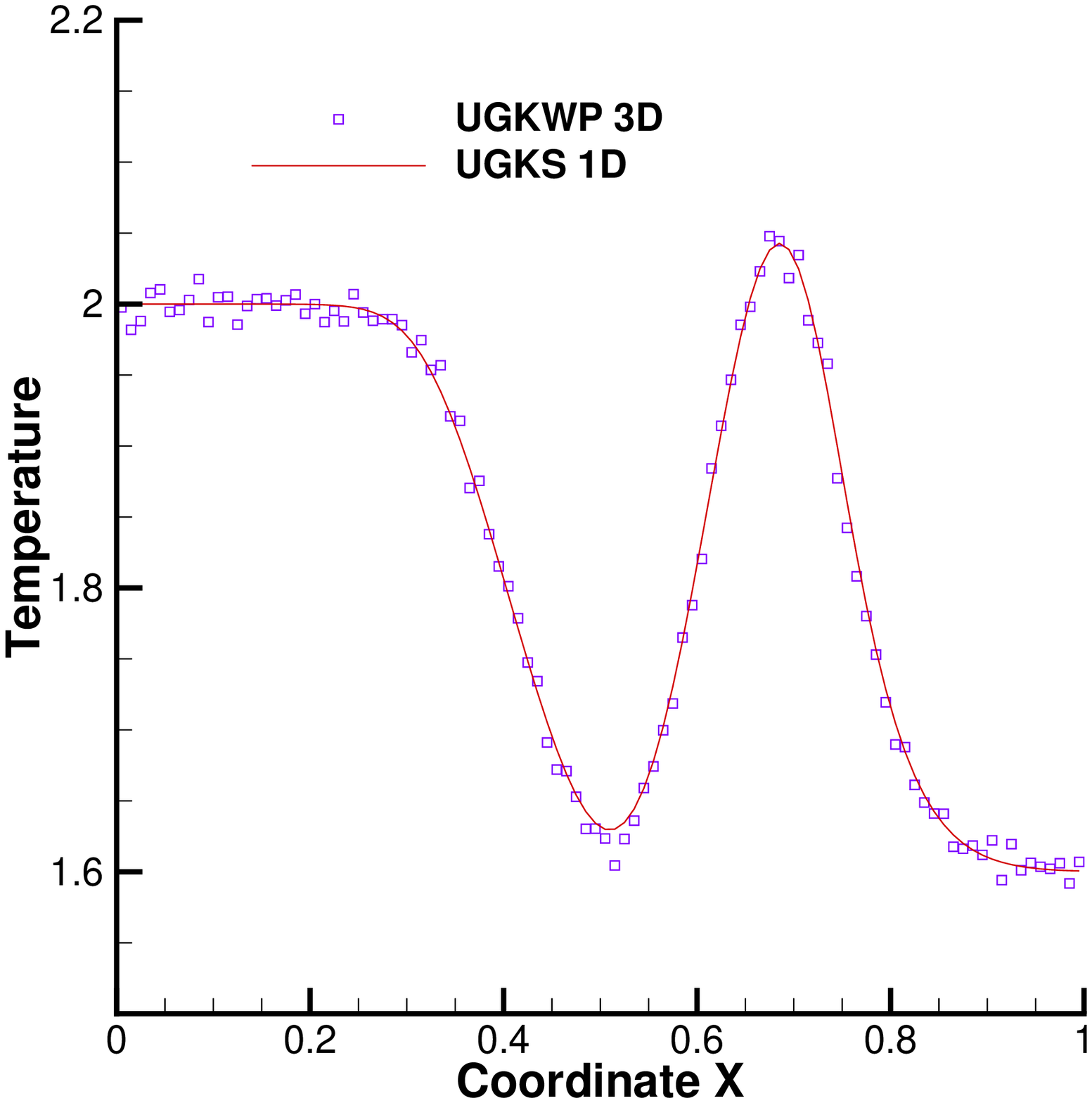}
        \caption{}
        \label{fig:SodShockTube_kn1e-2_Temperature}
    \end{subfigure}
    \caption{Sod shock tube at $Kn = 10^{-2}$. (a) Density, (b) X-Velocity U, and (c) Temperature.}\label{fig:SodShockTube_kn1e-2}
\end{figure}

\begin{figure}[htbp!]
    \centering
    \begin{subfigure}[b]{0.3\textwidth}
        \includegraphics[width=\textwidth]{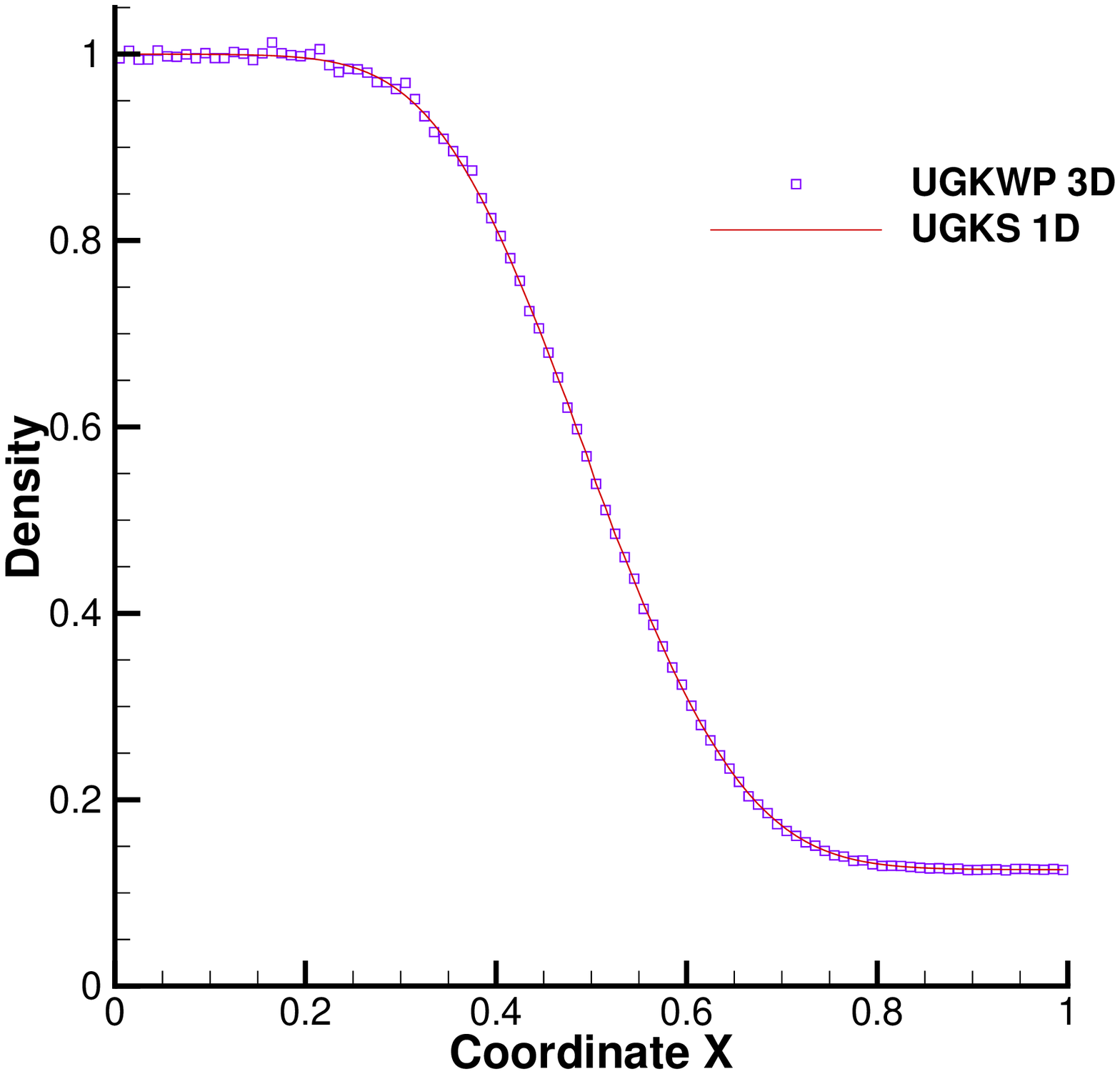}
        \caption{}
        \label{fig:SodShockTube_kn1e-1_Density}
    \end{subfigure}
    \begin{subfigure}[b]{0.3\textwidth}
        \includegraphics[width=\textwidth]{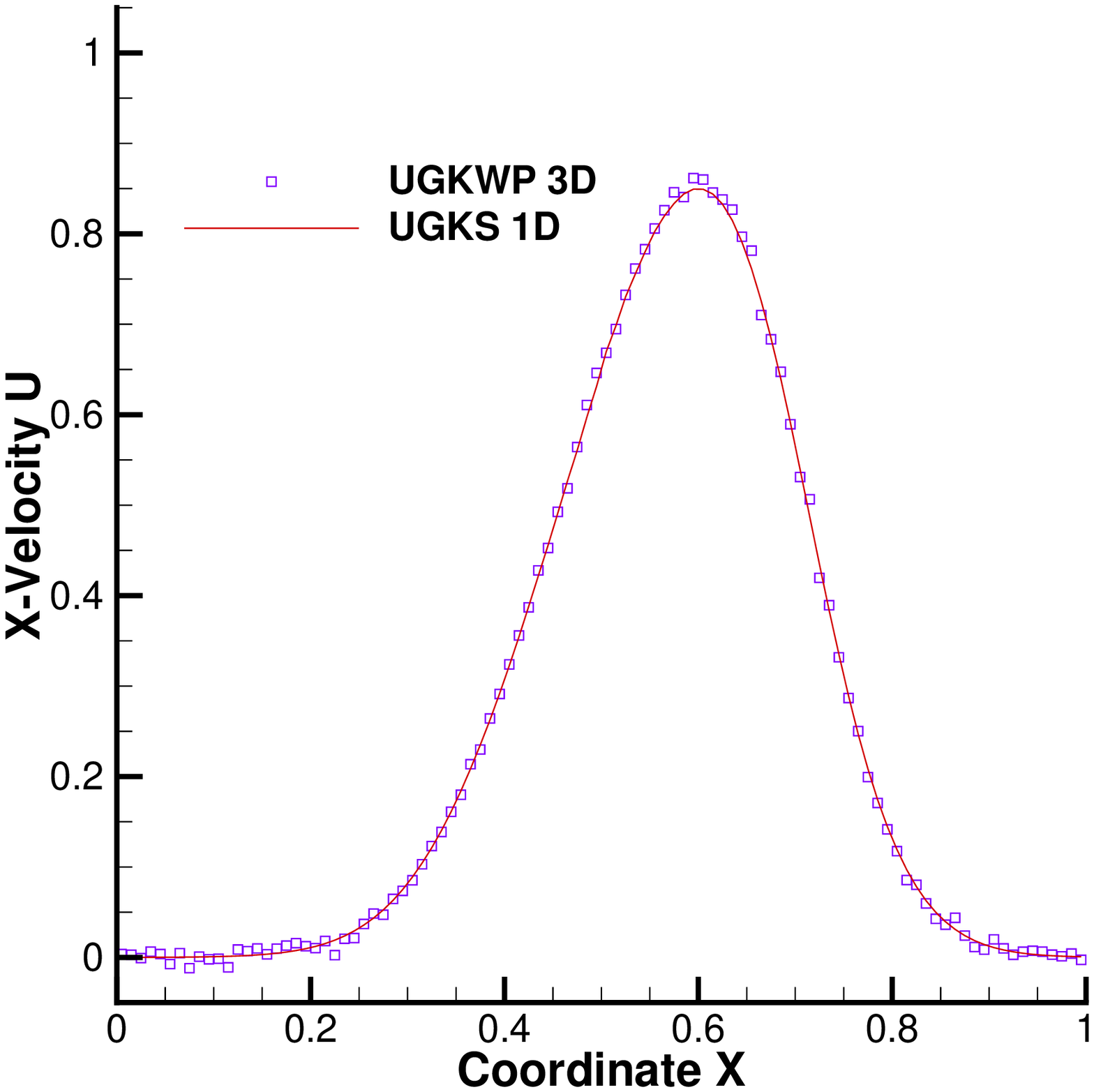}
		\caption{}
        \label{fig:SodShockTube_kn1e-1_U}
    \end{subfigure}
    \begin{subfigure}[b]{0.3\textwidth}
        \includegraphics[width=\textwidth]{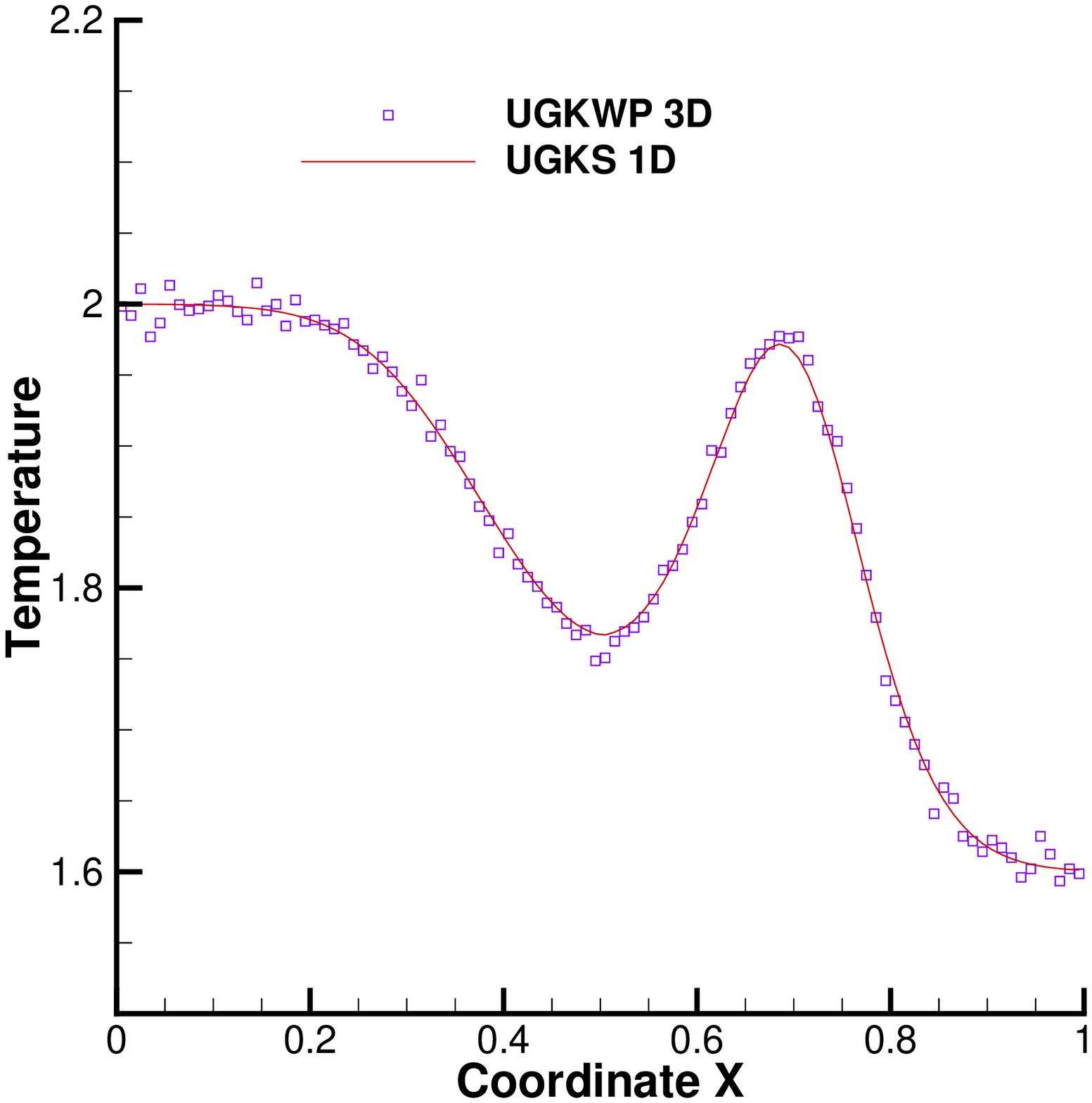}
        \caption{}
        \label{fig:SodShockTube_kn1e-1_Temperature}
    \end{subfigure}
    \caption{Sod shock tube at $Kn = 0.1$. (a) Density, (b) X-Velocity U, and (c) Temperature.}\label{fig:SodShockTube_kn1e-1}
\end{figure}

\begin{figure}[htbp!]
    \centering
    \begin{subfigure}[b]{0.3\textwidth}
        \includegraphics[width=\textwidth]{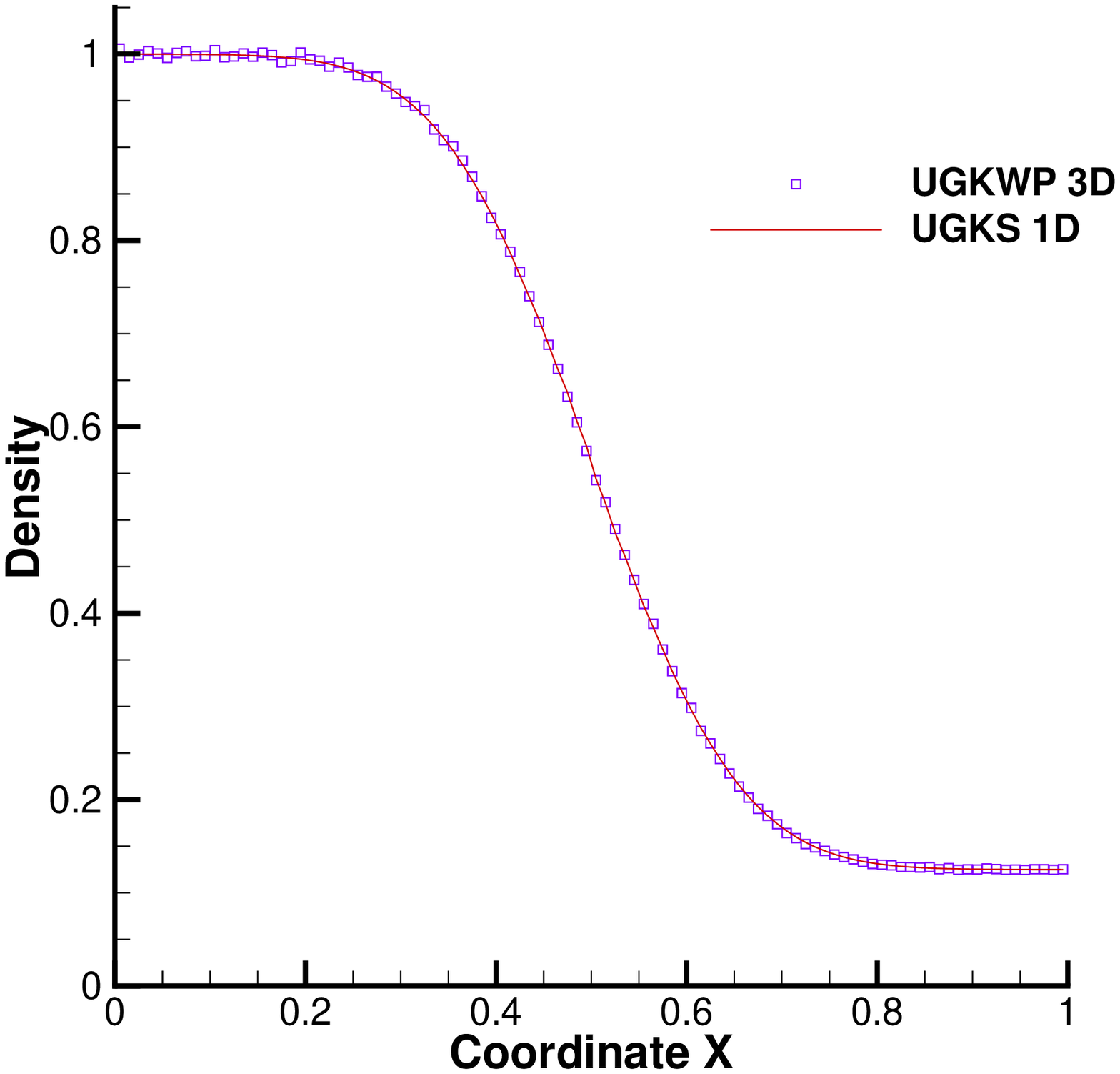}
        \caption{}
        \label{fig:SodShockTube_kn1e0_Density}
    \end{subfigure}
    \begin{subfigure}[b]{0.3\textwidth}
        \includegraphics[width=\textwidth]{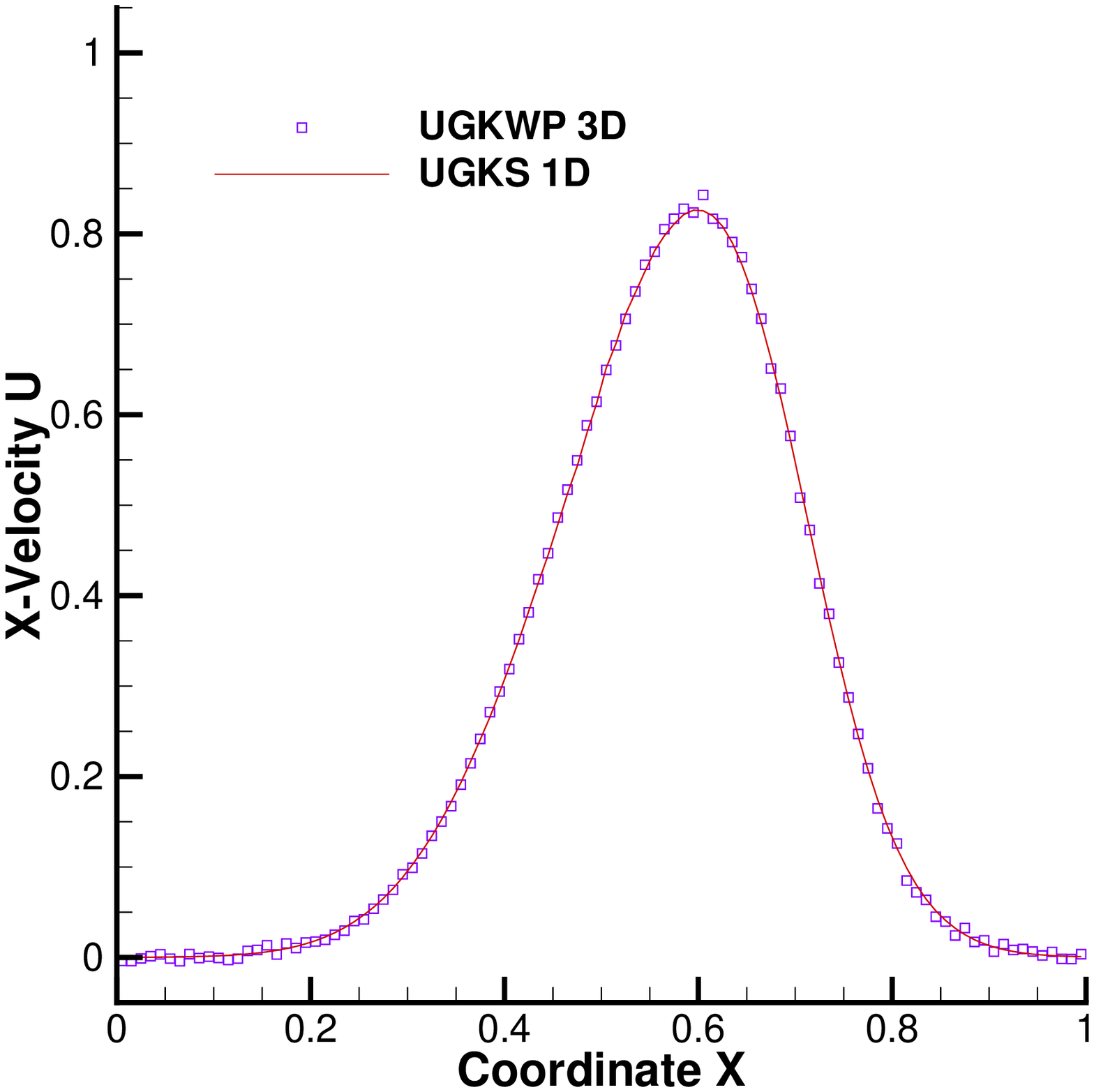}
		\caption{}
        \label{fig:SodShockTube_kn1e0_U}
    \end{subfigure}
    \begin{subfigure}[b]{0.3\textwidth}
        \includegraphics[width=\textwidth]{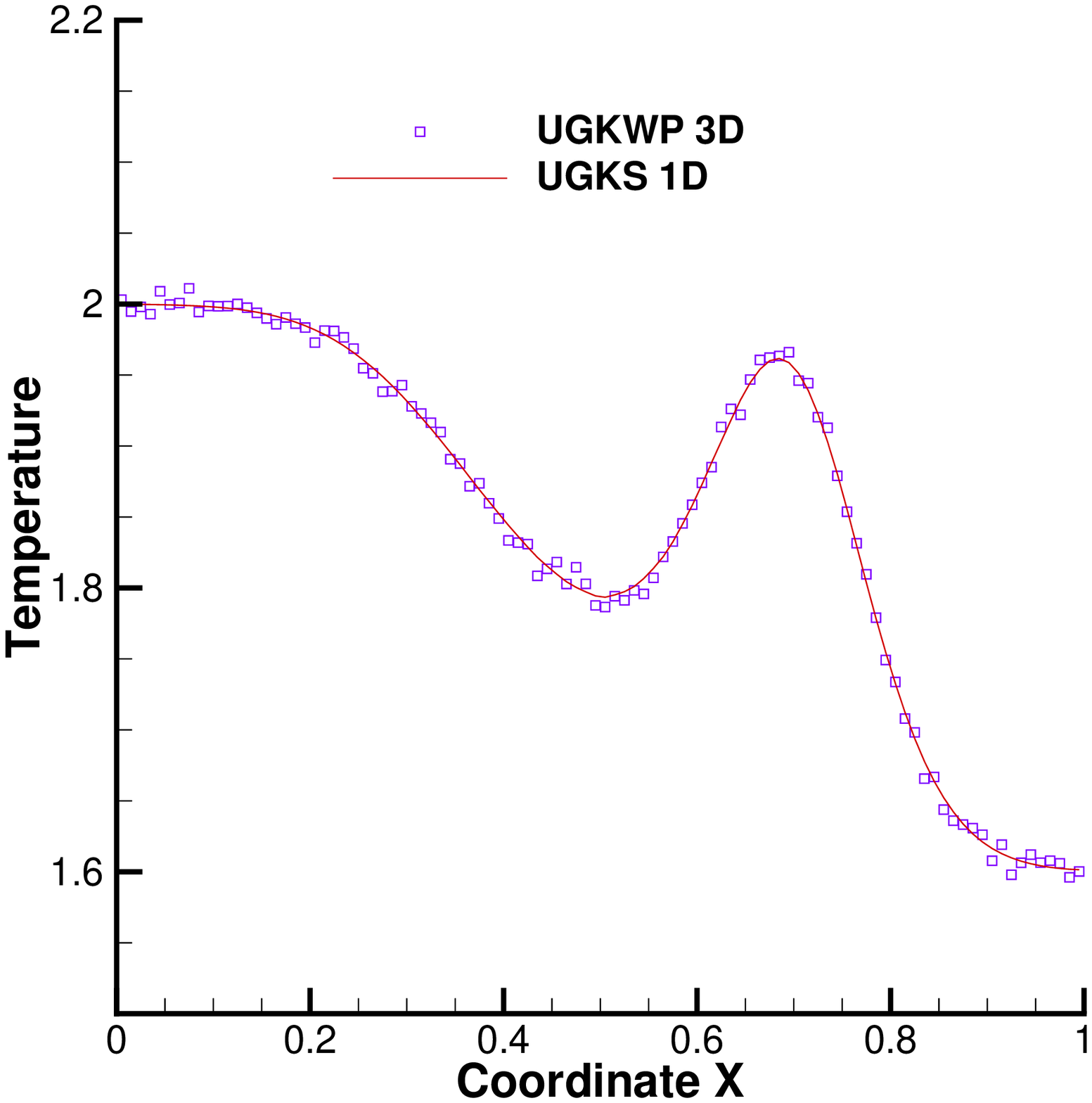}
        \caption{}
        \label{fig:SodShockTube_kn1e0_Temperature}
    \end{subfigure}
    \caption{Sod shock tube at $Kn = 1$. (a) Density, (b) X-Velocity U, and (c) Temperature.}\label{fig:SodShockTube_kn1e0}
\end{figure}

\begin{figure}[htbp!]
    \centering
    \begin{subfigure}[b]{0.3\textwidth}
        \includegraphics[width=\textwidth]{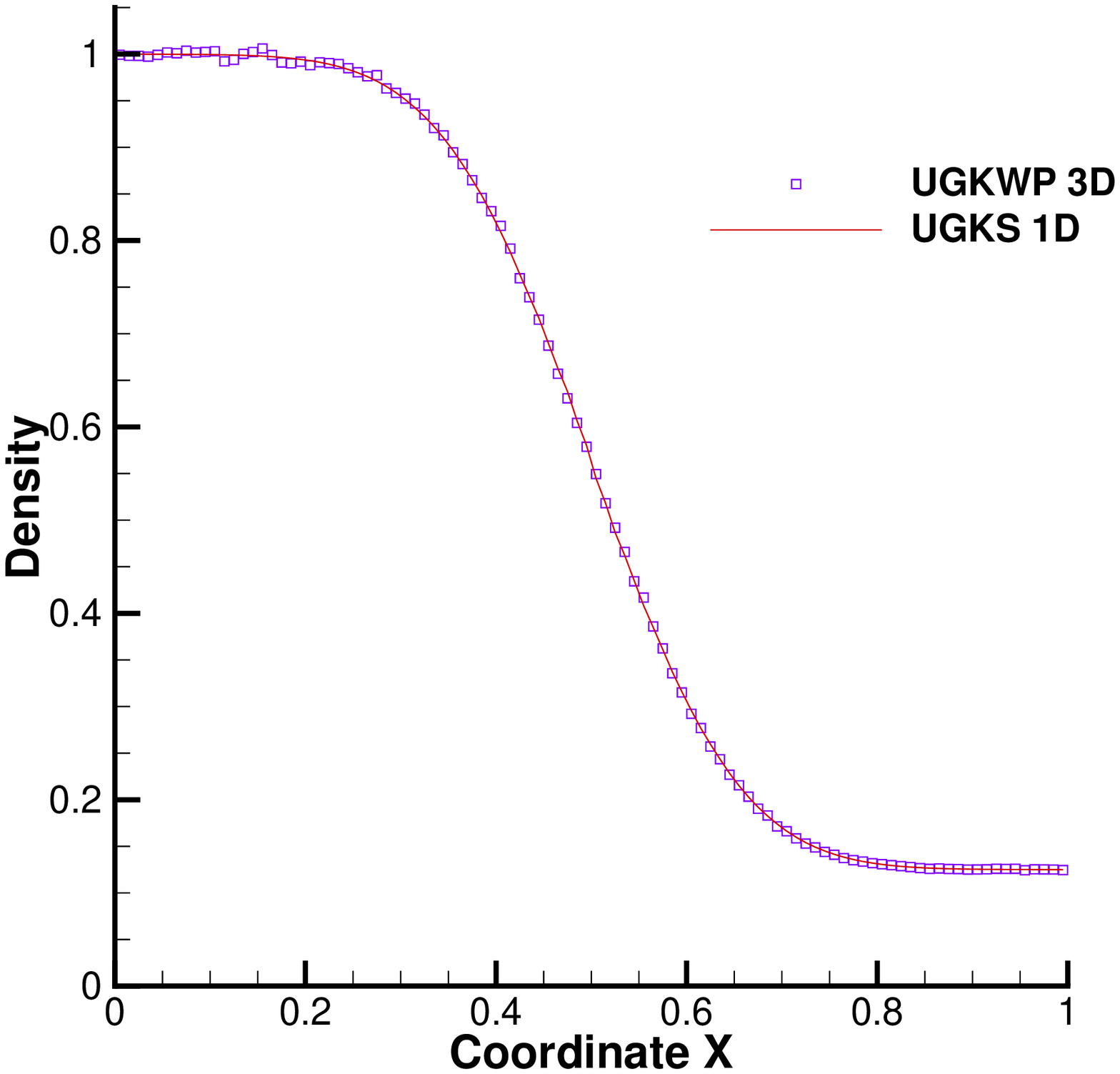}
        \caption{}
        \label{fig:SodShockTube_kn1e1_Density}
    \end{subfigure}
    \begin{subfigure}[b]{0.3\textwidth}
        \includegraphics[width=\textwidth]{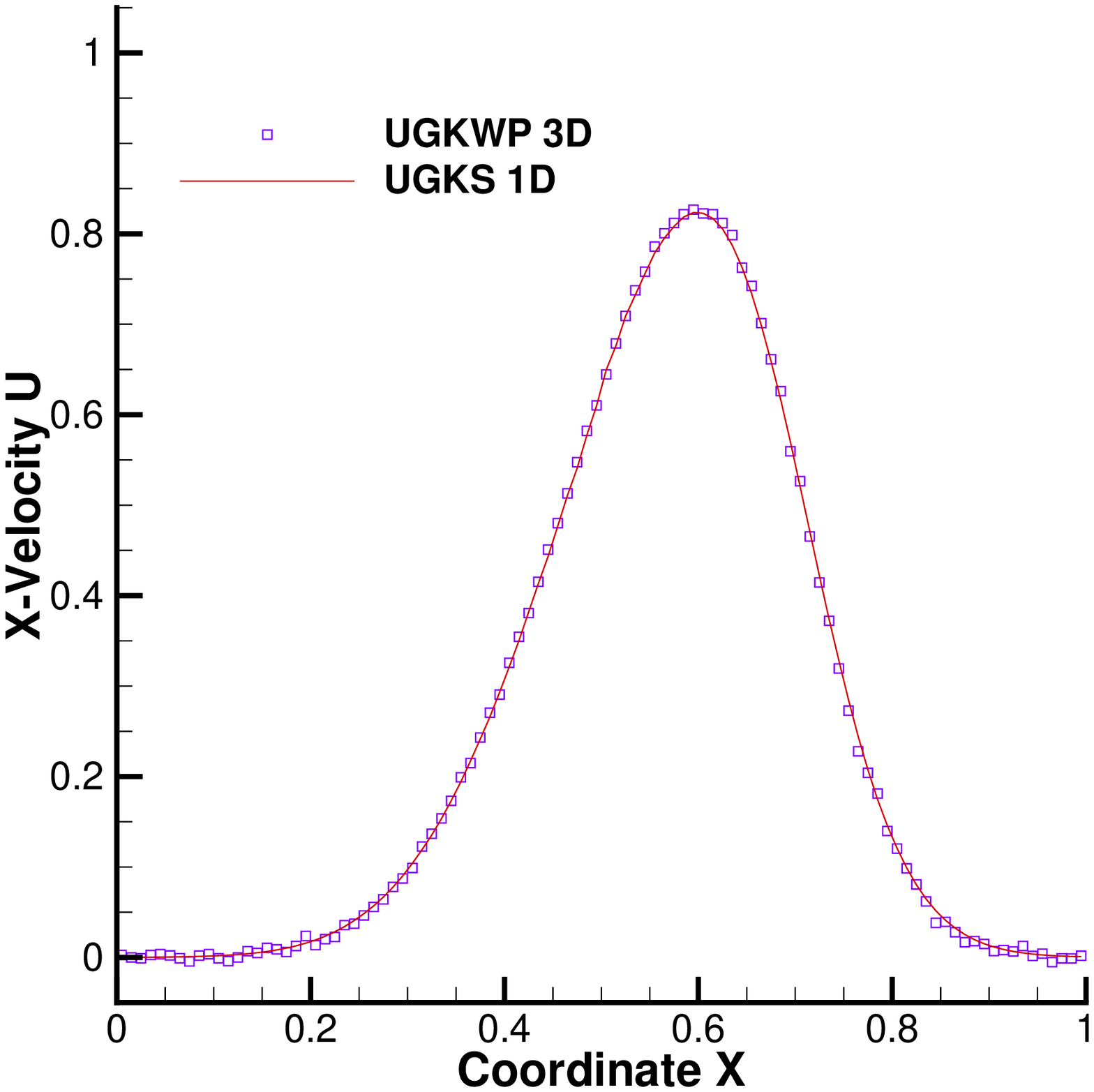}
		\caption{}
        \label{fig:SodShockTube_kn1e1_U}
    \end{subfigure}
    \begin{subfigure}[b]{0.3\textwidth}
        \includegraphics[width=\textwidth]{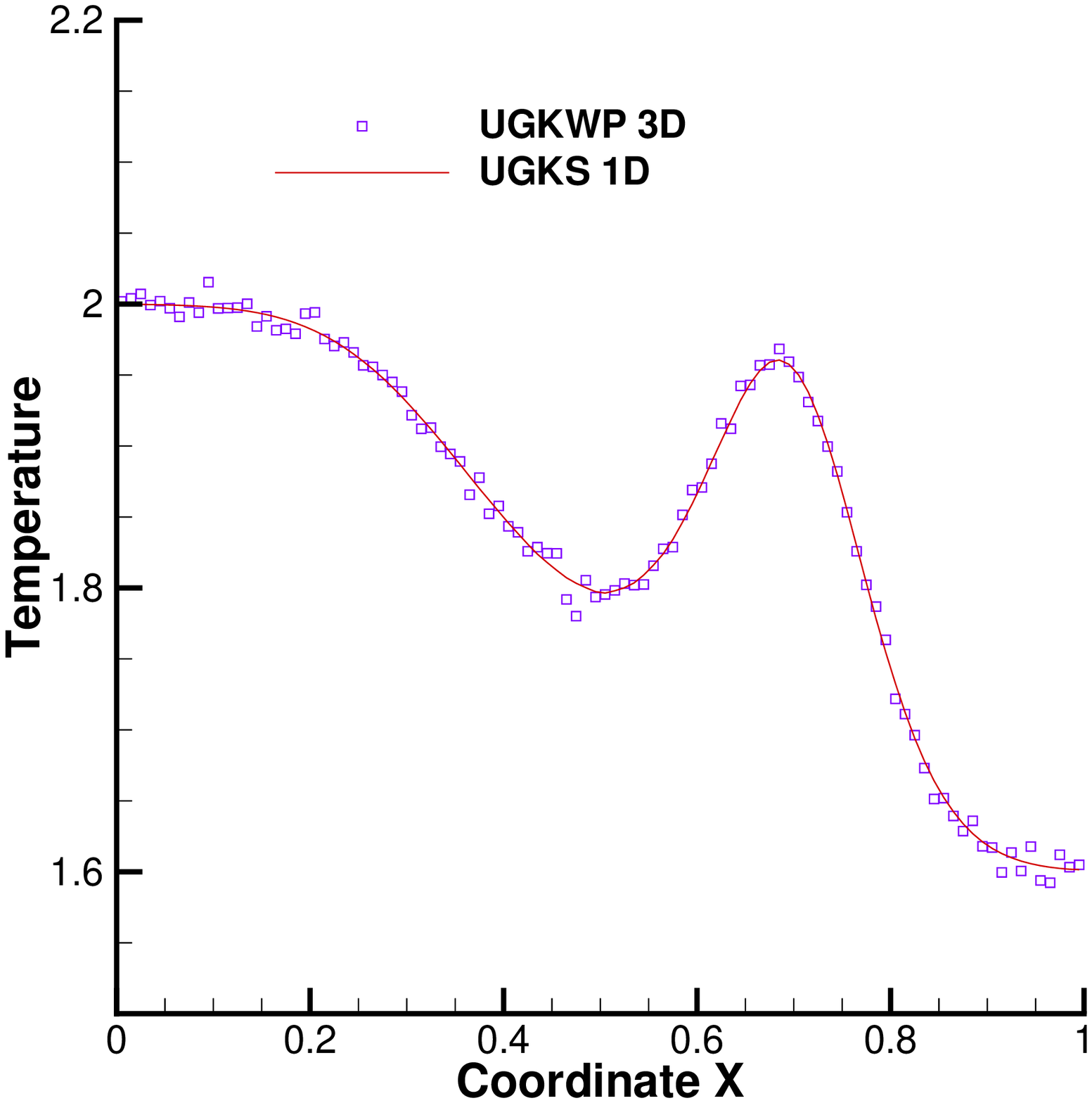}
        \caption{}
        \label{fig:SodShockTube_kn1e1_Temperature}
    \end{subfigure}
    \caption{Sod shock tube at $Kn = 10$. (a) Density, (b) X-Velocity U, and (c) Temperature.}\label{fig:SodShockTube_kn1e1}
\end{figure}

The density, velocity, and temperature obtained by the UGKS and the UGKWP method at different Knudsen numbers are plotted in \Cref{fig:SodShockTube_kn1e-4,fig:SodShockTube_kn1e-3,fig:SodShockTube_kn1e-2,fig:SodShockTube_kn1e-1,fig:SodShockTube_kn1e0,fig:SodShockTube_kn1e1}, where the three-dimensional flow field computed by UGKWP is projected to one-dimensional along the x-direction by taking ensemble average over the cells on y-z plane. No time averaging is applied, and the statistical noise is satisfactory for this unsteady flow simulation. For all the cases in different flow regimes, the 3D UGKWP solutions agree well with the 1D UGKS data. The slight difference is due to different limiters, i.e., van Leer limiter for UGKS and Venkatakrishnan limiter for UGKWP. The capability of the UGKWP method for numerical simulations in both continuum and rarefied regime is confirmed.

The distinguishable feature of multi-efficiency\cite{liu2020unified} can also be demonstrated here. For UGKS, the computational costs for all Knudsen number cases will be on the same order since its discretization of the particle velocity space is the same. While for the UGKWP method, the computational cost is reduced at small Knudsen number, e.g., $Kn=10^{-4}$ in near continuum regime, where the hydrodynamic wave is dominant, and few particles are sampled and tracked. The computational cost of UGKWP for 3D simulation is admissible because only a few hundred or thousand particles are enough to adaptively discretize the velocity space, whereas it becomes possible that $101^3$ mesh points in the velocity space may be required in DVM-based UGKS for high speed flow. For steady-state simulation, the number of particles can be reduced further since the statistical noise can be reduced through the temporal ensemble.

\subsection{Lid-driven cubic cavity flow}\label{subsec:Lid-driven_cubic_cavity_flow}
For low-speed flow, the UGKWP method is applied to study the three-dimensional lid-driven cubic cavity flow in the transition regime, and the results are compared with the solution predicted by dugksFoam\cite{zhu2017dugksfoam}.

The side length of the cubic cavity is $L = 1 m$ with the computational domain $[0,1]\times[0,1]\times[0,1]$, which is divided non-uniformly into $40^3$ hexahedrons with the cell size gradually increased towards to the cavity center. The ratio of the cell size in the center and the boundary is about $2$. The lid (top boundary) of the cavity moves in the positive x-direction with a constant velocity $U_w = 50 m/s$, while the other walls are kept fixed. All sidewalls have the diffusive boundary condition and keep a uniform temperature $T_w = 273 K$. The cavity is assumed to consist of monatomic argon gas with molecular mass $m = 6.63 \times 10^{-26} kg$ and diameter $d=4.17 \times 10^{-10} m$.
The Knudsen number is $Kn = \lambda/L = 0.075$, where the mean free path $\lambda$ is calculated from the initial uniform gas density by $\lambda = m/(\sqrt{2}\pi d^2 \rho)$. The gas viscosity depends on the temperature by Eq. \eqref{eq:tau_via_T_ref} with reference temperature $T_{ref} = T_w = 273 K$ and reference viscosity $\mu_{ref}$ given by variable hard sphere (VHS) model with $\omega = 0.81$.

Since it is a low-speed flow with small temperature variance, $N_{ref} = 5000$ reference number of simulation particles is used. The time-averaging is starting from $1000$ steps in order to reduce the statistical noises of high moments quantities, such as the temperature. The CFL number is set to be $0.95$, and the least square reconstruction with Venkatakrishnan limiter is employed for the gradient calculation. Physical space parallelization with 48 cores is adopted for UGKWP.

In the dugksFoam simulation, the three-dimensional velocity space is discredited using $28$ half-range Gauss-Hermit quadrature points in each direction. The CFL number is set to be $0.8$. The gradients are calculated by least Square. The Prandtl number is fixed as $Pr = 1.0$ in DUGKS simulation to eliminate the model difference since the BGK model is used in the construction of UGKWP. The velocity space decomposition approach is adopted for dugksFoam with 48 cores on the same machine.

\begin{figure}[htbp!]
    \centering
    \includegraphics[width=\textwidth]{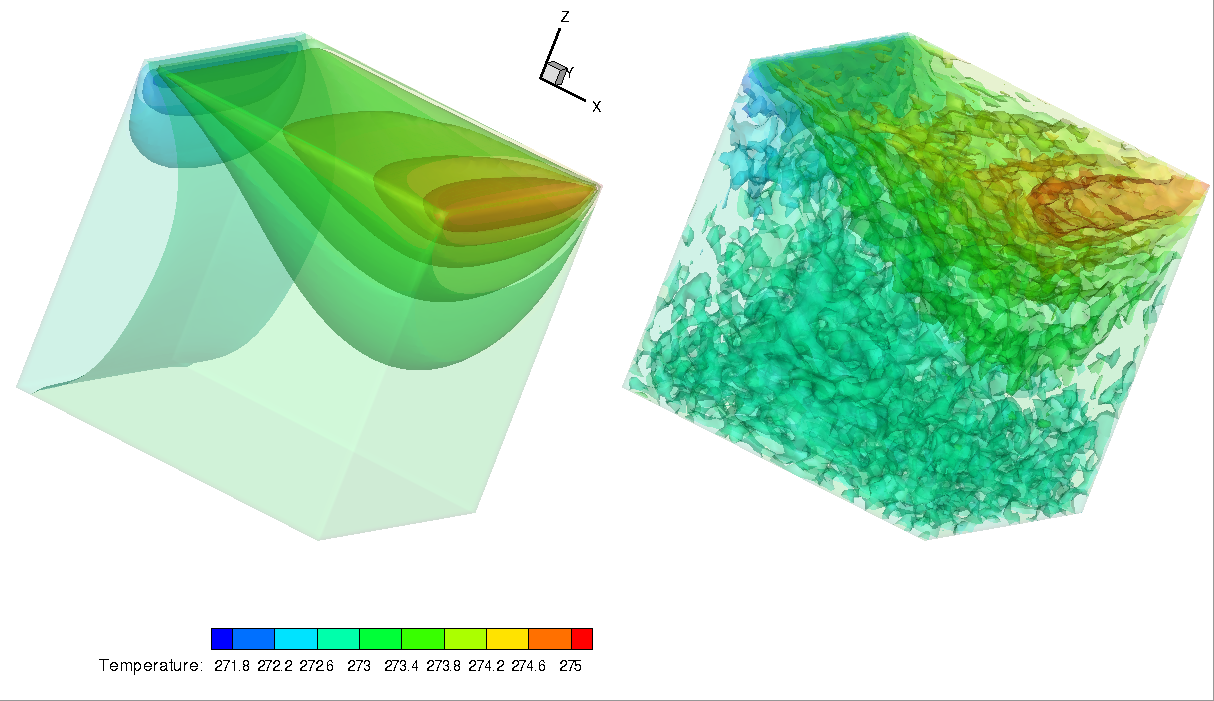}
    \caption{Comparison of the temperature iso-surfaces predicted by DUGKS (left) and UGKWP (right).}\label{fig:cavity3D_3D_Kn_75e-3_iso_Temperature}
\end{figure}

\begin{figure}[htbp!]
    \centering
    \begin{subfigure}[b]{0.49\textwidth}
        \includegraphics[width=\textwidth]{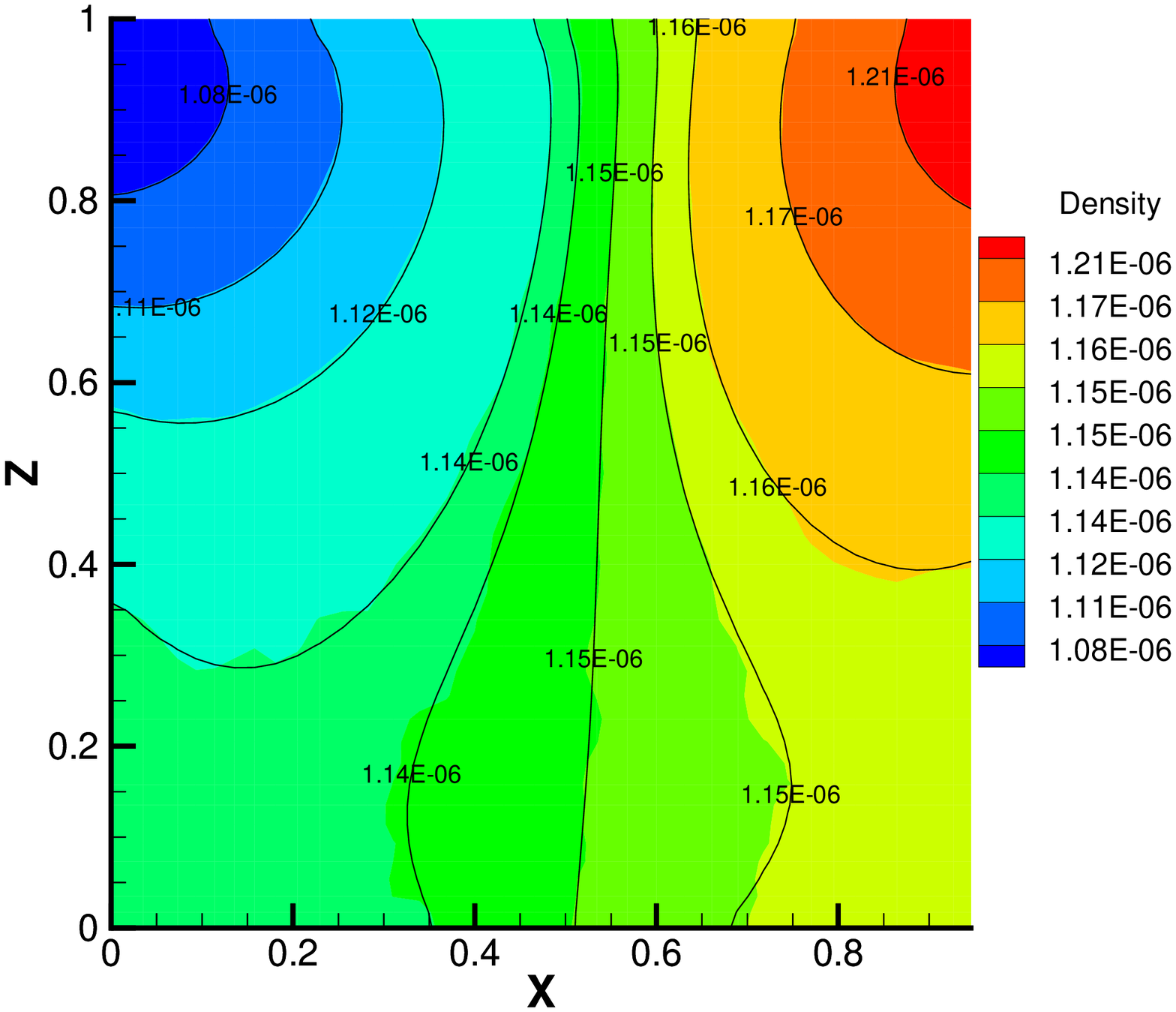}
        \caption{}
        \label{fig:cavity3D_Y-plane_Kn_75e-3_Density}
    \end{subfigure}
    \begin{subfigure}[b]{0.49\textwidth}
        \includegraphics[width=\textwidth]{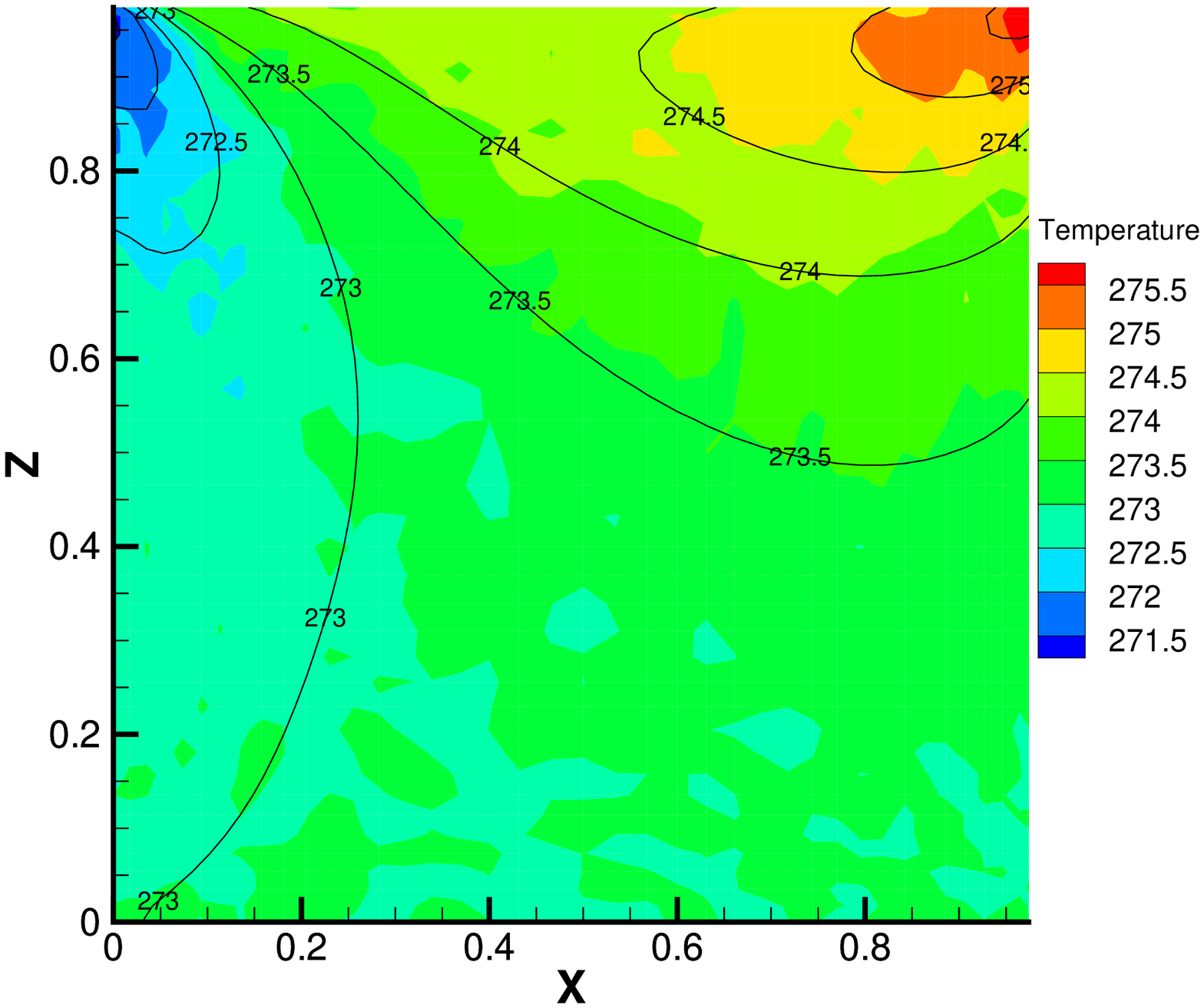}
		\caption{}
        \label{fig:cavity3D_Y-plane_Kn_75e-3_Temperature}
    \end{subfigure}\\
    \begin{subfigure}[b]{0.49\textwidth}
        \includegraphics[width=\textwidth]{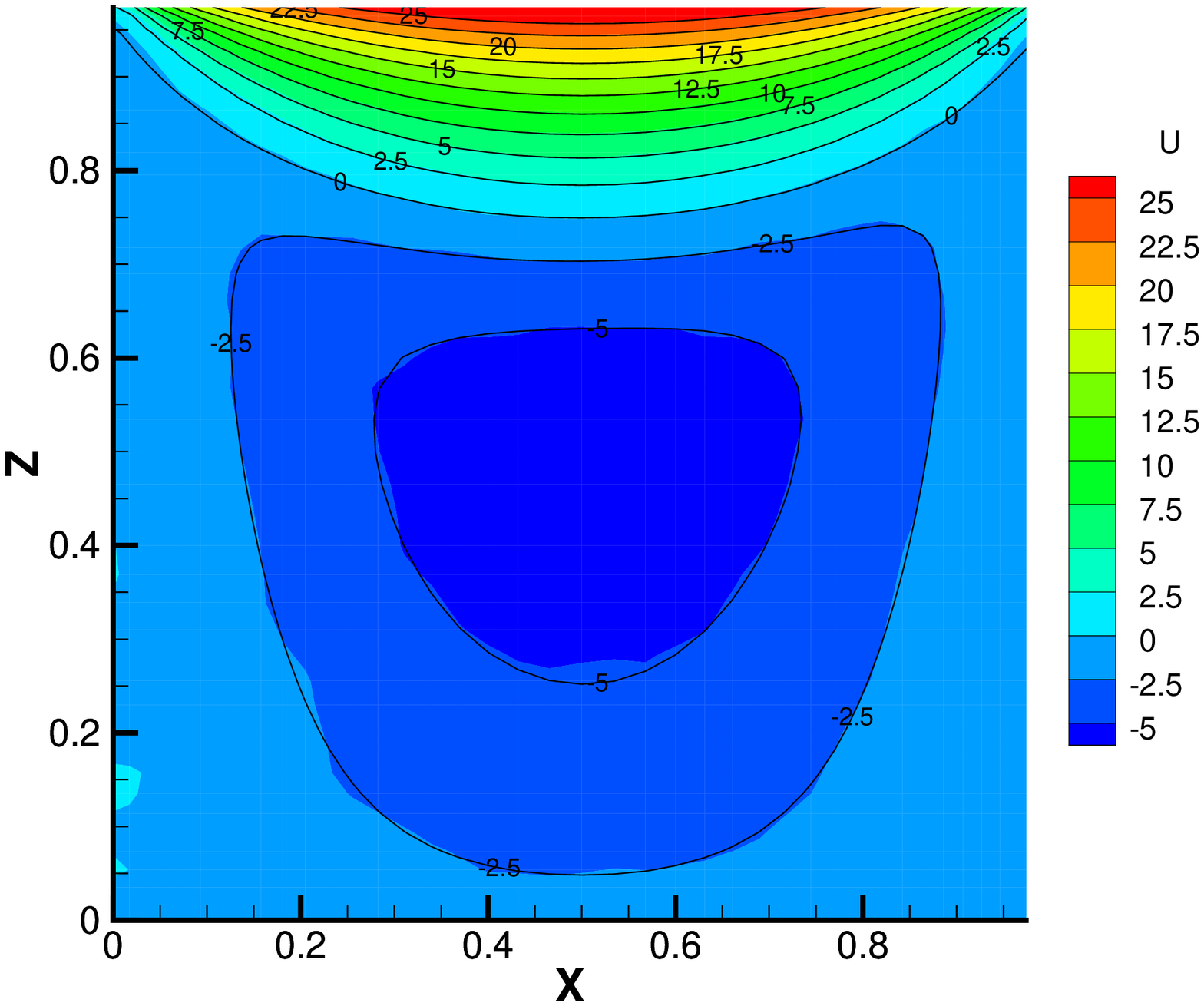}
        \caption{}
        \label{fig:cavity3D_Y-plane_Kn_75e-3_U}
    \end{subfigure}
    \begin{subfigure}[b]{0.49\textwidth}
        \includegraphics[width=\textwidth]{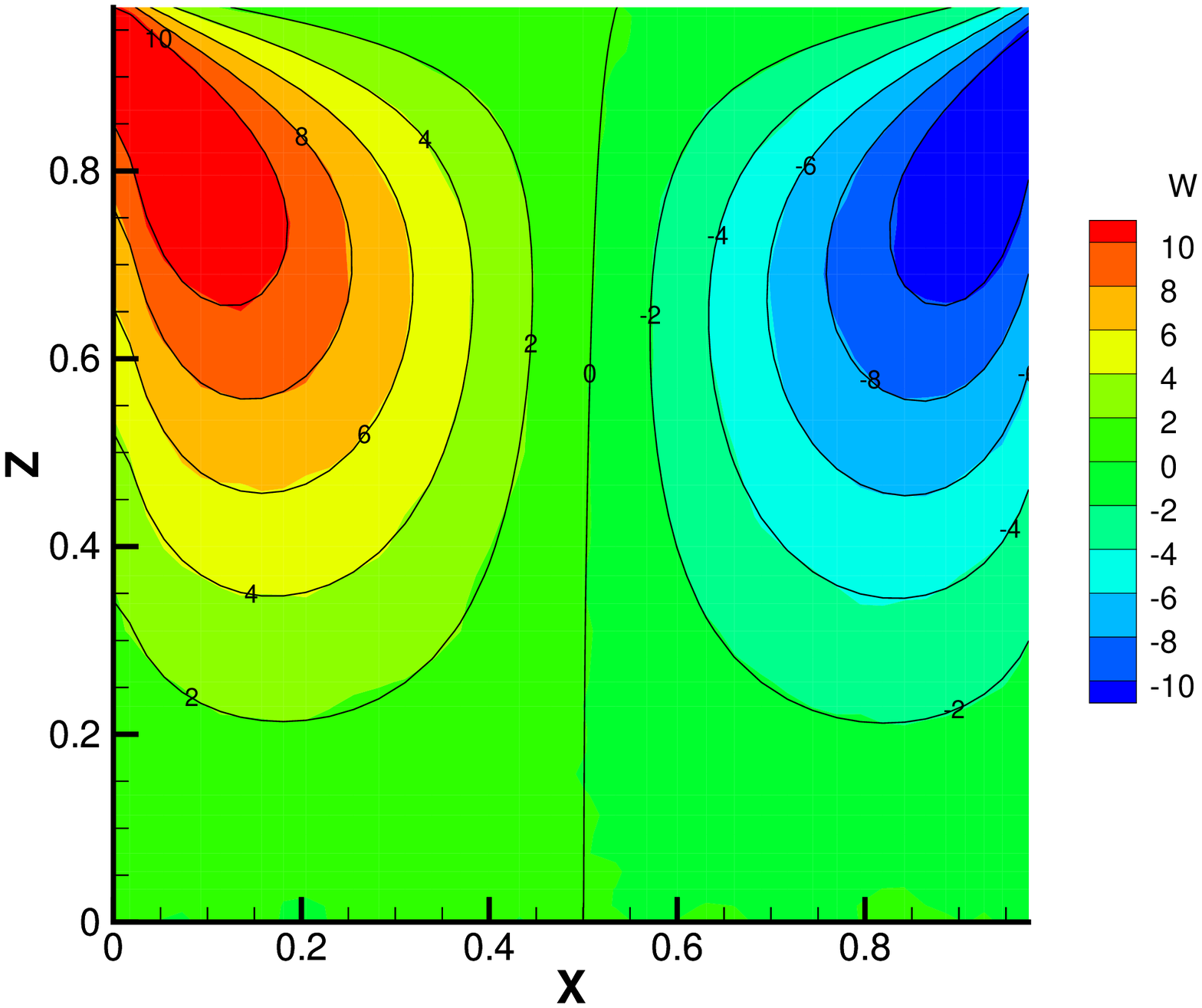}
        \caption{}
        \label{fig:cavity3D_Y-plane_Kn_75e-3_W}
    \end{subfigure}
    \caption{Symmetric X-Z cut-plane contour of cavity flow at $Kn = 0.075$. Background: UGKWP; Black lines with label: dugksFoam. (a) Density contour, (b) Temperature contour, (c) U-velocity contour, (d) W-velocity contour.}\label{fig:cavity3D_Y-plane_Kn_75e-3}
\end{figure}

\Cref{fig:cavity3D_3D_Kn_75e-3_iso_Temperature} presented the temperature iso-surfaces predicted by UGKWP and dugksFoam. Even though the UGKWP solution exhibits strong fluctuation, the two results agree well in general. To compare the solutions more precisely, the contours on the symmetric X-Z plane are shown in \Cref{fig:cavity3D_Y-plane_Kn_75e-3}, where the low order quantities between these two schemes, such as density, X and Y components of the velocity ($U$ and $V$), match well. For the temperature, as a higher moment quantity, the UGKWP solutions generally agree with that of dugksFoam, but still exhibit relatively large statistical noise, although a long time averaging has been performed in UGKWP. It is also noteworthy that the noise incurred by three-dimensional particles in real three-dimensional simulation is larger than that in the two-dimensional simulation with particles without Z-direction velocity, e.g., 2D Cavity flow\cite{zhu2019unified}.

The computational time for dugksFoam is around $154.1$ hours with $5000$ iterations to reach a velocity residual of $2.4\times10^{-7}$. The UGKWP solution takes $93.2$ hours, including $23000$ steps of averaging. The total memory consumption of dugksFoam reaches $205$ GB, whereas UGKWP is $70.1$ GB. For the low-speed flow calculation in the transition regime, the UGKWP method is as expensive as the explicit DUGKS. Techniques such as low variance DSMC\cite{baker2005variance,homolle2007low} can be incorporated in the UGKWP method to improve its efficiency for the low speed flow. However, as the Knudsen number decreases further to the continuum regime, the computational cost of UGKWP method approaches to the gas-kinetic scheme (GKS)\cite{xu2001gas} for the Navier-Stokes solutions, which has the similar efficiency as a standard NS solver.

\subsection{Flow passing through a cube in rarefied and transition regimes}\label{subsec:Flow_pass_a_cube}
\subsubsection{Supersonic flow in rarefied regime}
The first case is a supersonic rarefied gas flow passing through a cube at $Ma = 2$ and $Kn = 1$. The cube center is located at $(0,0,0)$, and the cube volume is $1 m^3$. The surfaces of the cube are diffusive wall boundary condition with a constant temperature $T_w = 273K$. Due to the symmetry, only a quadrant of the cube is simulated by UGKWP. The computational domain $[-6,8]\times[0,8]\times[0,8]$ is discretized by $(32+14+34) \times (7+34) \times (7+34) = 80 \times 41 \times 41$ cells with uniformly distributed grids on the surface of the cube. The cell size is stretched from the cube surface with a ratio of $1.0764$ up to the front side and a ratio of $1.083$ at the rear and lateral sides of the cube. The inflow is monatomic argon gas with molecular mass $m = 6.63 \times 10^{-26} kg$ and diameter $d=4.17 \times 10^{-10} m$. The CFL number for UGKWP simulation is $0.9$, and the reference viscosity is given by the variable hard sphere (VHS) model with $\omega = 0.81$. To capture the lowest temperature that appears at the rear of the cube caused by the expanding flow, a large number of particles $N_{ref} = N_{min} = 5000$ is required. The simulation is carried out with first-order spatial accuracy to reduce the noises caused by unreliable wave reconstruction. The time-averaging starts from $1400$ steps with an initial field computed by $1000$ steps GKS. The simulation runs $120.8$ hours with 48 cores and consumes $183$ GB memory, including $8600$ steps of averaging.

\begin{figure}[htbp!]
    \centering
    \begin{subfigure}[b]{0.49\textwidth}
        \includegraphics[width=\textwidth]{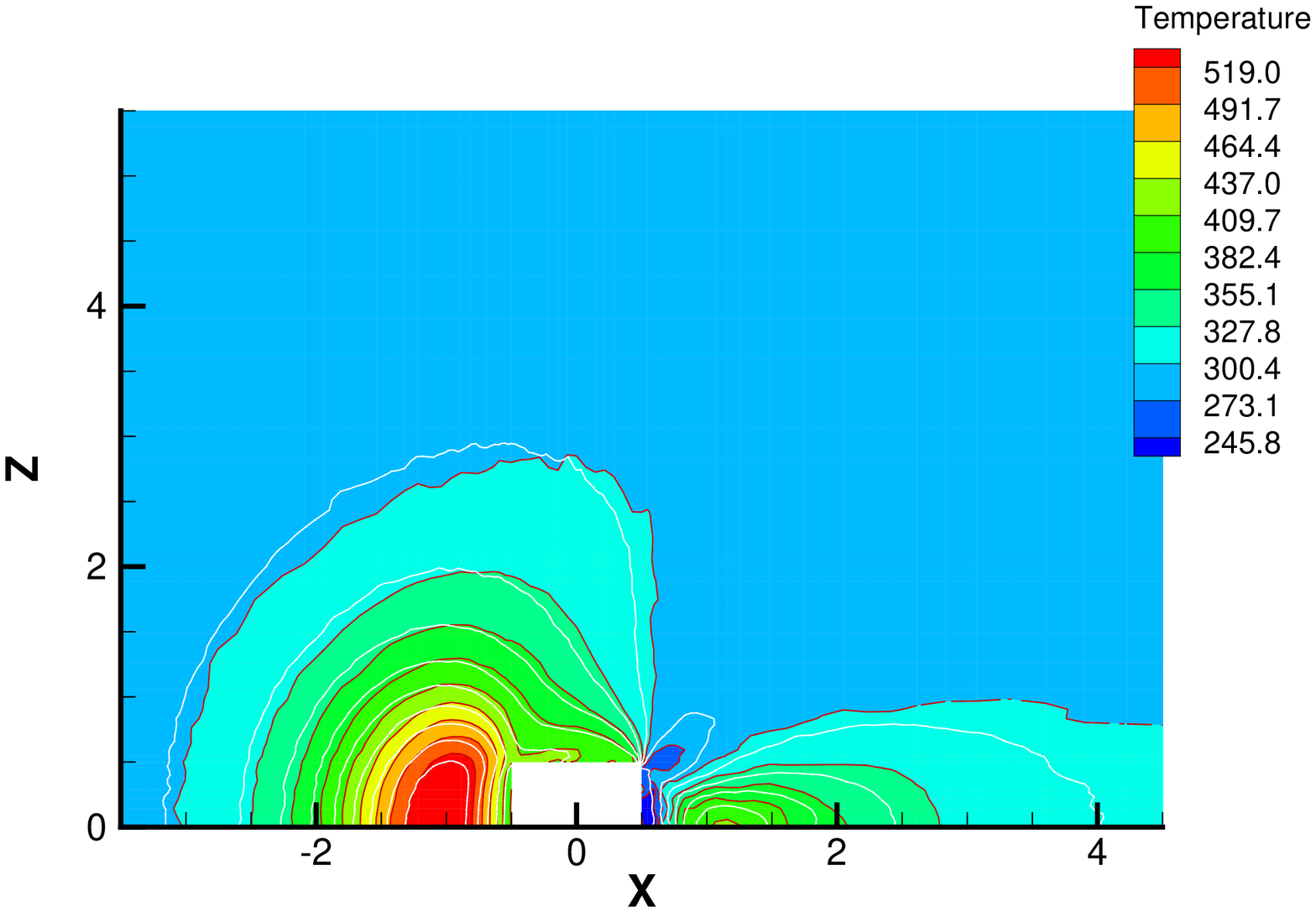}
        \caption{}
        \label{fig:cube3D_Y-plane_Ma2_Kn1_Temperature_UGKWP_DSMC}
    \end{subfigure}
        \begin{subfigure}[b]{0.49\textwidth}
        \includegraphics[width=\textwidth]{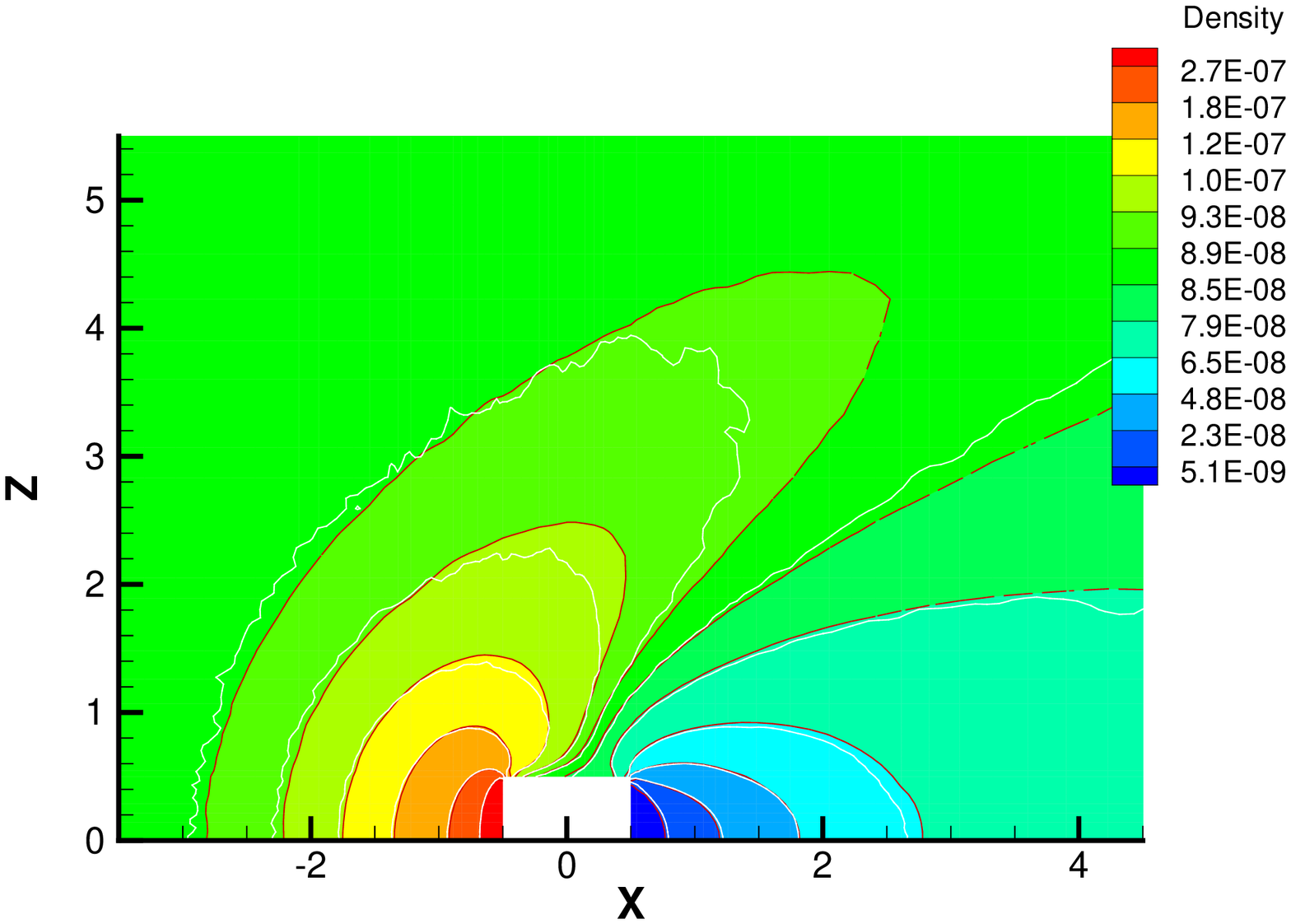}
        \caption{}
        \label{fig:cube3D_Y-plane_Ma2_Kn1_Density_UGKWP_DSMC}
    \end{subfigure}\\
    \begin{subfigure}[b]{0.49\textwidth}
        \includegraphics[width=\textwidth]{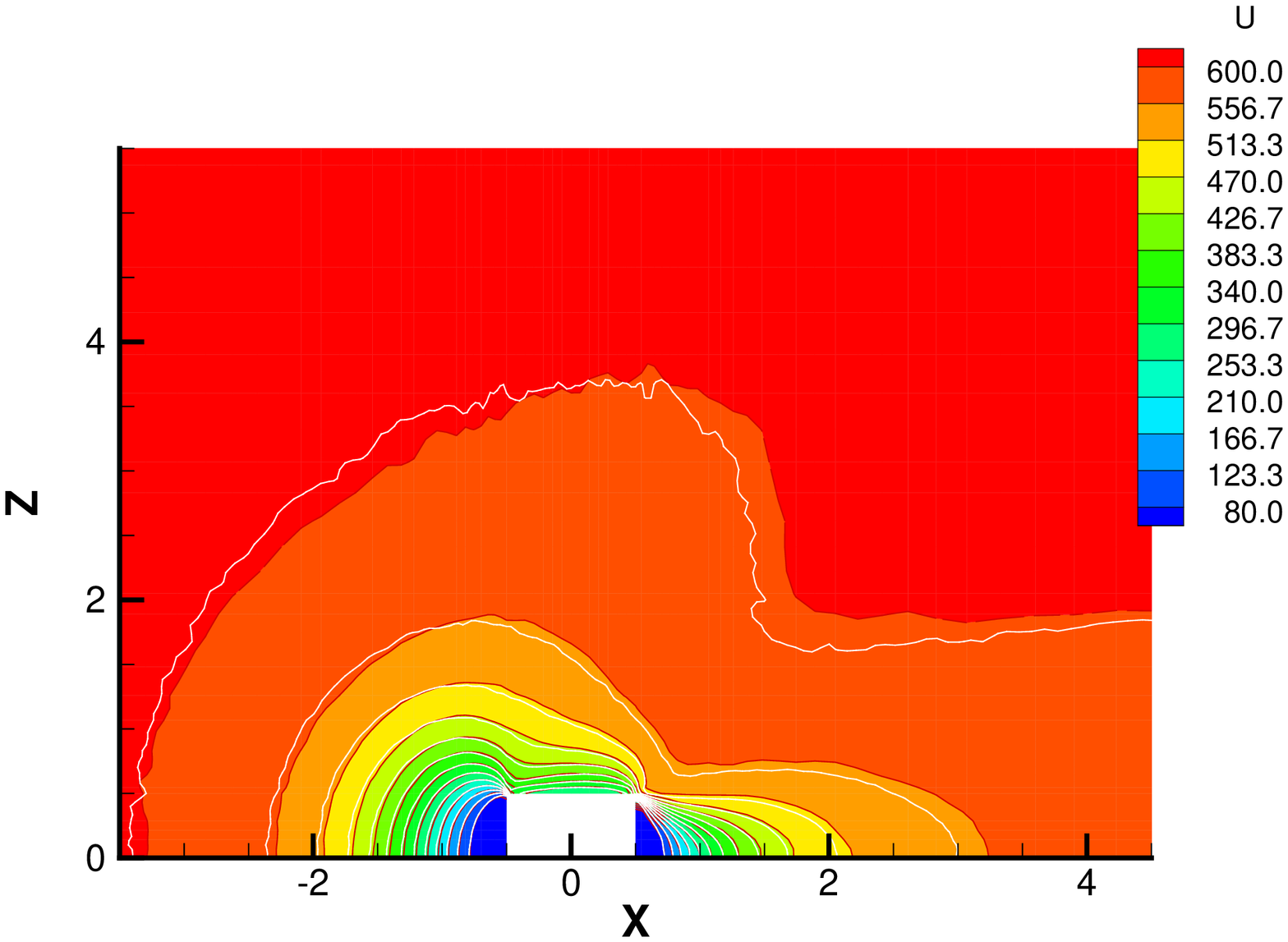}
		\caption{}
        \label{fig:cube3D_Y-plane_Ma2_Kn1_U_UGKWP_DSMC}
    \end{subfigure}
    \begin{subfigure}[b]{0.49\textwidth}
        \includegraphics[width=\textwidth]{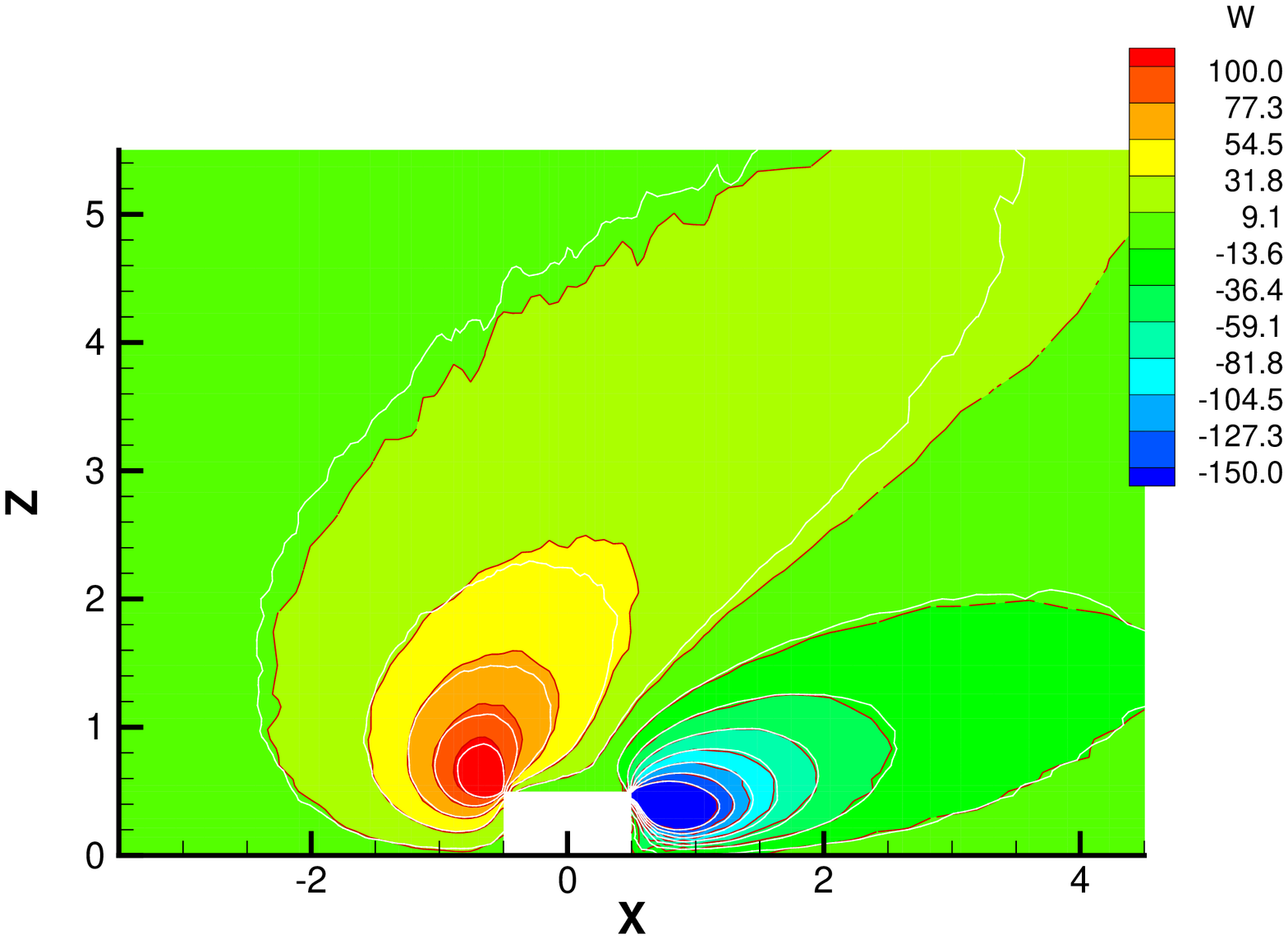}
        \caption{}
        \label{fig:cube3D_Y-plane_Ma2_Kn1_W_UGKWP_DSMC}
    \end{subfigure}
    \caption{Comparison of distributions between UGKWP and DSMC on the X-Z symmetric cut-plane at $Ma = 2$ and $Kn = 1$. Dashed red lines with colored background represent the UGKWP result, and the solid white lines denote the DSMC solution. (a) temperature contour, (b) density contour, (c) contour of $U$ (X-component velocity) and (d) contour of $W$ (Z-component velocity).}\label{fig:cube3D_Y-plane_Ma2_Kn1_UGKWP_DSMC}
\end{figure}

\Cref{fig:cube3D_Y-plane_Ma2_Kn1_UGKWP_DSMC} presents the temperature, density, and velocities distributions on the X-Z symmetry plane which are compared with the benchmark DSMC result, which is obtained using dsmcFoam\cite{scanlon2010open} \cite{zhu2019gpu}. Similar to UGKWP simulation, a quadrant of the cube is simulated in DSMC with a much finer physical grid of $191 \times 91 \times 91$. Each cell has 50 particles on average, and the time step is $2.0 \times 10^{-7}s$. The averaging begins from $1000$ steps and continues for $68000$ steps, which take $128.5$ hours on $128$ CPU cores (Xeon E5-2680v3 (Haswell) @2.5 GHz). The results have a satisfactory agreement overall, especially the flow field near the cube wall. However, regarding temperature contour, visible differences can be observed at the front of the bow shock and at the rear part of the cube. The differences come from different kinetic models in UGKWP and DSMC. The UGKWP uses the BGK model and the DSMC solves the full Boltzmann collision term.
Same as UGKS \cite{liu2016unified}, more realistic models, such as Shakhov and the full Boltzmann collision term can be used in the
 construction of UGKWP. The research in this direction is under investigation.

Another notable point is that UGKWP requests at least $5000$ particles per cell in the rear part of the cube to get the temperature field at such a low-density region. In contrast, to use roughly $500$ particles per cell in UGKWP is enough to capture the nonequilibrium shock structure in front of the cube. The reason is that in the case of $\tau\gg\Delta t$, the simulation particle increases, and the stochastic noise becomes significant. Instead of choosing collision pairs in DSMC, UGKWP re-sample the collisional particles according to the cell averaged temperature and macroscopic velocity. When the temperature has a small variance, such as in the low-speed cavity flow, the inadequate particle number could deteriorate the temperature with noise. Then, the re-sampled collisional particles even inherit the inaccuracy and poison the low-temperature area at the rear part of the cube, where artificial heating with over-estimated temperature appears. Even though with the above weakness, UGKWP can perform simulation on a coarse mesh than that used in the DSMC. Consequently, UGKWP and DSMC have comparable computational cost in the rarefied regime.

\begin{figure}[htbp!]
    \centering
    \begin{subfigure}[b]{0.49\textwidth}
        \includegraphics[width=\textwidth]{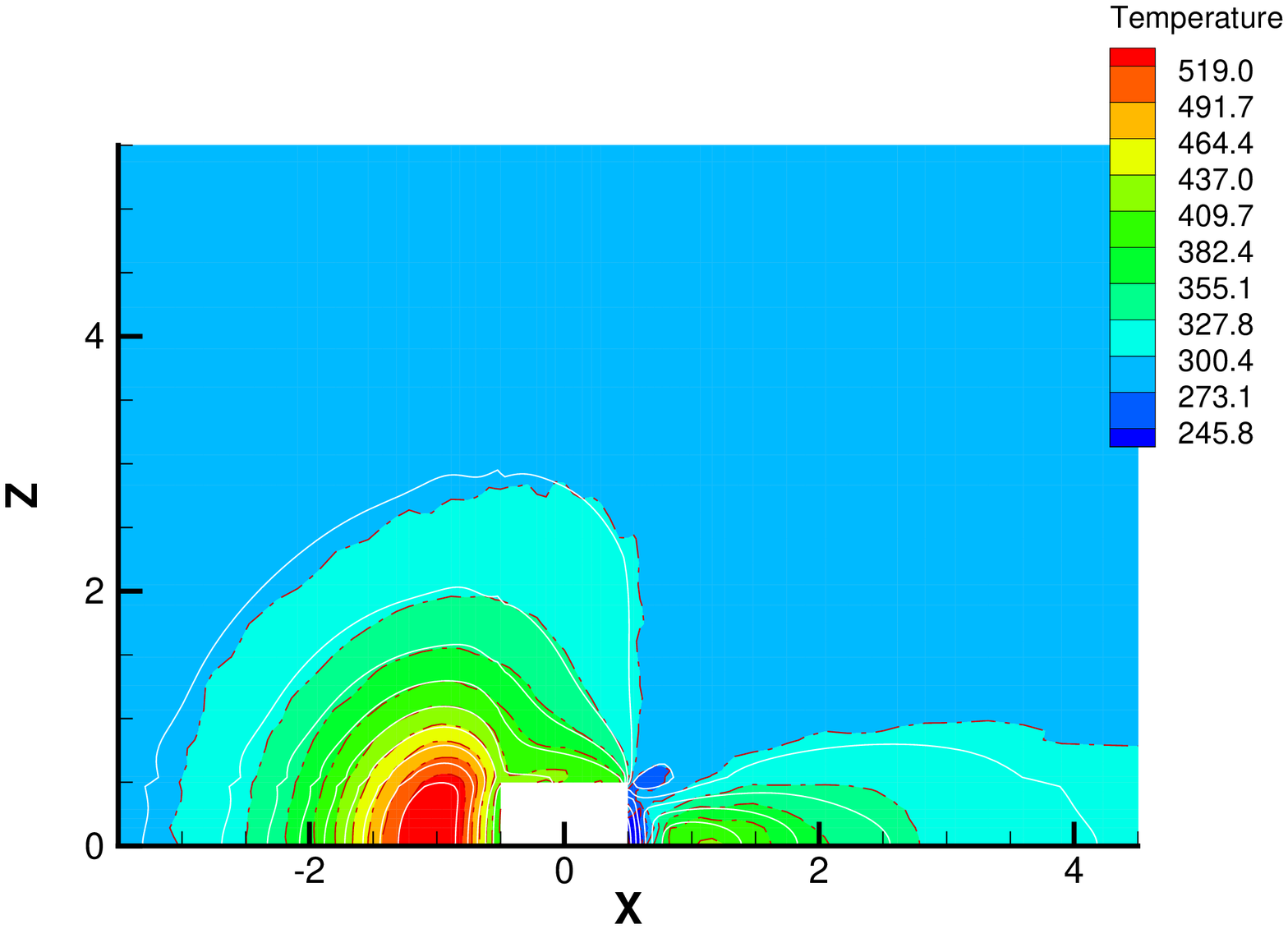}
        \caption{}
        \label{fig:cube3D_Y-plane_Ma2_Kn1_Temperature_UGKWP_DVM}
    \end{subfigure}
    \begin{subfigure}[b]{0.49\textwidth}
        \includegraphics[width=\textwidth]{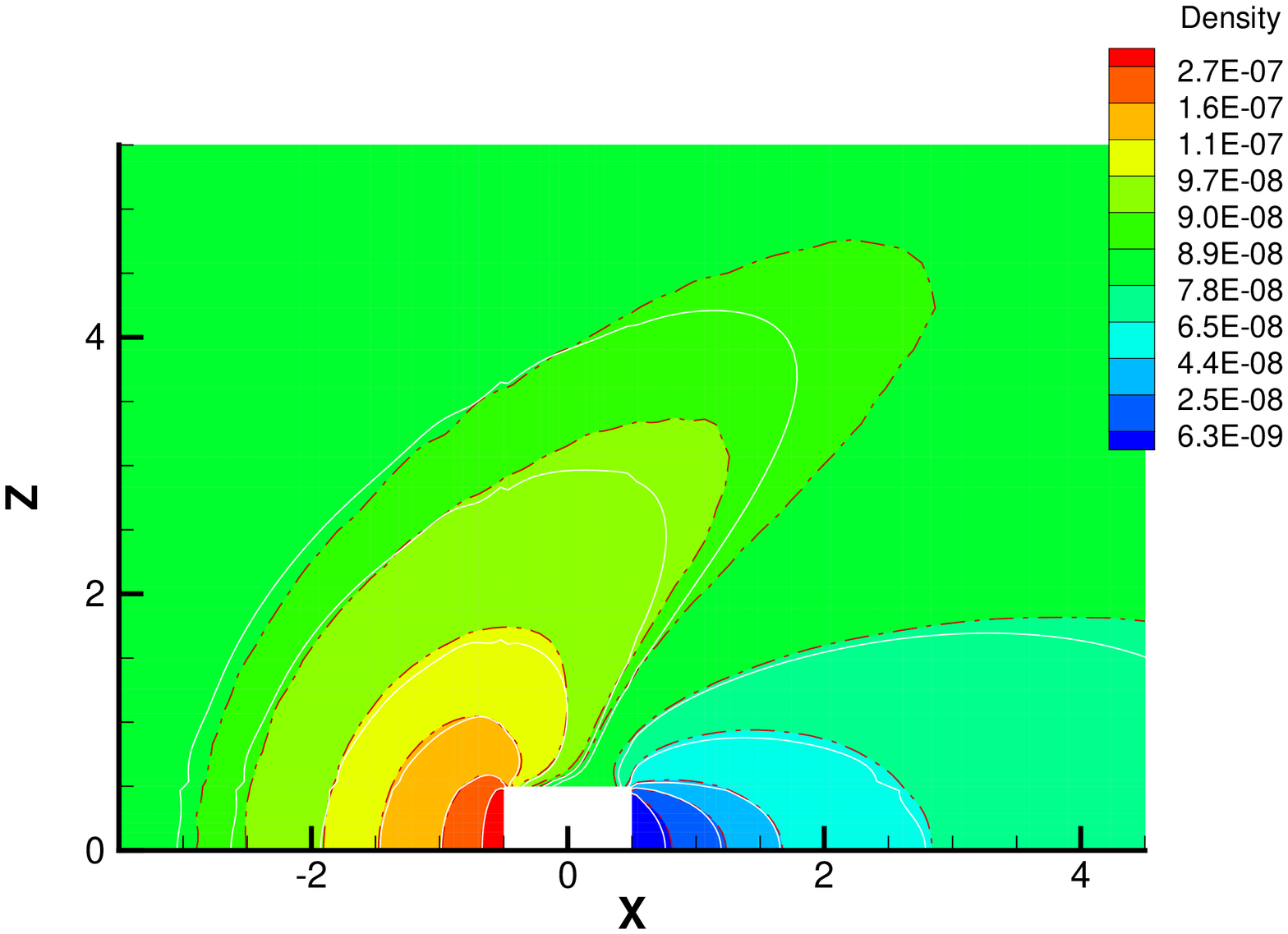}
        \caption{}
        \label{fig:cube3D_Y-plane_Ma2_Kn1_Density_UGKWP_DVM}
    \end{subfigure}\\
    \begin{subfigure}[b]{0.49\textwidth}
        \includegraphics[width=\textwidth]{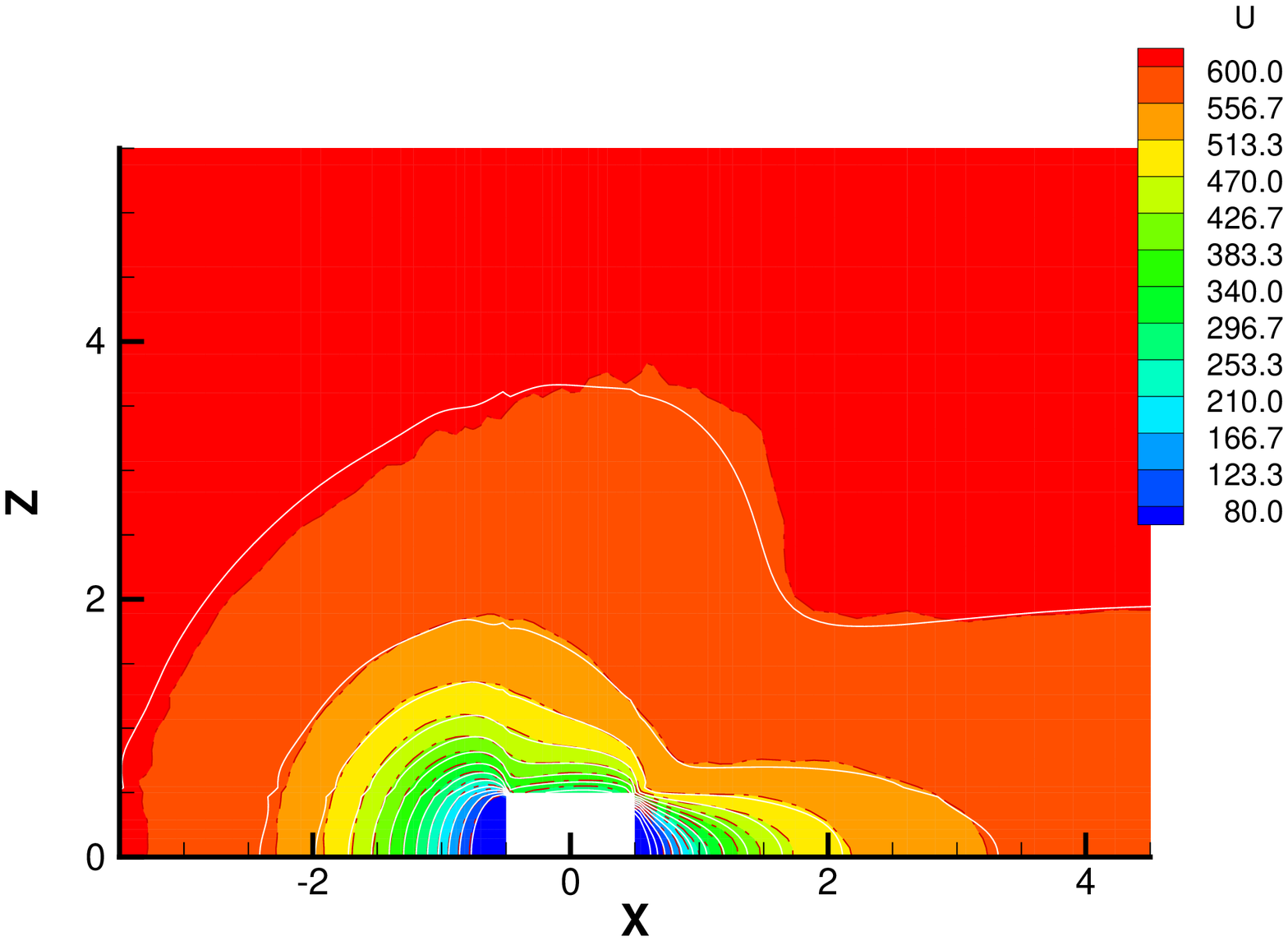}
		\caption{}
        \label{fig:cube3D_Y-plane_Ma2_Kn1_U_UGKWP_DVM}
    \end{subfigure}
    \begin{subfigure}[b]{0.49\textwidth}
        \includegraphics[width=\textwidth]{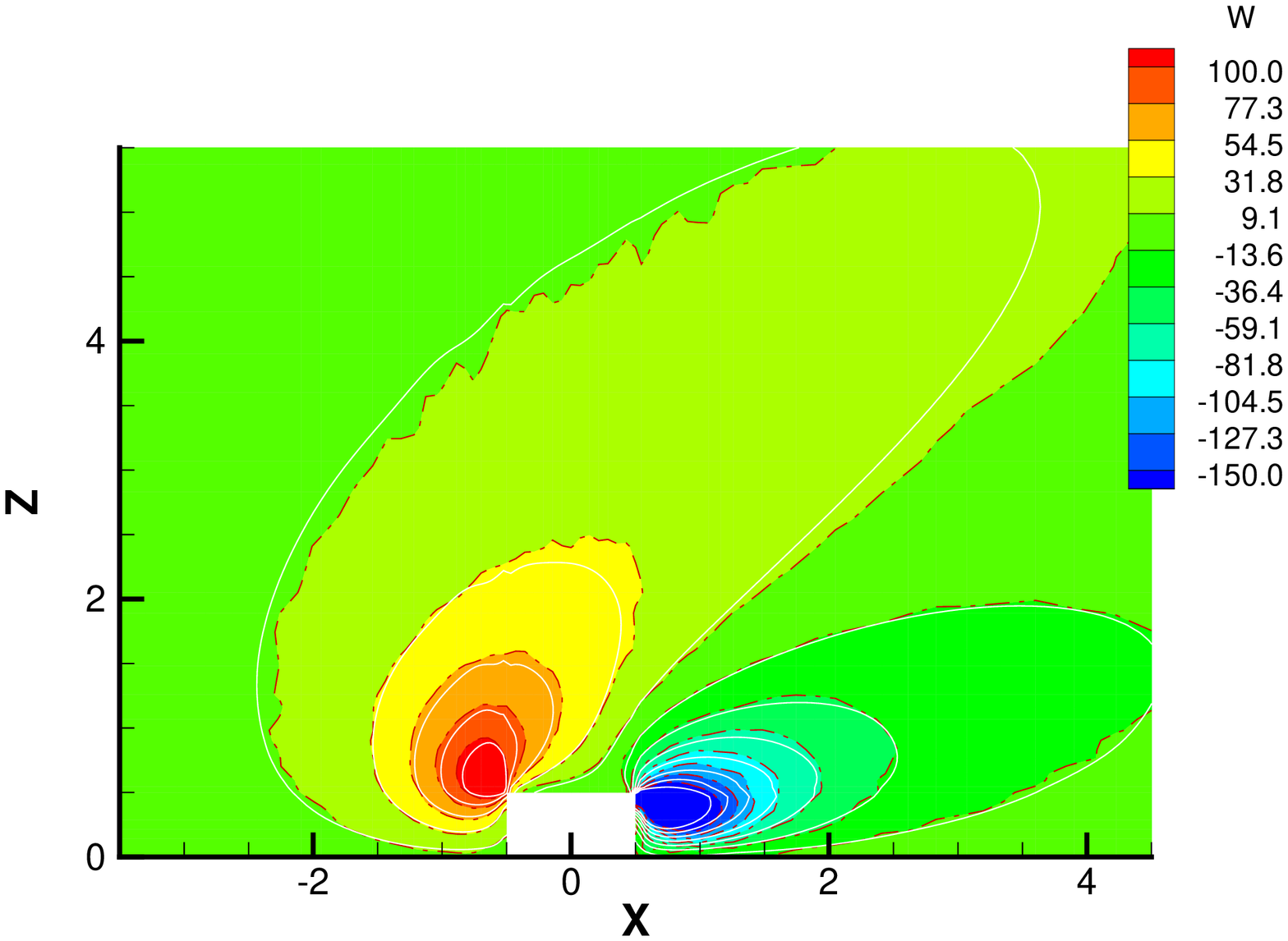}
        \caption{}
        \label{fig:cube3D_Y-plane_Ma2_Kn1_W_UGKWP_DVM}
    \end{subfigure}
    \caption{Comparison of distributions between UGKWP and DVM on the X-Z symmetric cut-plane at $Ma = 2$ and $Kn = 1$. Dashed red lines with colored background represent the UGKWP result, and the solid white lines denote the DVM solution. (a) temperature contour, (b) density contour, (c) contour of $U$ (X-component velocity) and (d) contour of $W$ (Z-component velocity).}\label{fig:cube3D_Y-plane_Ma2_Kn1_UGKWP_DVM}
\end{figure}

In DVM method \cite{zhu2019gpu}, the implicit discretization with memory reduction technique on GPU is implemented. The full cube is simulated with a physical grid $191 \times 181 \times 181$ and a velocity grid $48^3$. The velocity points are distributed uniformly to cover a range of $[-4\sqrt{2RT_w},4\sqrt{2RT_w}]^3$, and the trapezoidal rule is used to calculate the moments. The simulation takes approximately $20$ hours, with $41$ iteration steps on the Tesla K40 GPU. \Cref{fig:cube3D_Y-plane_Ma2_Kn1_UGKWP_DVM} shows the detailed comparisons of the temperature, density, and velocities distributions on the X-Z symmetry plane with DVM solutions. Similar to the comparison with DSMC, the shock thickness and the separation distance between the density and temperature profiles have small variations between UGKWP and implicit DVM solutions, where the Shakhov model is used in DVM. Again, the UGKWP uses the BGK model. The differences in the shock structure solution between the BGK and Shakhov model have been presented in \cite{xu2011improved}. As for computational time, owing to the implicit treatment and implementation of GPU acceleration, the DVM simulation is around an order faster than the current UGKWP simulation at such a low Mach number $Ma=2$.

\subsubsection{Hypersonic flow in transition regime}
\begin{figure}[htbp!]
    \centering
    \begin{subfigure}[b]{0.49\textwidth}
        \includegraphics[width=\textwidth]{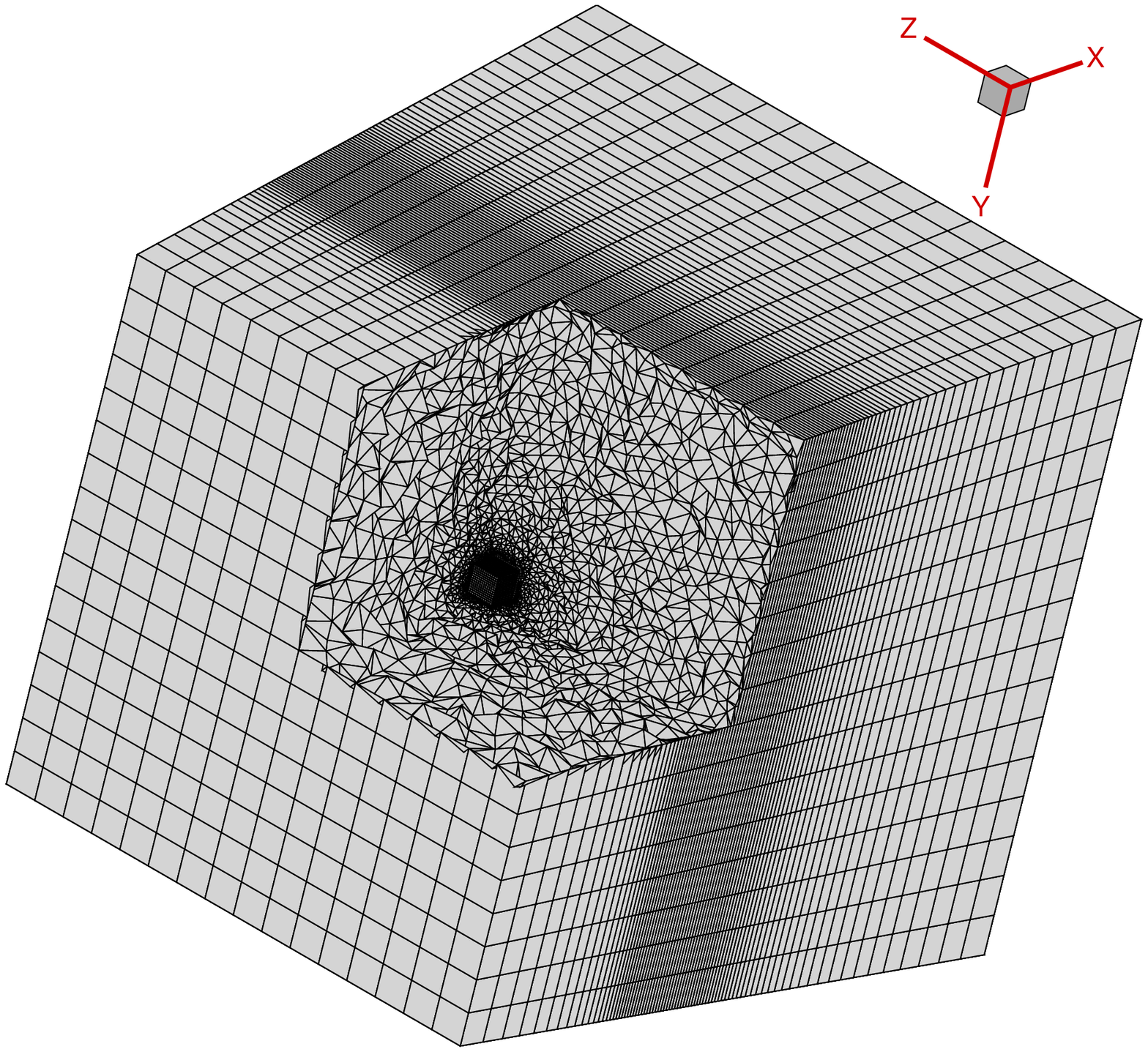}
        \caption{}
        \label{fig:cube3D_Ma20_Kn005_full_mesh_UGKWP}
    \end{subfigure}
    \begin{subfigure}[b]{0.49\textwidth}
        \includegraphics[width=\textwidth]{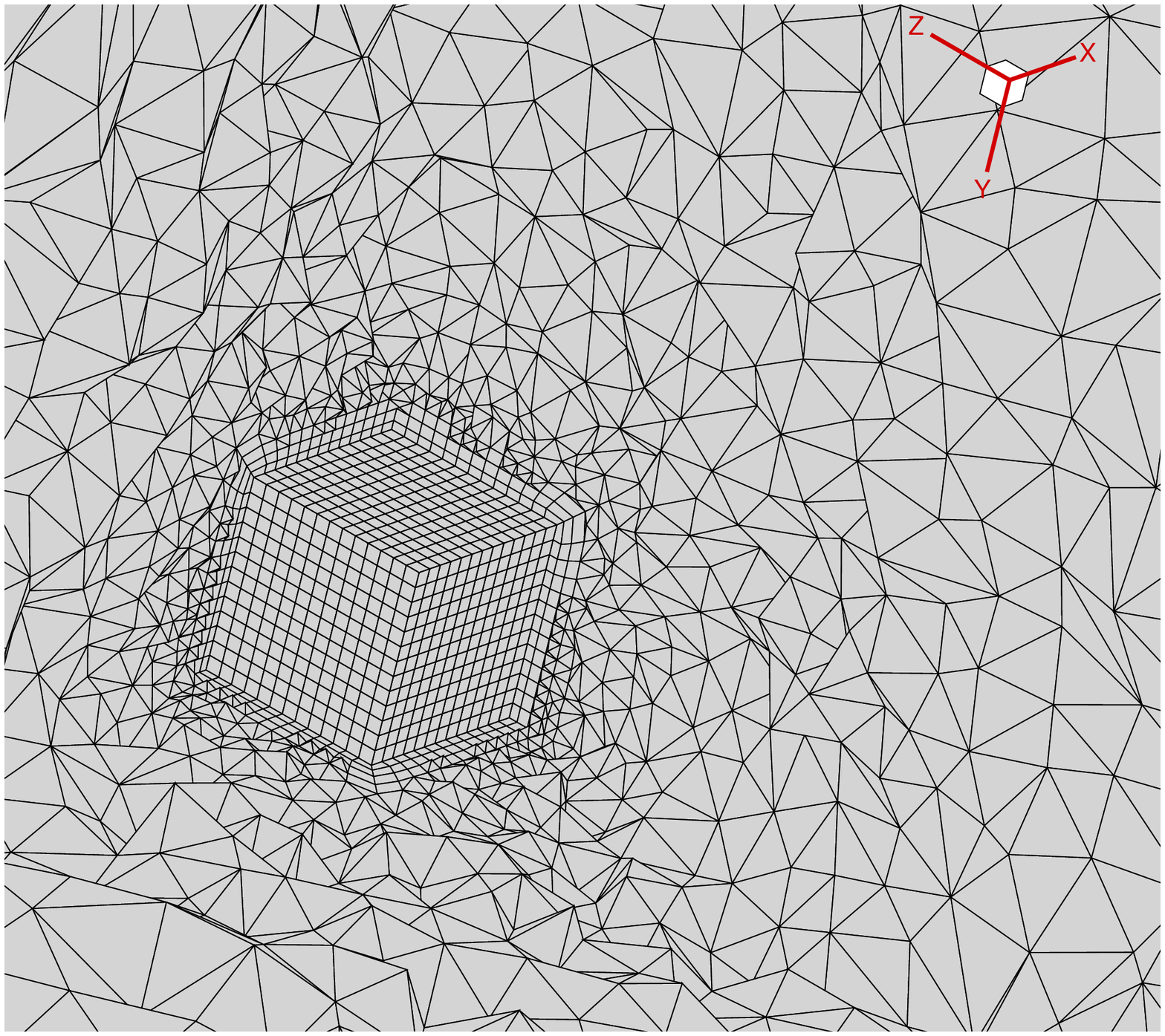}
        \caption{}
        \label{fig:cube3D_Ma20_Kn005_local_mesh_UGKWP}
    \end{subfigure}
    \caption{Unstructured mesh configuration at $Ma = 20$ and $Kn = 0.05$ (a) Full view and (b) local enlargement}\label{fig:cube3D_Ma20_Kn005_mesh_UGKWP}
\end{figure}

To highlight the efficiency and capability of UGKWP, hypersonic flow at $Ma=20$ is simulated in the transition regime ($Kn = 0.05$). All the parameters are the same as the previous case except that the inflow temperature is $56K$, and the wall temperature is $300K$. The cube is contained in a volume with base and top side lengths of $a=16m, b=10m$, and height $h=14m$.
As shown in \Cref{fig:cube3D_Ma20_Kn005_mesh_UGKWP}, a unstructured mesh with total $420702$ cells is generated, which is composed of $8305$ hexahedra, $52$ prisms, $25030$ pyramids and $387315$ tetrahedra with a minimum cell height $0.0248m$ near the cube wall. Distinguishable from the rarefied case, the number of particles required drops dramatically. Here, the reference and minimum number of particles per cell $N_{ref} = N_{min} = 400$ are used. The simulation is conducted with the least square reconstruction and Barth and Jespersen limiter. An initial field is firstly computed with $5000$ steps by GKS, and after $8000$ steps of UGKWP calculation the time averaging of the flow field starts for the steady-state solution.  The simulation runs $59.9$ hours with $48$ cores and consumes $68.1$ GB memory, including $7000$ steps of averaging.

\begin{figure}[htbp!]
    \centering
    \begin{subfigure}[b]{0.49\textwidth}
        \includegraphics[width=\textwidth]{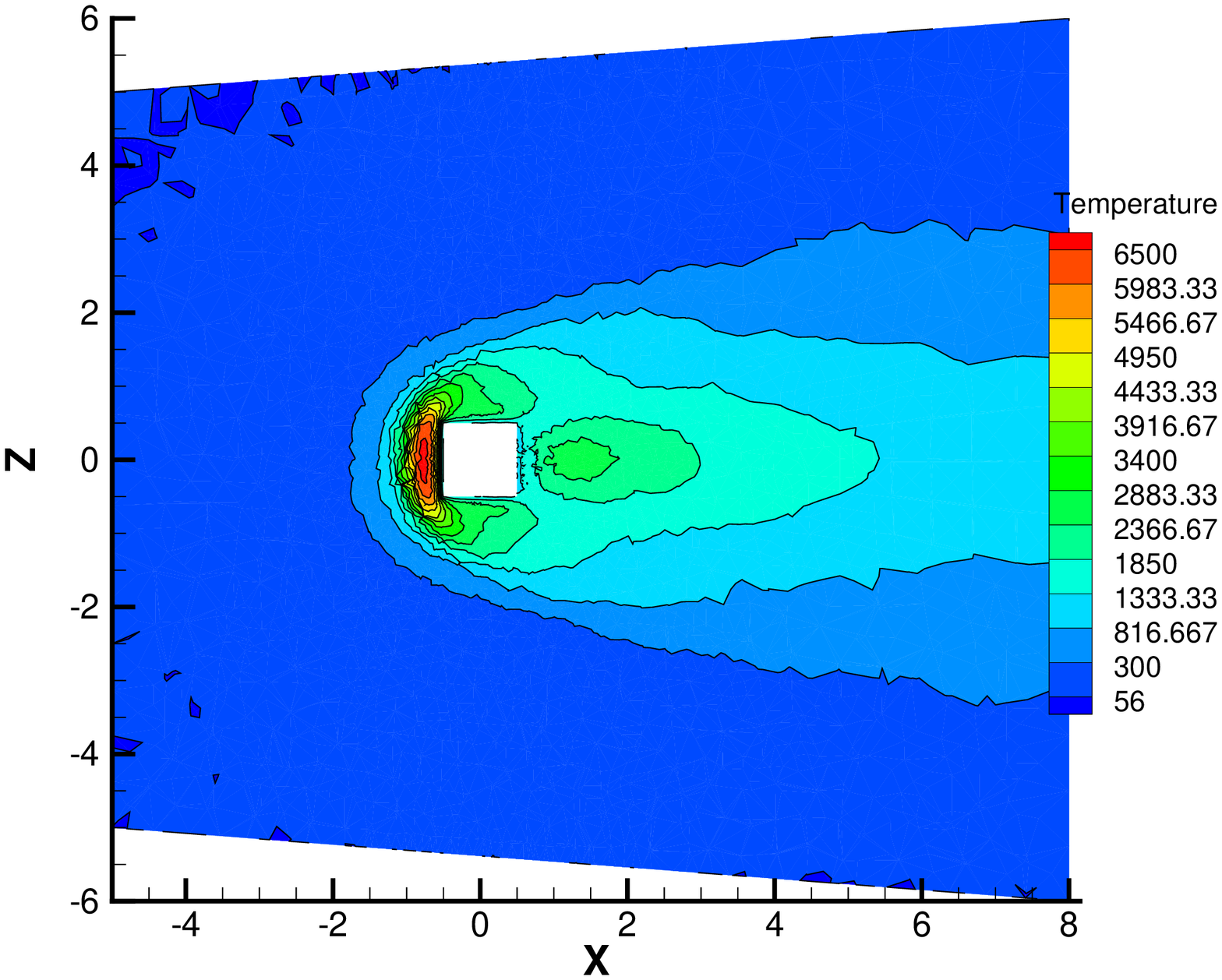}
        \caption{}
        \label{fig:cube3D_Y-plane_Ma20_Kn005_Temperature_UGKWP}
    \end{subfigure}
    \begin{subfigure}[b]{0.49\textwidth}
        \includegraphics[width=\textwidth]{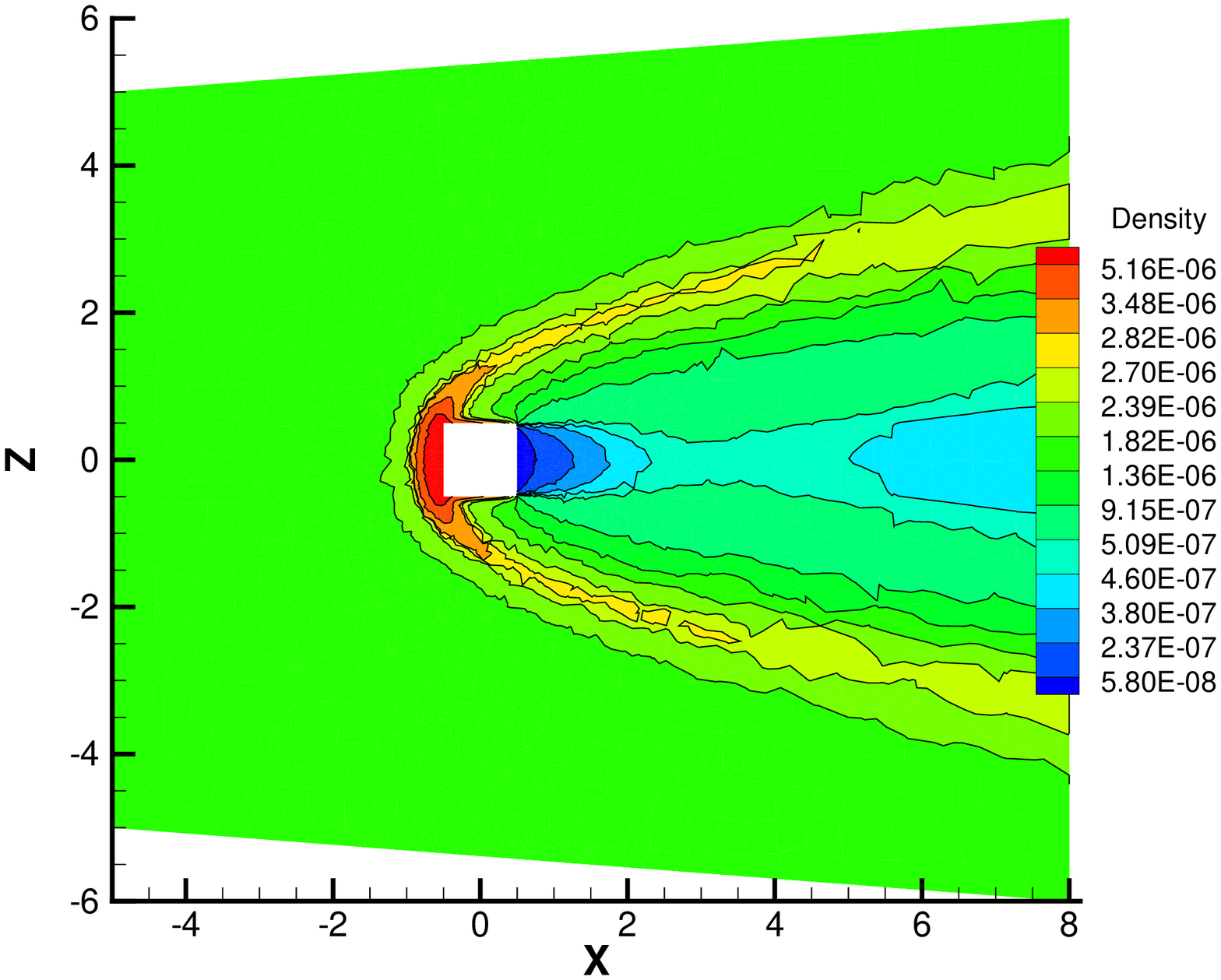}
        \caption{}
        \label{fig:cube3D_Y-plane_Ma20_Kn005_Density_UGKWP}
    \end{subfigure}\\
    \begin{subfigure}[b]{0.49\textwidth}
        \includegraphics[width=\textwidth]{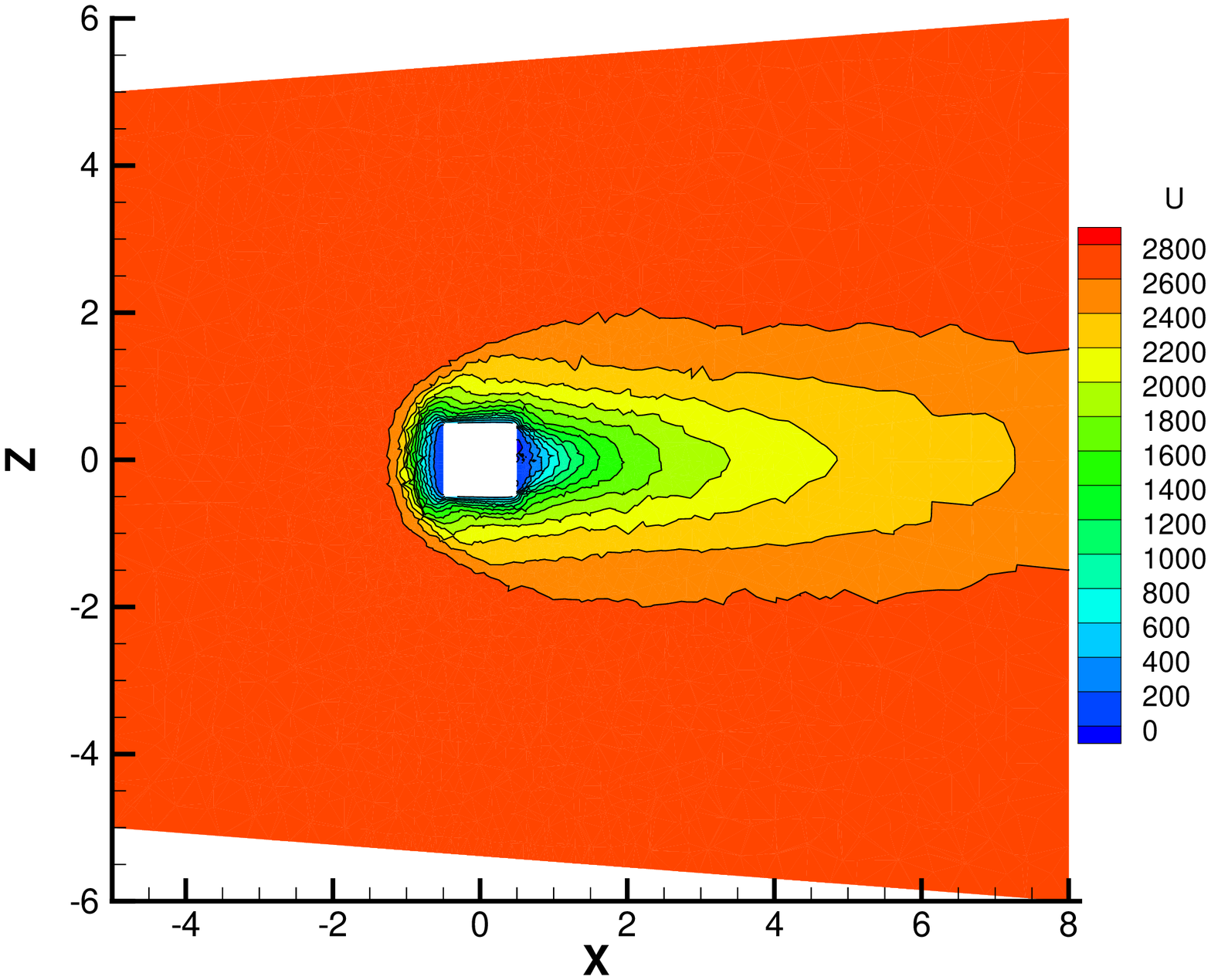}
		\caption{}
        \label{fig:cube3D_Y-plane_Ma20_Kn005_U_UGKWP}
    \end{subfigure}
    \begin{subfigure}[b]{0.49\textwidth}
        \includegraphics[width=\textwidth]{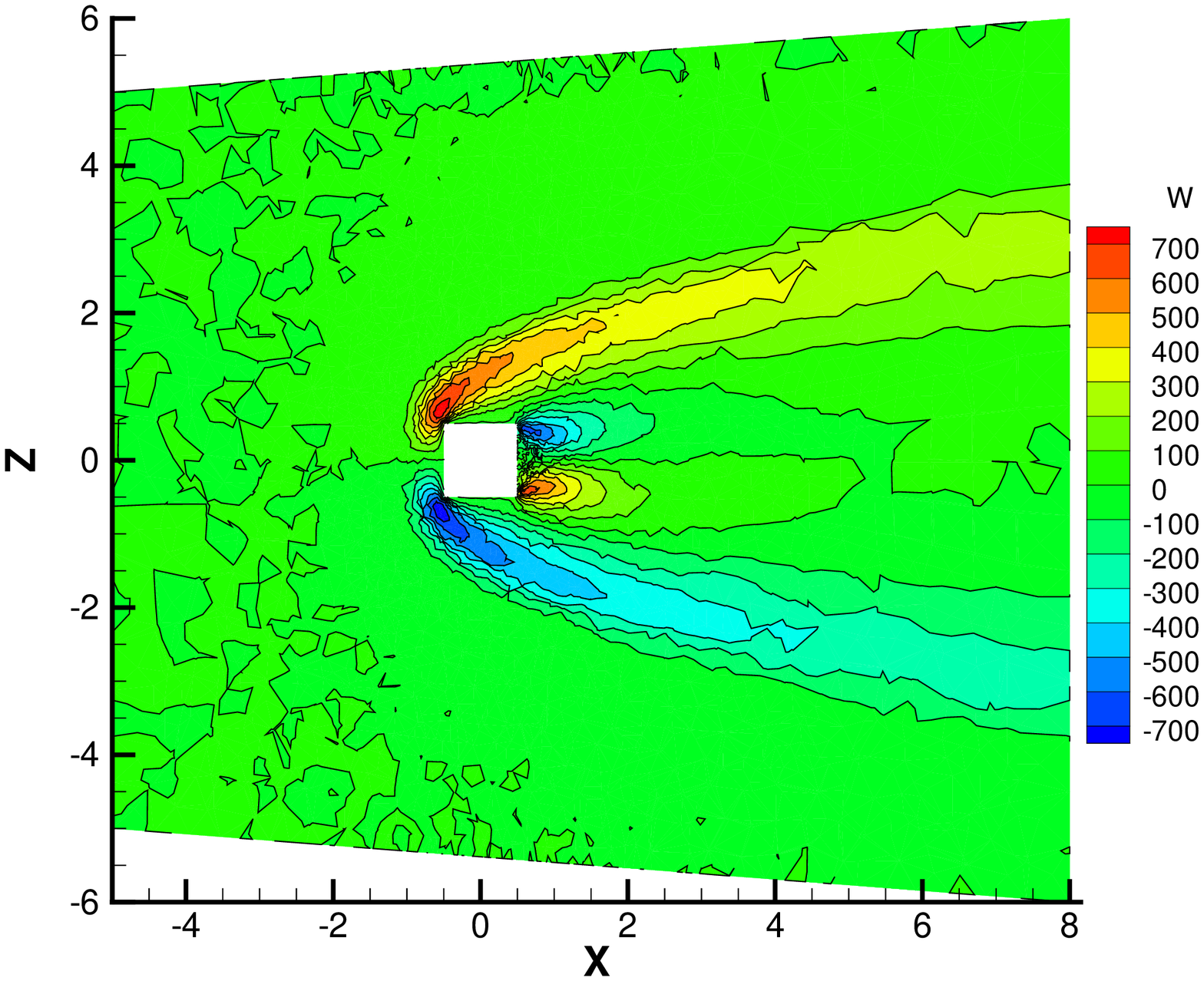}
        \caption{}
        \label{fig:cube3D_Y-plane_Ma20_Kn005_W_UGKWP}
    \end{subfigure}
    \caption{Symmetric X-Z cut-plane contour of various flow fields at at $Ma = 20$ and $Kn = 0.05$. (a) temperature contour, (b) density contour, (c) contour of $U$ (X-component velocity) and (d) contour of $W$ (Z-component velocity).}\label{fig:cube3D_Y-plane_Ma20_Kn005_UGKWP}
\end{figure}

\Cref{fig:cube3D_Y-plane_Ma20_Kn005_UGKWP} shows the distributions of temperature, density, and velocities on the X-Z symmetry plane. The hypersonic flow computation in the transition regime is a challenge for both stochastic and deterministic methods. For the DSMC, an extremely fine mesh in physical space is required. For the DVM-type deterministic solvers, a tremendous amount of discrete velocity points becomes necessary. The UGKWP is an idealized method for the hypersonic flow in all flow regimes. Due to the high Mach number, even with a low inflow temperature of $56K$, the maximum temperature inside the shock region can get to $6500K$ and over. In the future, the physics associating with high temperature, such as ionization and chemical reaction, will be added in UGKWP.

\begin{figure}[htbp!]
    \centering
    \begin{subfigure}[b]{0.49\textwidth}
        \includegraphics[width=\textwidth]{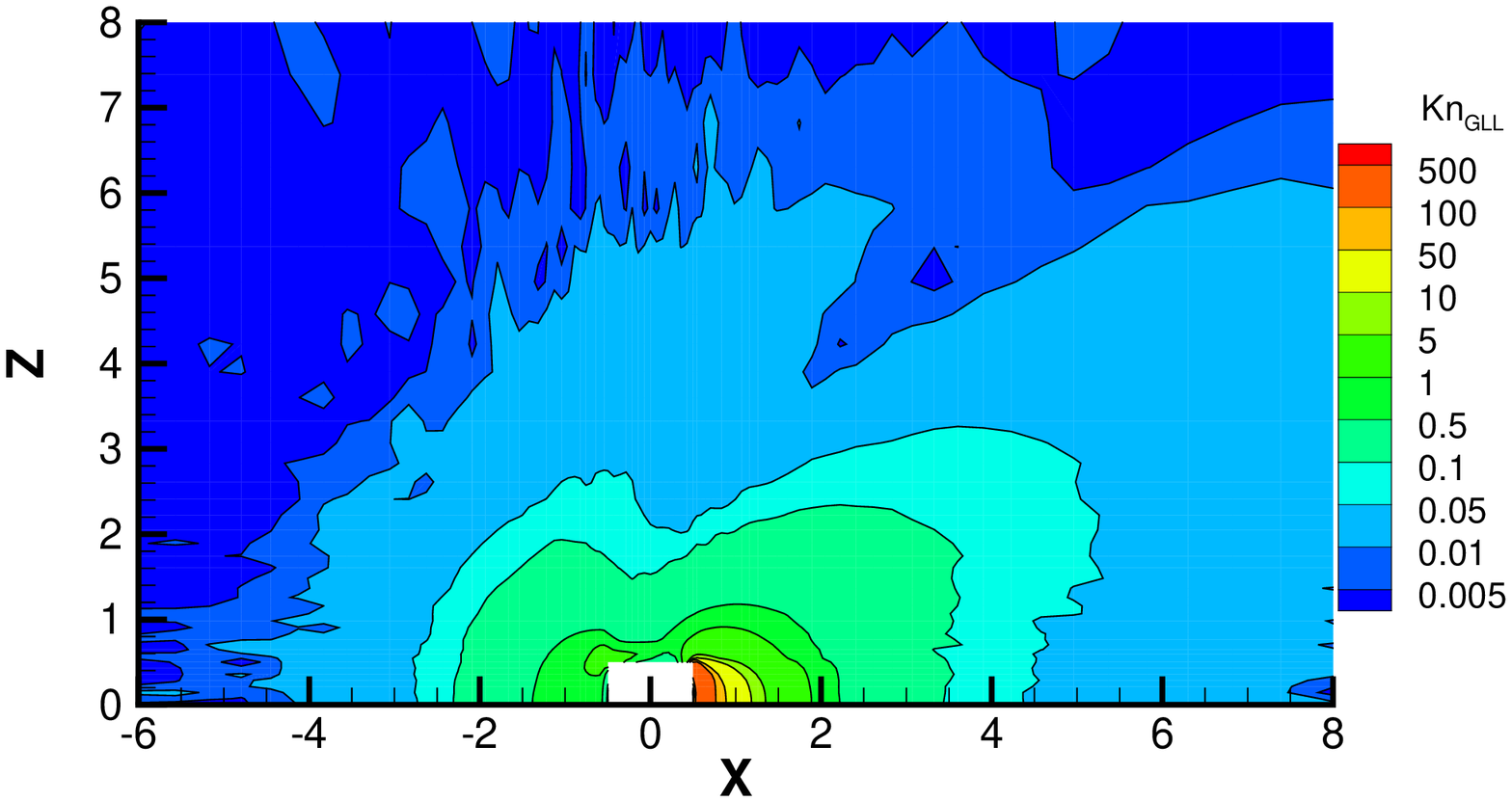}
        \caption{}
        \label{fig:cube3D_Y-plane_Ma2_Kn1_Kn_UGKWP}
    \end{subfigure}
    \begin{subfigure}[b]{0.49\textwidth}
        \includegraphics[width=\textwidth]{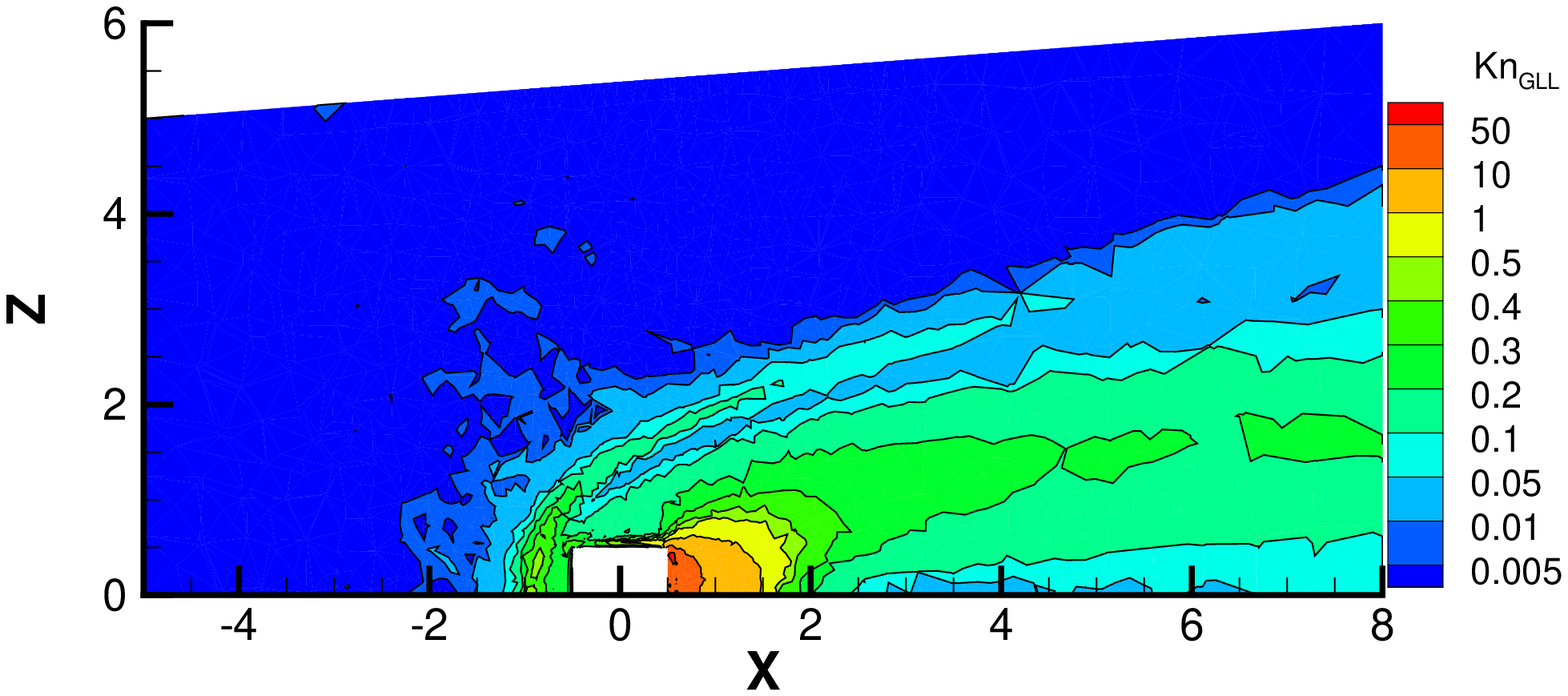}
        \caption{}
        \label{fig:cube3D_Y-plane_Ma20_Kn005_Kn_UGKWP}
    \end{subfigure}
    \caption{Local Knudsen number contour on symmetric X-Z cut-plane at (a)$Ma = 2$ and $Kn = 1$ and (b)$Ma = 20$ and $Kn = 0.05$}\label{fig:cube3D_Y-plane_Kn_UGKWP}
\end{figure}

To further illustrate the multiscale nature of the simulation, the local Knudsen number \cite{lofthouse2008velocity} based on the gradient  $Kn_{GLL} = l|\nabla \rho|/\rho$ for the above two cases are presented in \cref{fig:cube3D_Y-plane_Kn_UGKWP}, which presents five orders of magnitude difference.

\begin{figure}[htbp!]
    \centering
    \begin{subfigure}[b]{0.49\textwidth}
        \includegraphics[width=\textwidth]{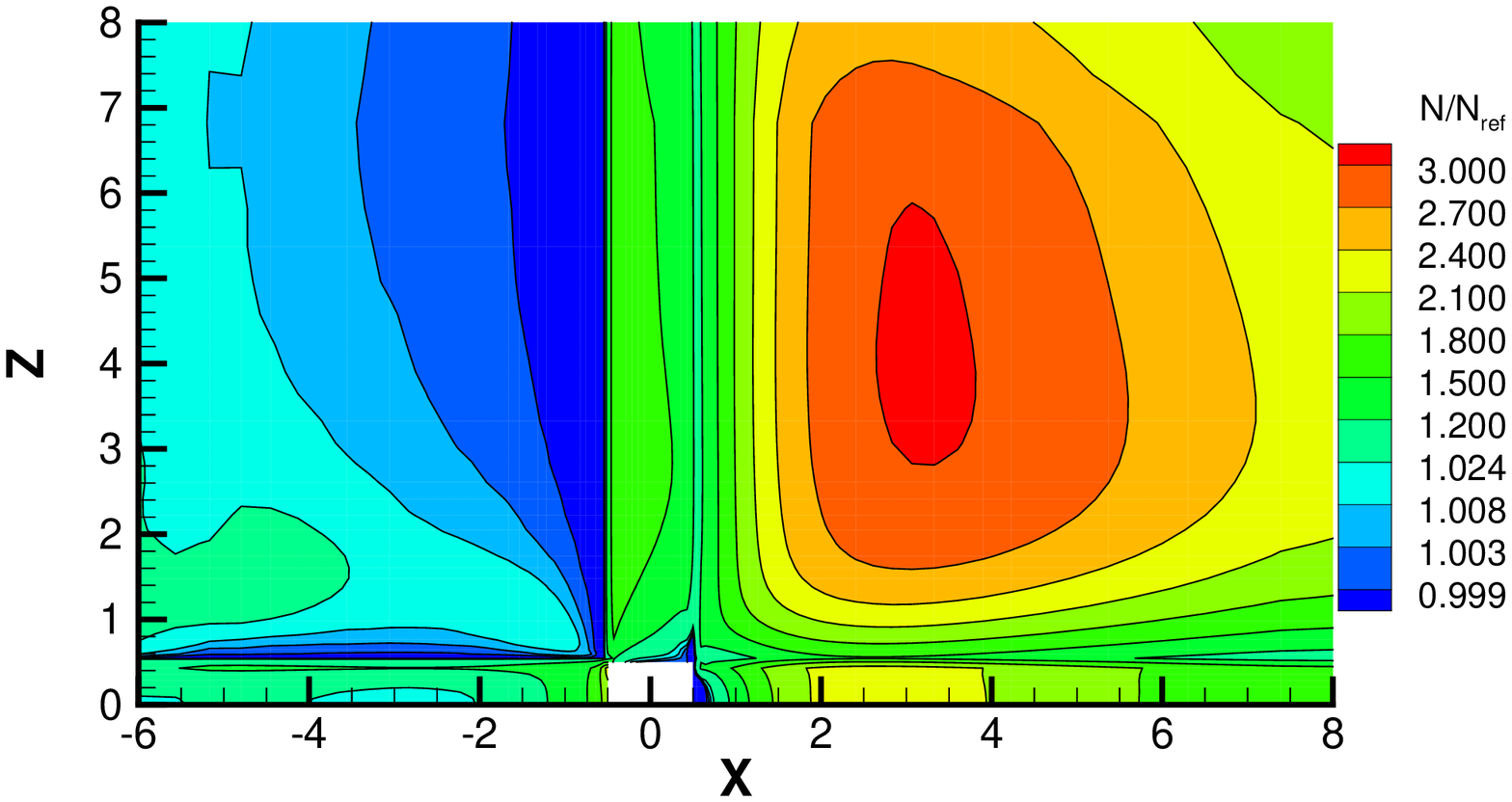}
        \caption{}
        \label{fig:cube3D_Y-plane_Ma2_Kn1_num_UGKWP}
    \end{subfigure}
    \begin{subfigure}[b]{0.49\textwidth}
        \includegraphics[width=\textwidth]{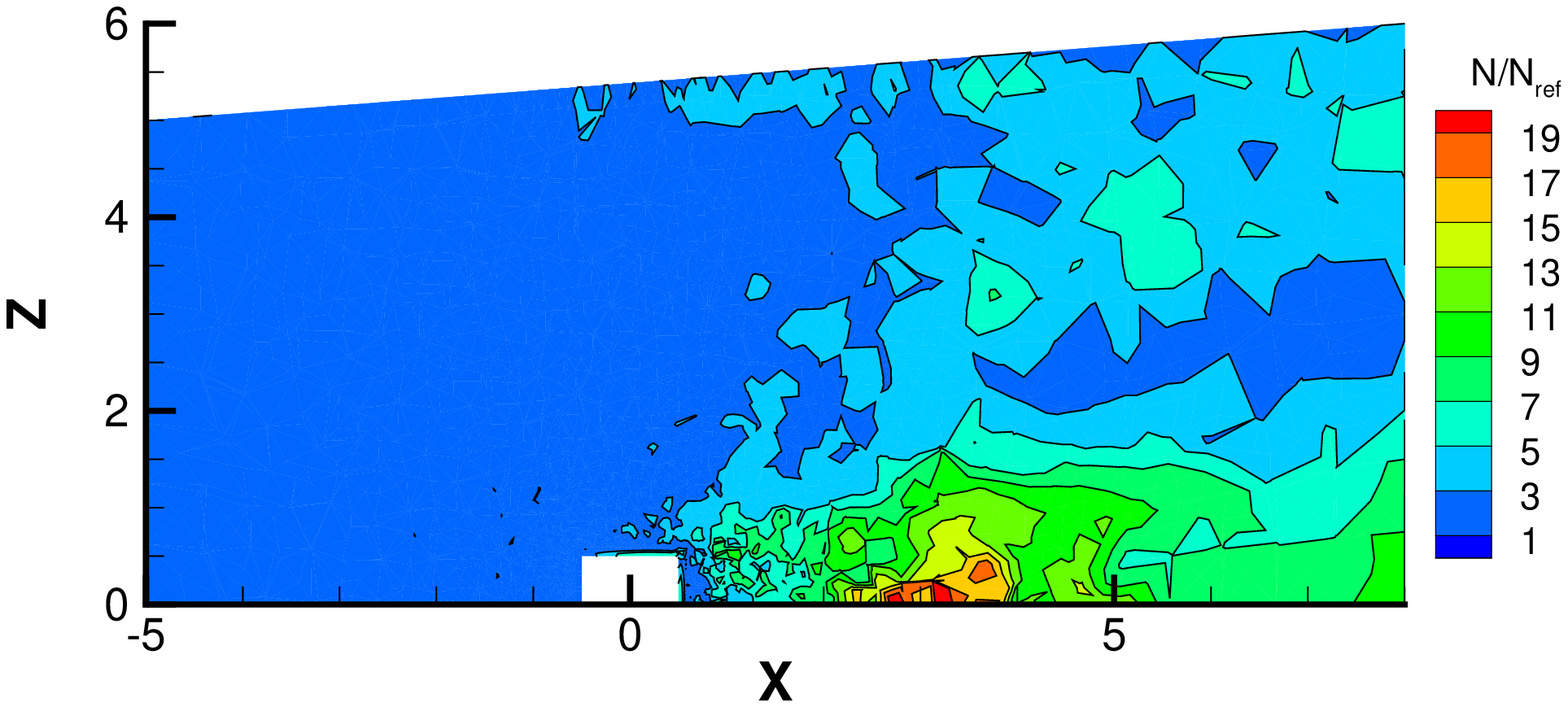}
        \caption{}
        \label{fig:cube3D_Y-plane_Ma20_Kn005_num_UGKWP}
    \end{subfigure}
    \caption{Distribution of normalized particle number per cell on symmetric X-Z cut-plane at (a)$Ma = 2$ and $Kn = 1$ and (b)$Ma = 20$ and $Kn = 0.05$}\label{fig:cube3D_Y-plane_num_UGKWP}
\end{figure}

The decline of parallel efficiency in the rarefied regime, as presented in \cref{subsec:Parallelization}, can be visualized through the averaged number of particles per cell in the simulation. \Cref{fig:cube3D_Y-plane_num_UGKWP} shows the distribution of normalized particle number per cell $N/N_{ref}$ on symmetric X-Z cut-plane at $Ma=2, Kn=1$ and $Ma=20, Kn=0.05$. The probability of particle collision in the cell becomes lower with the increment of $\tau/\Delta t$. Therefore, the particle tends to keep free streaming and concentrate in the rearward of the computational domain, especially in the cell with a large volume. This mechanism causes an imbalance in the distributions of particles across different CPU cores. Nevertheless, this problem can be mitigated by implementing dynamic load balancing as used in the DSMC implementation.

\subsection{Hypersonic flow over a space vehicle}\label{subsec:Hypersonic_flows_over_X-38_like_vehicle}
\begin{figure}[htbp!]
    \centering
    \begin{subfigure}[b]{0.49\textwidth}
        \includegraphics[width=\textwidth]{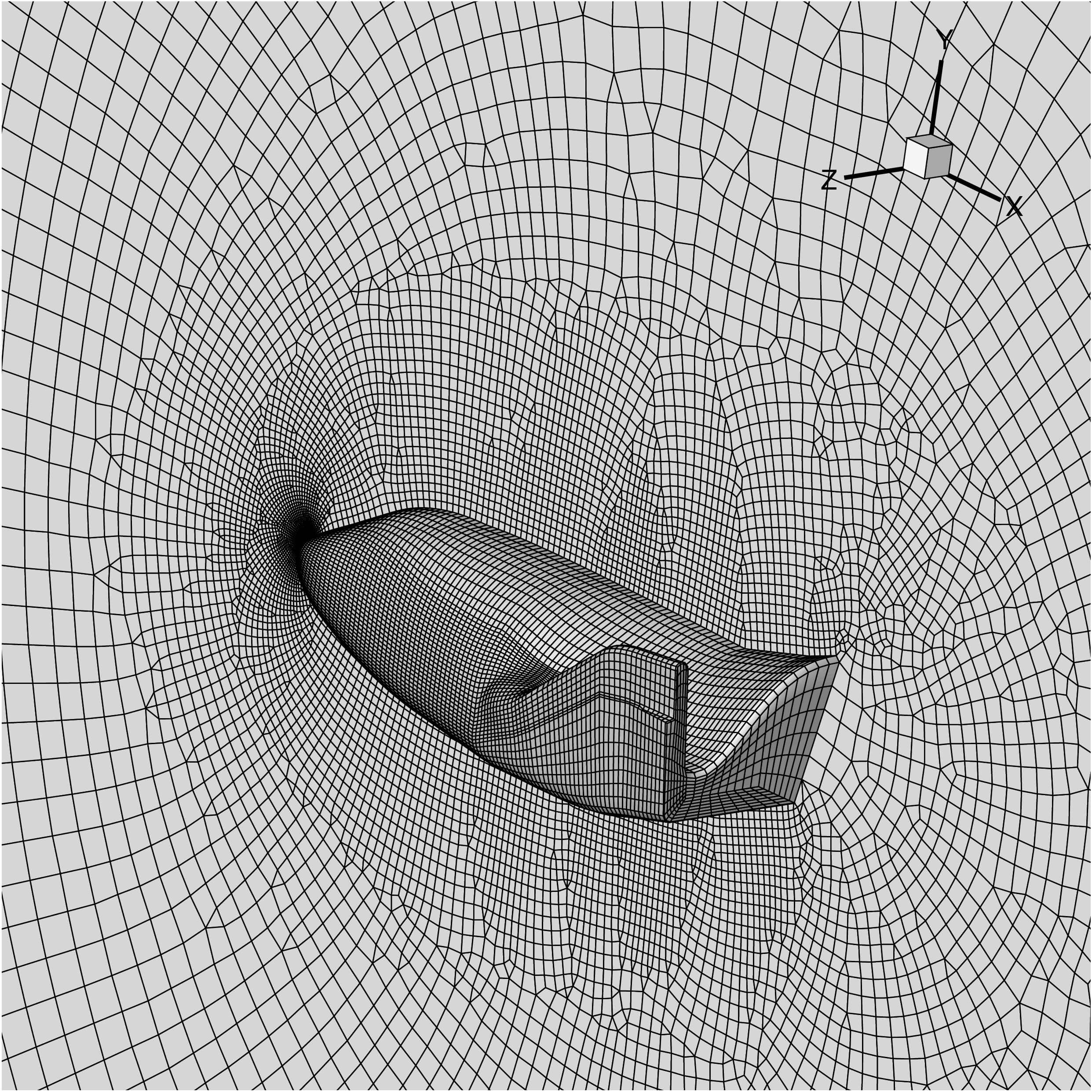}
        \caption{}
        \label{fig:X-38_local_mesh}
    \end{subfigure}
    \begin{subfigure}[b]{0.49\textwidth}
        \includegraphics[width=\textwidth]{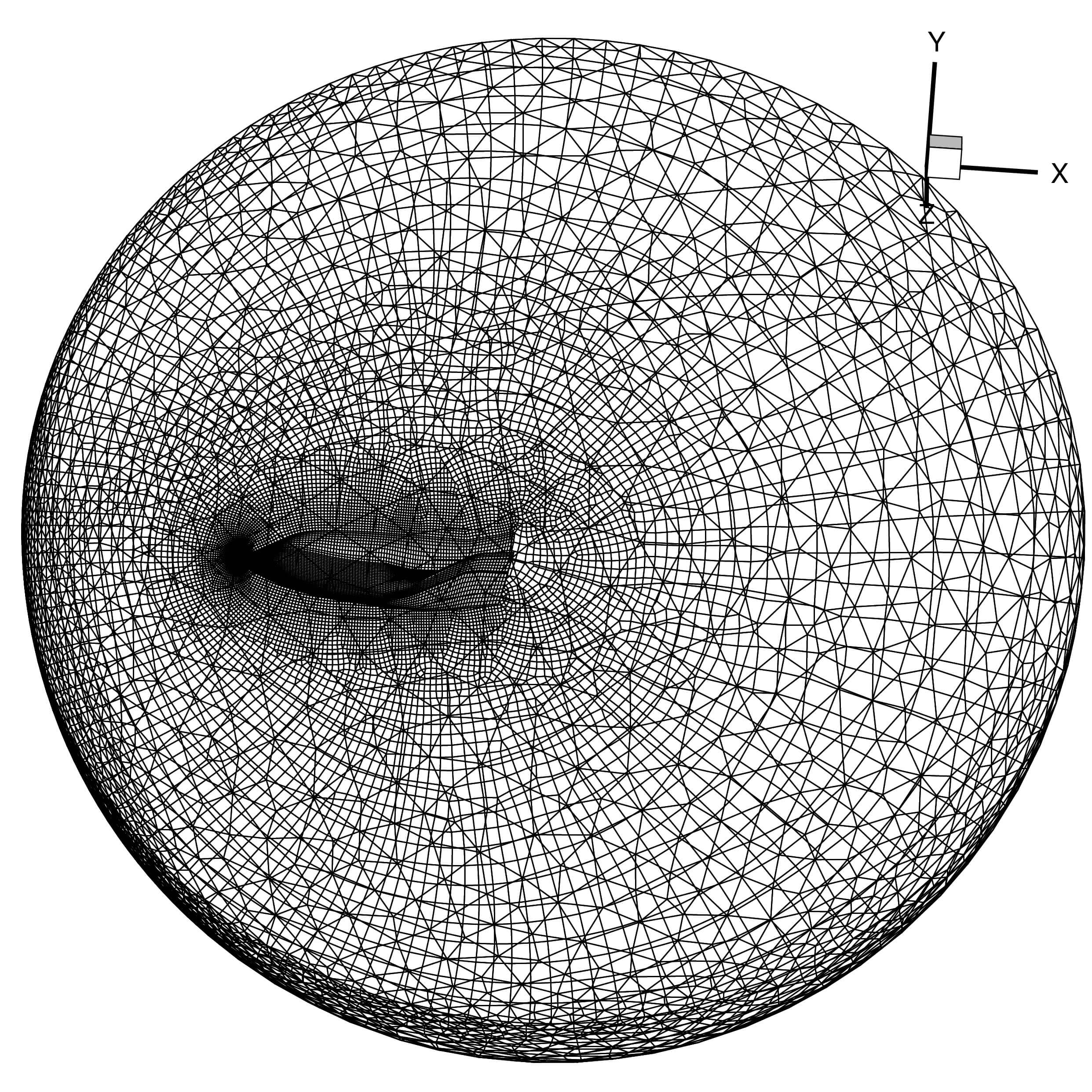}
        \caption{}
        \label{fig:X-38_global_mesh}
    \end{subfigure}
    \caption{Surface mesh of space vehicle (a) local enlargement, (b) global view} \label{fig:X-38_mesh}
\end{figure}

The last example is hypersonic flow at Mach numbers $6$ and $10$ over a space vehicle in the transition regimes $Kn = 10^{-3}$. This case shows the efficiency and capability of UGKWP for simulating three-dimension hypersonic flow over complex geometry configuration. The angle of attack is $20^{\circ}$ degrees in this case. As seen in \cref{fig:X-38_mesh}, the unstructured mesh of $560593$ cells consists of $15277$ pyramids and $545316$ tetrahedra with minimum cell height $0.001 L$ near the front of the vehicle surface. The reference length for the definition of Knudsen number is $L=0.28m$. The boundary condition on the vehicle surface is a diffusive one, on which the temperature maintains at $T_w = 300K$. Due to the symmetry, only half of the vehicle is simulated. The inflow is monatomic argon gas with molecular mass $m=6.63 \times 10^{-26} kg$ and diameter $d=4.17 \times 10^{-10} m $ at $T_\infty =300K$. The CFL number for the simulation is $0.95$, and the reference viscosity is given by the variable hard sphere (VHS) model with $\omega = 0.81$. The least square reconstruction with Venkatakrishnan limiter is used in the simulation.

\begin{figure}[htbp!]
    \centering
    \begin{subfigure}[b]{0.49\textwidth}
        \includegraphics[width=\textwidth]{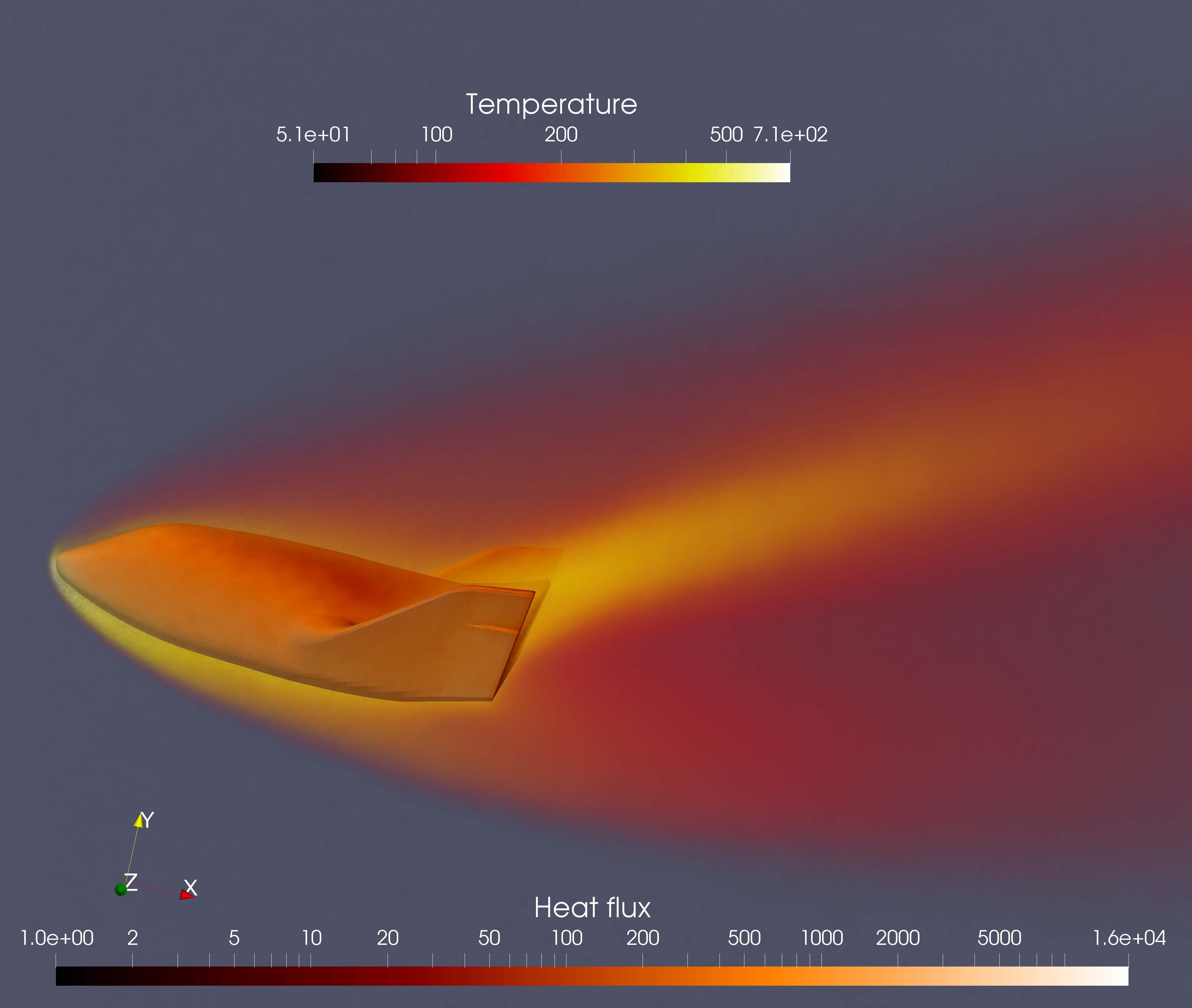}
        \caption{}
        \label{fig:paraview_X-38_Ma6_Kn1e-3_Temperature}
    \end{subfigure}
    \begin{subfigure}[b]{0.49\textwidth}
        \includegraphics[width=\textwidth]{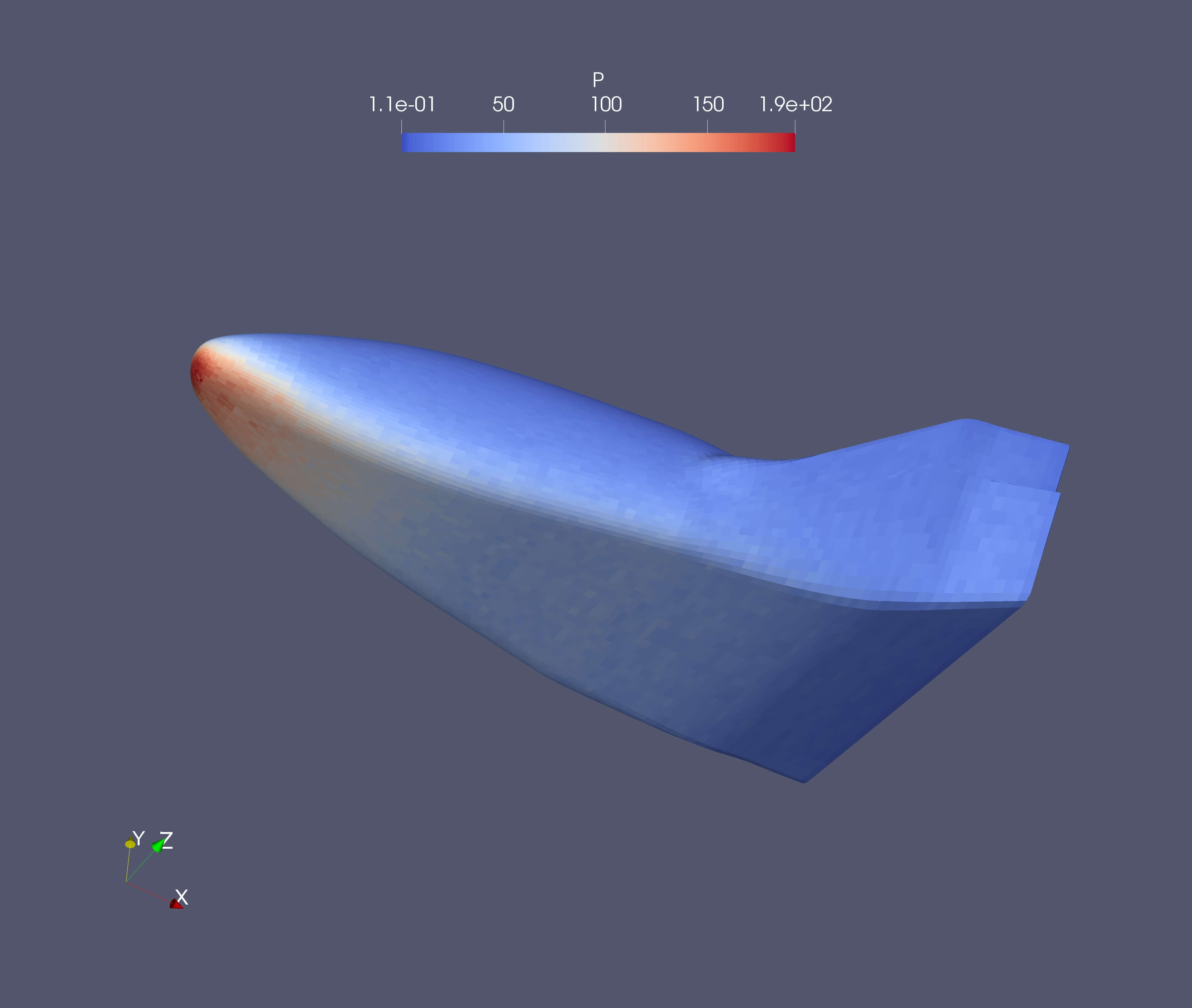}
        \caption{}
        \label{fig:paraview_X-38_Ma6_Kn1e-3_P}
    \end{subfigure}\\
    \begin{subfigure}[b]{0.49\textwidth}
        \includegraphics[width=\textwidth]{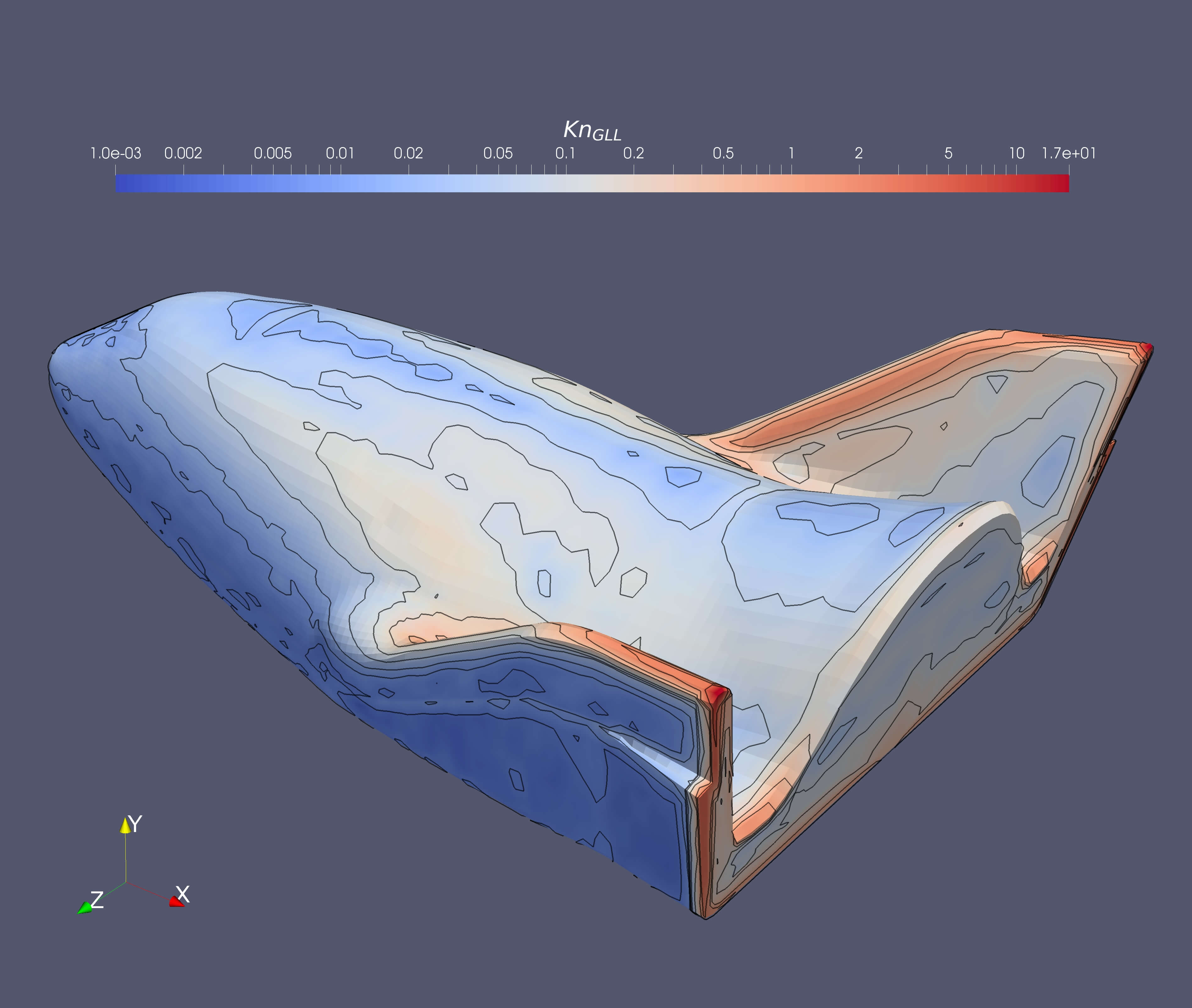}
        \caption{}
        \label{fig:paraview_X-38_Ma6_Kn1e-3_kn}
    \end{subfigure}
    \begin{subfigure}[b]{0.49\textwidth}
        \includegraphics[width=\textwidth]{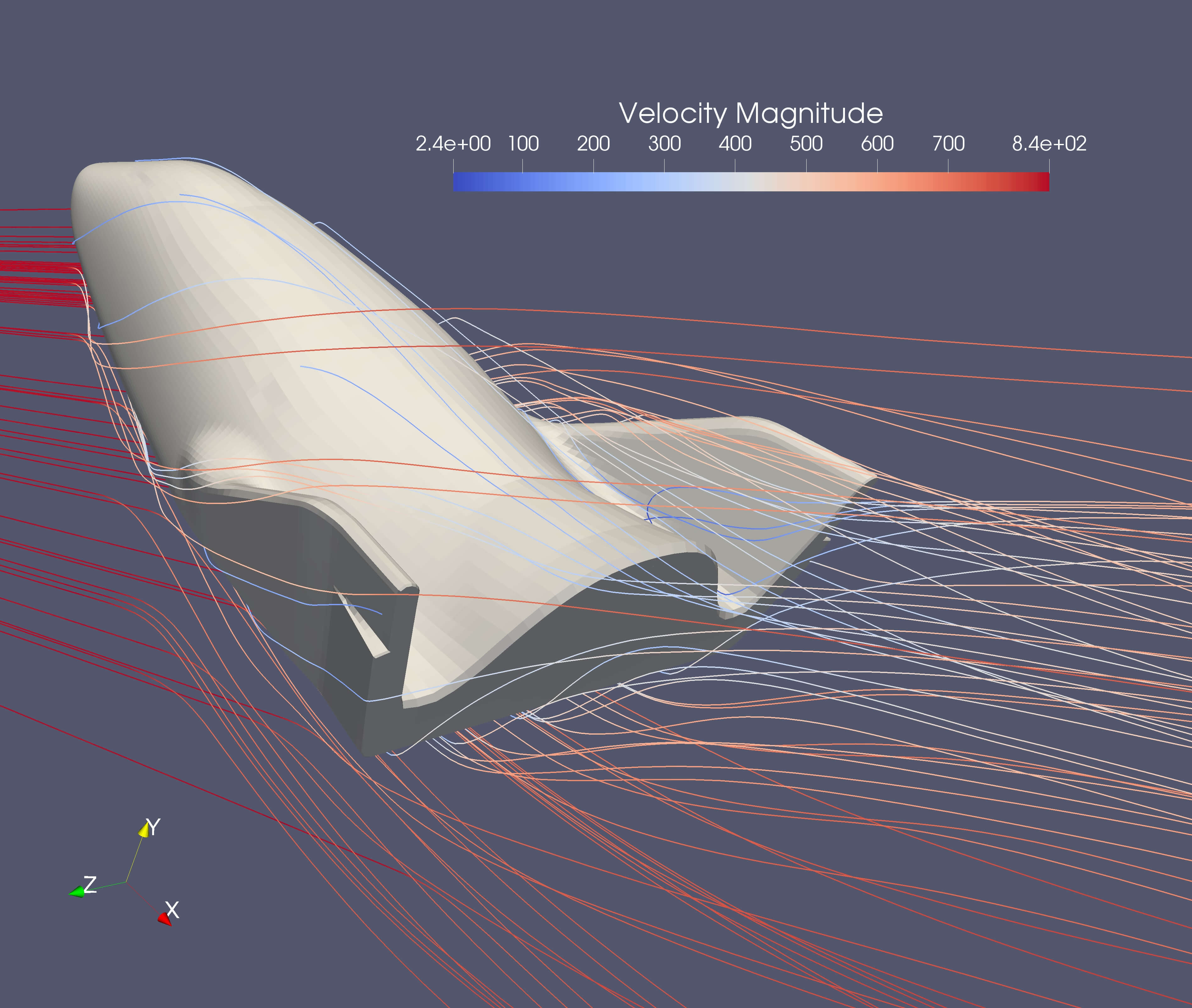}
        \caption{}
        \label{fig:paraview_X-38_Ma6_Kn1e-3_Streamlines}
    \end{subfigure}
    \caption{Space vehicle at $Ma = 6$ and $Kn = 10^{-3}$. (a) Temperature and surface distribution of heat flux, (b) Pressure distribution, (c) Kundsen number distribution, (d) Streamlines color by magnitude of velocity.}\label{fig:paraview_X-38_Ma6_Kn1e-3}
\end{figure}

\begin{figure}[htbp!]
    \centering
    \begin{subfigure}[b]{0.49\textwidth}
        \includegraphics[width=\textwidth]{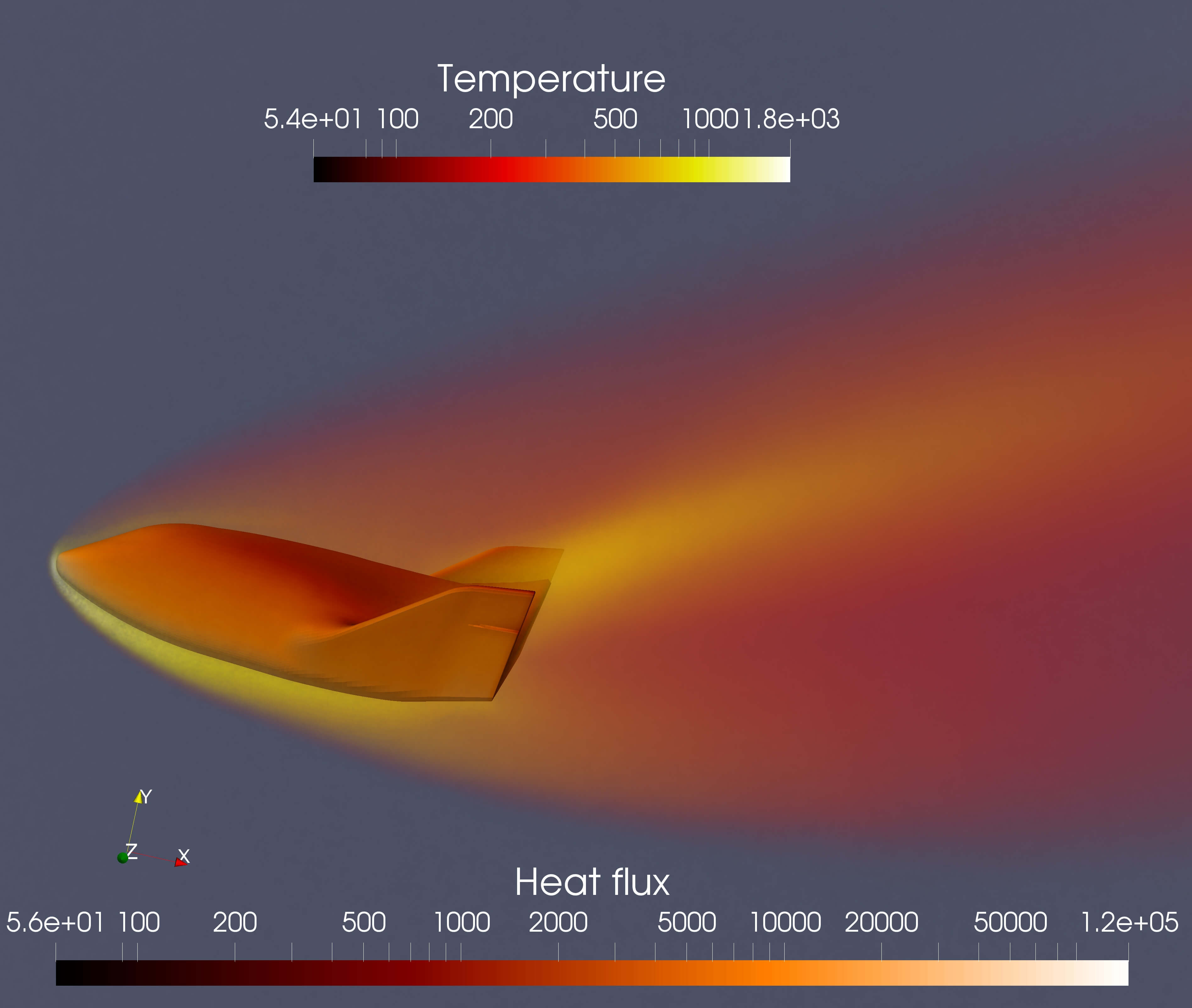}
        \caption{}
        \label{fig:paraview_X-38_Ma10_Kn1e-3_Temperature}
    \end{subfigure}
    \begin{subfigure}[b]{0.49\textwidth}
        \includegraphics[width=\textwidth]{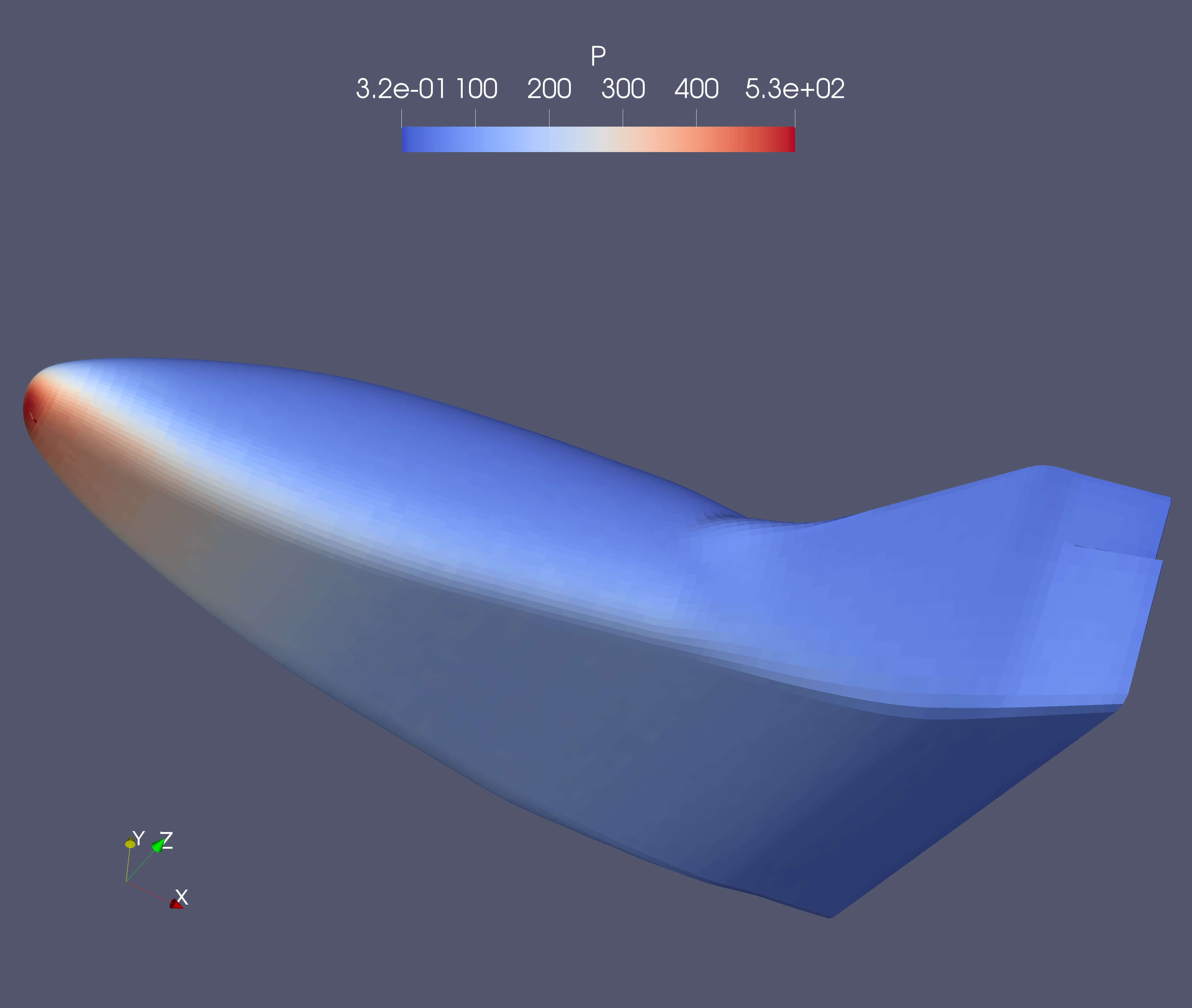}
        \caption{}
        \label{fig:paraview_X-38_Ma10_Kn1e-3_P}
    \end{subfigure}\\
    \begin{subfigure}[b]{0.49\textwidth}
        \includegraphics[width=\textwidth]{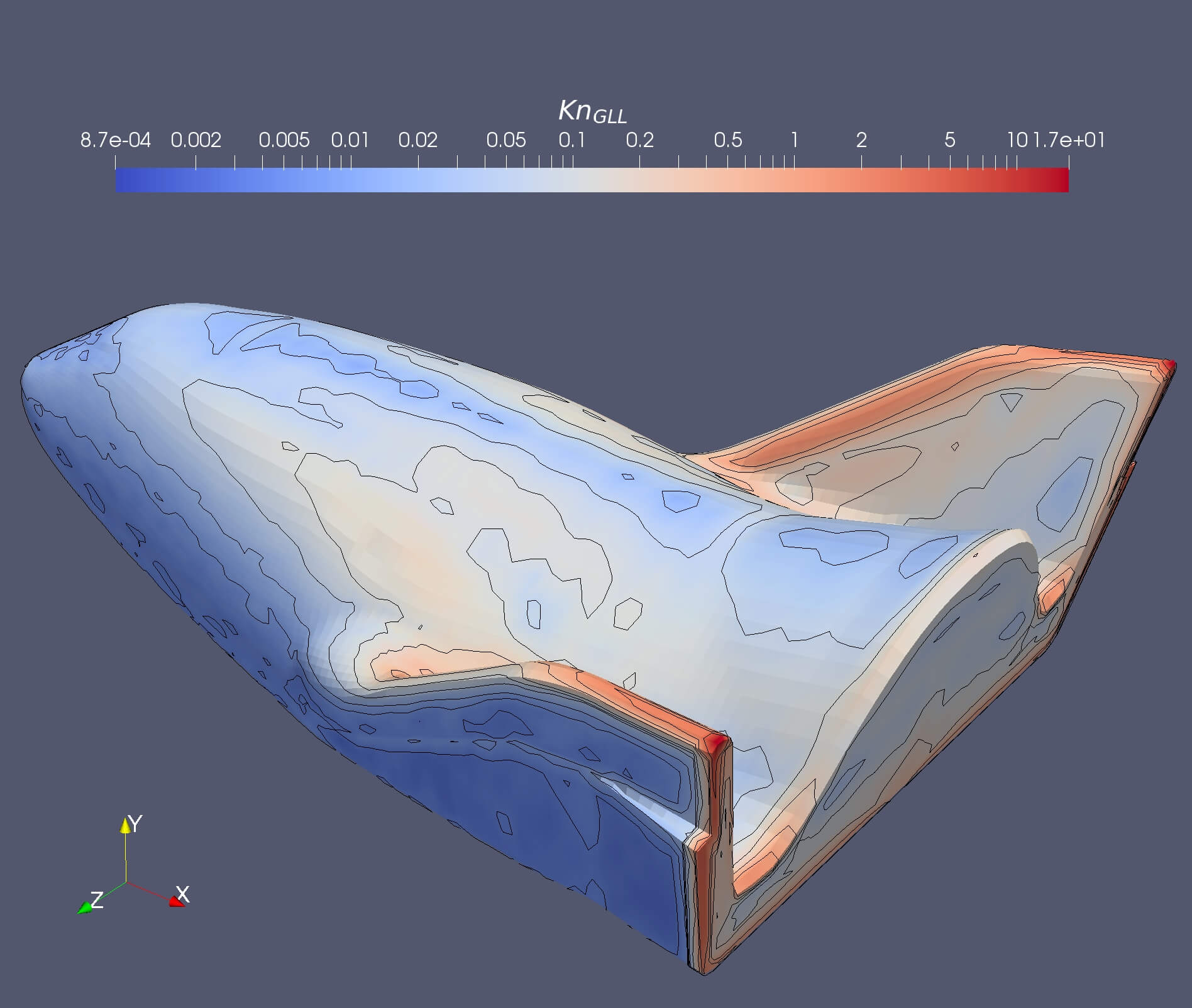}
        \caption{}
        \label{fig:paraview_X-38_Ma10_Kn1e-3_kn}
    \end{subfigure}
    \begin{subfigure}[b]{0.49\textwidth}
        \includegraphics[width=\textwidth]{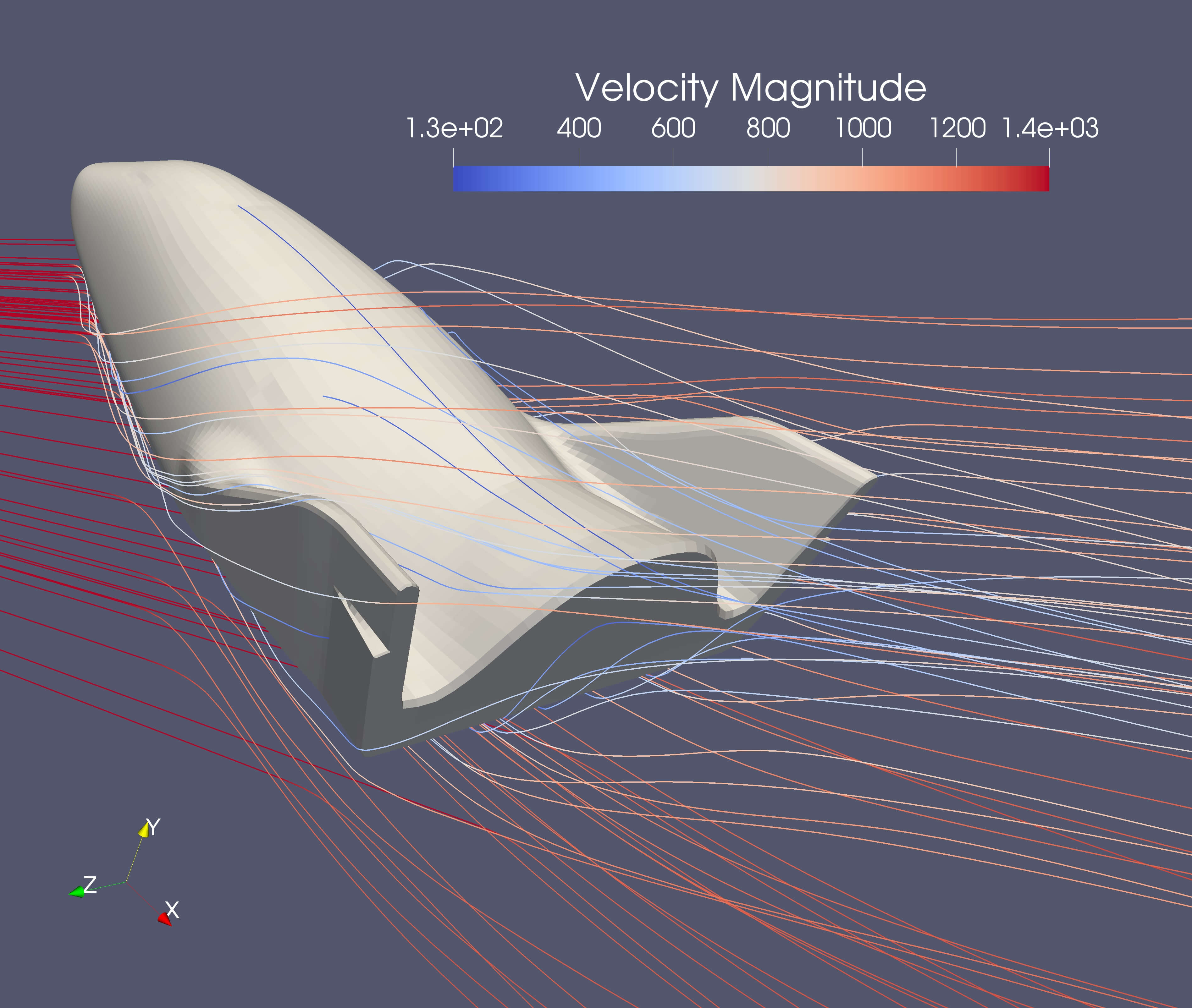}
        \caption{}
        \label{fig:paraview_X-38_Ma10_Kn1e-3_Streamlines}
    \end{subfigure}
    \caption{Space vehicle at $Ma = 10$ and $Kn = 10^{-3}$. (a) Temperature and surface distribution of heat flux, (b) Pressure distribution, (c) Kundsen number distribution, (d) Streamlines color by magnitude of velocity.}\label{fig:paraview_X-38_Ma10_Kn1e-3}
\end{figure}

\Cref{fig:paraview_X-38_Ma6_Kn1e-3} presents the distribution of temperature, heat flux, pressure, local Knudsen number, and streamlines around the vehicle at Mach number $6$. \Cref{fig:paraview_X-38_Ma10_Kn1e-3} shows the solutions at Mach number $10$. Even the free-stream Knudsen number is relatively small, no vortex flow is observed in the rear part of the vehicle, see   \cref{fig:paraview_X-38_Ma6_Kn1e-3_Streamlines,fig:paraview_X-38_Ma10_Kn1e-3_Streamlines}, which is observed in the simulation of near continuum flow \cite{jiang2019implicit}. Meanwhile, from \cref{fig:paraview_X-38_Ma6_Kn1e-3_kn,fig:paraview_X-38_Ma10_Kn1e-3_kn}, the density-based local Knudsen number $Kn_{GLL}$ can cover a wide range of values with five orders of magnitude difference. Therefore, a multi-scale method, like UGKWP, is necessary to capture the flow physics in different regimes correctly. As presented in \cref{fig:paraview_X-38_Ma6_Kn1e-3_Temperature,fig:paraview_X-38_Ma10_Kn1e-3_Temperature}, a high-temperature region is detected at the leeward side despite the low intensity of heat exchange upon vehicle surface. This is mainly caused by particle collisions in the strong recompression region with a relatively low free-stream Knudsen number.

As for the computational cost, for the $Ma = 6$ case, the initial field is obtained by GKS with $6000$ local time stepping, and the time-averaging starts after $12000$ steps of UGKWP computation. The simulation runs $24.87$ hours with $48$ cores and consumes $35$ GB memory, including $8000$ steps of averaging. For the case of $Ma = 10$, $N_{ref} = N_{min} = 400$ particles is used. The simulation is conducted on Tianhe-2 with $8$ nodes or $192$ cores, and it takes $22.8$ hours, including $5000$ steps GKS calculation with local time stepping for the initial field, and $9000$ steps of time averaging after $10000$ steps UGKWP calculation for the steady-state solution.

\section{Conclusion and further improvements}\label{sec:Conclusion and further improvements}
In this paper, a unified gas-kinetic wave-particle (UGKWP) method is constructed on three-dimensional unstructured mesh with parallel computing on supercomputer. The scheme is validated for flow simulation in both continuum and rarefied regimes at different flow speeds. Compared with other popular flow solvers, such as the DSMC method and the deterministic DVM-based Boltzmann solver, the UGKWP has multiscale property and is efficient in simulating 3D supersonic/hypersonic flow, especially in the transition and near continuum flow regime. However, for practical engineering applications, further optimization and extension have to be implemented in the code to make it more efficient and comprehensive. It is expected that the dynamic load balancing implementation can enhance the parallel efficiency in rarefied regime considerably as in conventional DSMC implementation. Moreover, coalescing of small weight particles can also save the memory substantially in the rarefied regime. Besides, the improved sampling technique can be employed to moderate the noise for low-speed or small temperature variance simulations. On the physical modeling side, instead of the BGK, Shakhov or ellipsoidal statistical model for monatomic gas and Rykov model\cite{rykov1975model} for diatomic gas can also be used in the construction of UGKWP in 3D computation. Further extensions can be the coupling between the UGKWP particle re-sampling with DSMC collision model, and the including of complex physical processes, such as ionization and chemical reaction, for the high speed and high temperature flow.

\section*{Author's contributions}

All authors contributed equally to this work.

\section*{Acknowledgments}
This work was supported by Hong Kong research grant council (16206617), National Natural Science Foundation of China (Grant Nos. 11772281, 91852114), and the National Numerical Windtunnel project.

\section*{Data Availability}

The data that support the findings of this study are available from the corresponding author upon reasonable request.





\bibliographystyle{elsarticle-num}
\section*{\refname}
\bibliography{ugkwp3D}







\end{document}